\documentclass[11pt]{article}
\newcommand{\jmp}{J. Math. Phys.~}

\newcommand{\pra}{Phys. Rev. A~}

\usepackage{young}
\usepackage[latin1]{inputenc}
\usepackage{appendix}
\usepackage{epstopdf}
\usepackage{xcolor}
\usepackage{subfigure}
\usepackage[T1]{fontenc}
\usepackage[sc]{mathpazo}
\usepackage{amsmath}
\usepackage{amssymb}
\usepackage{graphicx,color}
\usepackage{framed}
\usepackage{multirow}
\usepackage{enumerate}
\usepackage{amsthm}% theorems and proofs
\usepackage{amsfonts,mathrsfs}
\usepackage{geometry} % allows easy specifying of the page layout
\usepackage{eepic}
\usepackage{ifthen}
\usepackage[vcentermath]{youngtab}
\usepackage[unicode=true,pdfusetitle, bookmarks=true,bookmarksnumbered=false,bookmarksopen=false, breaklinks=false,pdfborder={0 0 0},backref=false,colorlinks=false] {hyperref}
\hypersetup{
colorlinks,linkcolor=myurlcolor,citecolor=myurlcolor,urlcolor=myurlcolor}
\definecolor{myurlcolor}{rgb}{0,0,0.7}% Color highlighted for numbers in citations and theorems

\usepackage{color}

\newcommand{\blue}{\textcolor{blue}}

\newcommand{\proj}[1]{| #1\rangle\!\langle #1 |}

\usepackage{hyperref}
\hypersetup{pdfpagemode=UseNone}

\newcommand{\tinyspace}{\mspace{1mu}}

\newcommand{\norm}[1]{\left\lVert\tinyspace #1 \tinyspace\right\rVert}

\def\ee{\mathrm{e}}

\def \dif {\mathrm{d}}
\def \diag {\mathrm{diag}}

\def\I{\mathbb{1}}

\def\bdlambda{\boldsymbol{\lambda}}
\def\bdalpha{\boldsymbol{\alpha}}
\def\bdbeta{\boldsymbol{\beta}}

\newenvironment{mylist}[1]{\begin{list}{}{
    \setlength{\leftmargin}{#1}
    \setlength{\rightmargin}{0mm}
    \setlength{\labelsep}{2mm}
    \setlength{\labelwidth}{8mm}
    \setlength{\itemsep}{0mm}}}
    {\end{list}}

\renewcommand{\baselinestretch}{1.3}

\def\ot{\otimes}

\newcommand{\out}[2]{| #1\rangle\langle #2 |}
\newcommand{\Inner}[2]{\left\langle #1 , #2\right\rangle}
\newcommand{\Innerm}[3]{\left\langle #1 \left| #2 \right| #3 \right\rangle}

\newcommand{\Pa}[1]{\left(#1\right)}

\newcommand{\Br}[1]{\left[#1\right]}
\newcommand{\set}[1]{\{#1\}}
\newcommand{\Set}[1]{\left\{#1\right\}}

\newcommand{\ket}[1]{|#1\rangle}

\DeclareMathOperator{\trace}{Tr}

\newcommand{\Ptr}[2]{\trace_{#1}\Pa{#2}}

\newcommand{\Tr}[1]{\Ptr{}{#1}}

\def\cG{\mathcal{G}}\def\cH{\mathcal{H}}\def\cJ{\mathcal{J}}

\def\cR{\mathcal{R}}

\def\bsE{\boldsymbol{E}}

\def\bsP{\boldsymbol{P}}\def\bsT{\boldsymbol{T}}
\def\bsU{\boldsymbol{U}}
\def\bsZ{\boldsymbol{Z}}
\def\bsa{\boldsymbol{a}}\def\bsb{\boldsymbol{b}}\def\bsc{\boldsymbol{c}}\def\bse{\boldsymbol{e}}
\def\bsh{\boldsymbol{h}}
\def\bsk{\boldsymbol{k}}
\def\bsp{\boldsymbol{p}}\def\bsq{\boldsymbol{q}}\def\bsr{\boldsymbol{r}}
\def\bsu{\boldsymbol{u}}\def\bsv{\boldsymbol{v}}

\def\rD{\mathrm{D}}

\def\sS{\mathscr{S}}

\def\F{\textsf{F}}                  

\def\Q{\textsf{Q}}\def\S{\textsf{S}}

\def\bbC{\mathbb{C}}\def\bbE{\mathbb{E}}
                  
\def\bbN{\mathbb{N}}

\def\sfQ{\mathsf{Q}}\def\sfS{\mathsf{S}}

\newtheorem{thrm}{Theorem}[section]
\newtheorem{lem}[thrm]{Lemma}
\newtheorem{prop}[thrm]{Proposition}
\newtheorem{cor}[thrm]{Corollary}
\theoremstyle{definition}

\newtheorem{remark}[thrm]{Remark}

\numberwithin{equation}{section}

\newcounter{questionnumber}

\begin{document}

\title{A sharp comparison of subentropy and linear-entropy gaps}
\author{\blue{Lin Zhang}$^{1}$\footnote{E-mail: godyalin@163.com}, \quad \blue{Wenlin Zhang}$^{2}$\footnote{E-mail: e1327962@u.nus.edu}, \quad\blue{Haobo Ma}$^{3}$\footnote{E-mail: auric@aelf.io}\\
  {\it\small $^{1}$School of Mathematical Sciences, Hangzhou Dianzi University, Hangzhou 310018, China}\\
{\it\small $^{2}$The Omega Institute, National University of Singapore, Singapore}\\
{\it\small $^{3}$The Omega Institute, ChronoAI Pte Ltd, Singapore}
}

\date{}
\maketitle

\begin{abstract}
We prove that, for every multipartite quantum state on an arbitrary finite collection of finite-dimensional Hilbert spaces, the subentropy subadditivity gap is at least one third of the linear-entropy subadditivity gap. The coefficient $1/3$ is optimal whenever at least two local dimensions are at least two, and equality in both inequalities holds for exactly the same product states: those with at most one mixed factor, or with exactly two mixed factors that are maximally mixed on two-dimensional supports while all other factors are pure. For probability tables, merging the two smallest rows, convexity at fixed marginals, endpoint estimates, and quantitative Schur concavity reduce the comparison to the quantum setting. We also establish the sharp multipartite linear-entropy comparison with coefficient one half, and complete the single-state comparison into a sharp two-sided sandwich with constant $2\log 2-1$ on the lower side and one half of the linear entropy on the upper side, the latter expressed through a concave function of the square root of purity. Finally, we determine the maximum subentropy difference at every fixed trace distance in each dimension at least two, classify all attaining pairs, and obtain a uniform dimension-free bound given by the trace distance minus half its square.
\end{abstract}

\newpage
\tableofcontents
\newpage

%===============================%
\section{Introduction}
%===============================%

Subentropy was introduced by Jozsa, Robb and Wootters as a tight lower bound on the accessible information of pure-state ensembles with a fixed average state \cite{Jozsa1994}. Nichols and Wootters later embedded subentropy within a one-parameter entropy family interpolating between subentropy and von Neumann entropy \cite{Nichols2003}. Subsequent work by Datta, Dorlas, Jozsa and Benatti established product-state subadditivity and a Fannes-type continuity bound for subentropy, and raised the question of whether such subadditivity extends to arbitrary bipartite states \cite{Datta2014}. General subadditivity for the Nichols-Wootters entropy family was proved in \cite{Zhang2026}, while Patil \cite[Theorem~2]{Patil2026} obtained only finite-dimensional subadditivity of subentropy via a measurement comparison argument with different method. In \cite[Theorems~3.12 and~3.15]{Zhang2026}, Zhang, Zhang, and Ma established, for arbitrary finite-dimensional settings, a nontrivial quantitative lower bound for the subadditivity of the Nichols-Wootters entropy family, which also yields a quantitative lower bound for subentropy subadditivity. They further obtained a sharp two-qubit bound in terms of the smallest marginal eigenvalues and a dimension-dependent lower bound in terms of the two marginal linear entropies. Note that the linear entropy is the order-two Tsallis entropy, whose subadditivity follows from \cite{Audenaert2007}; sharp dimension-dependent relations between joint and marginal linear entropies were derived in \cite{Morelli2020}.

Building on these developments, the present paper undertakes a comparative study of subentropy subadditivity and linear-entropy subadditivity in finite dimensions. For $\sigma\in\rD(\bbC^d)$, the set of all density matrices on $\bbC^d$, the subentropy and the linear entropy are defined, respectively, by
$$
\Q(\sigma) = -\Q(\bdlambda(\sigma)) = \sum^{d}_{i=1} \frac{\lambda^d_j}{\prod_{j\in\hat i}(\lambda_i-\lambda_j)}\log\lambda_j,\qquad \S_L(\sigma)=1-\Tr{\sigma^2}.
$$
Here $\bdlambda(\sigma)$ denotes the vector of eigenvalues of $\sigma$. Given $N\geqslant2$ and any state
$\rho$ on the finite tensor product $\bigotimes_{i=1}^N\cH_i$ , write $\rho_i=\Ptr{\hat i}{\rho}$, where $\hat i=\set{1,\ldots,N}\setminus\set{i}$. We introduce the subadditivity gaps $\Delta_{\Q}$ and
$\Delta_L$  for subentropy and linear entropy:
\begin{eqnarray}
\Delta_{\Q}:= \sum^N_{i=1}\Q(\rho_i) - \Q(\rho) ,\qquad \Delta_{L} :=  \sum^N_{i=1}\S_L(\rho_i) - \S_L(\rho).
\end{eqnarray}
Our main result establishes the inequality
\begin{eqnarray}
\Delta_{\Q}(\rho)\geqslant  \frac13\max\Set{\Delta_L(\rho), \Delta_L\Pa{\bigotimes_{i=1}^N\rho_i}},
\end{eqnarray}
where
\begin{eqnarray}
\Delta_L\Pa{\bigotimes_{i=1}^N\rho_i}=\sum_{i=1}^N \S_L(\rho_i) - 1 + \prod_{i=1}^N(1-\S_L(\rho_i))
\end{eqnarray}
reduces to $\S_L(\rho_A)\S_L(\rho_B)$ for $N=2$. The constant $1/3$ is optimal provided at least two local dimensions are no smaller than $2$, and the two terms inside the maximum share exactly the same multipartite equality set. The first term depends on the joint purity of $\rho$, while the second depends only on the purities of the individual marginal states.

Remarkably, subentropy subadditivity and linear-entropy subadditivity share identical saturation conditions. This coincidence motivates our main comparison, and it follows from two known results summarized below:
\begin{enumerate}[(i)]
\item \textbf{Saturation of Nichols-Wootters family of entropies  \cite{Zhang2026}:} For any bipartite state $\rho_{AB}\in\rD(\bbC^m\ot\bbC^n)$, and $2\leqslant k\leqslant d:=mn\leqslant D$, we have
\begin{eqnarray*}
\S^{(D)}_k(\rho_A) + \S^{(D)}_k(\rho_B) - \S^{(D)}_k(\rho_{AB})&\geqslant& \frac{k(k-1)}{D(D-1)}\Pa{\Q(\rho_A)+\Q(\rho_B)-\Q(\rho_{AB})},\\
\Q(\rho_A)+\Q(\rho_B)-\Q(\rho_{AB})&\geqslant& \frac{\S_L(\rho_A)\S_L(\rho_B)}{8d(d+1)(d+2)(d+3)}.
\end{eqnarray*}
Moreover $\S^{(d)}_k(\rho_A) + \S^{(d)}_k(\rho_B) = \S^{(d)}_k(\rho_{AB})$ if and only if $\Q(\rho_A)+\Q(\rho_B)=\Q(\rho_{AB})$, which is valid if and only if at least one of the marginals $\rho_A$ and $\rho_B$ is pure.
\item \textbf{Saturation of linear-entropy subadditivity \cite{Guo2020}:} For any bipartite state $\rho_{AB}\in\rD(\bbC^m\ot\bbC^n)$, we have
$$
1+\max\Set{\Tr{\rho^2_A},\Tr{\rho^2_B}}\Tr{\rho^2_{AB}}\geqslant
\Tr{\rho^2_A}+\Tr{\rho^2_B}.
$$
Moreover $1+\Tr{\rho^2_{AB}}= \Tr{\rho^2_A}+\Tr{\rho^2_B}$, equivalently $\S_L(\rho_A)+\S_L(\rho_B)=\S_L(\rho_{AB})$, if and
only if at least one of the marginals $\rho_A$ and $\rho_B$ is pure.
\end{enumerate}
The proof of our main inequality centers on Lemma~\ref{lem:diagonal-core}. We analyze the behaviour of the two subadditivity gaps under row merging on joint probability tables\footnote{A nonnegative table is a matrix (rectangular table of numbers) where every entry is nonnegative: $\bsT=(t_{ij})$ for $t_{ij}\geqslant0$ for all $i,j$. A probability table is a type of nonnegative table $\bsP=(p_{ij})$ with $\sum_{i,j}p_{ij}=1$.}. A determinant bound from \cite[Proposition~4]{Patil2026} controls the monotonicity of the subentropy gap under such merging, and a sharp estimate allows us to reduce general finite tables to the $2\times 2$ case, where equality conditions are analyzed in Section~\ref{sec:diagonal-proof}. The marginal-product bound relies on constrained minimization and finite-interval estimates collected in Appendix~\ref{app:sharp-marginal-product}. Induction extends this bipartite comparison to the full multipartite setting via Theorem~\ref{th:infinite-sharp}. The result is lifted from classical distributions to quantum states by dephasing in marginal eigenbases, exploiting the quantitative Schur concavity of
$\Q-\frac13\S_L$ (Proposition~\ref{prop:mixing}), which we refine with an explicit spectral margin. The limiting constants $1/3$ and $1/2$ are attained asymptotically near rank-2 uniform spectra and along uniform spectra of increasing rank, respectively.

We further establish three complementary structural results. First, Proposition~\ref{prop:single-state-sharp} provides sharp two-sided bounds
\begin{eqnarray}
(2\log2-1)\S_L(\rho)\leqslant\Q(\rho) \leqslant\Q^\circ\!\left(\sqrt{1-\S_L(\rho)}\right) \leqslant\frac12\S_L(\rho),
\end{eqnarray}
showing that subentropy and linear entropy of a single state are globally
equivalent up to a constant factor below $1.30$. By contrast, the ratio $\Delta_{\Q}/\Delta_L$ of their subadditivity gaps is bounded below by $1/3$ yet unbounded above. Second, Proposition~\ref{prop:linear-marginal-product} proves a sharp multipartite comparison for linear entropy:
\begin{eqnarray}
\Delta_L(\rho)\geqslant\frac12\Delta_L\Pa{\bigotimes^N_{i=1}\rho_i}.
\end{eqnarray}
Third, building on the extremal-spectrum analysis of Hanson and Datta \cite[Corollary~6.2]{HansonDatta2021}, Patil \cite[Section~4.1, Theorem~3]{Patil2026} characterized the maximal subentropy difference subject to a fixed trace-distance constraint; this maximum plateaus at $1-\frac1d$. Theorem~\ref{th:finite-continuity} determines the maximal value and the optimizing state pairs at every prescribed trace distance, including beyond this threshold. Majorization together with Lemma~\ref{lem:diagonal-core} yields a correction term governed by the overlap between ordered spectra. For distances larger than $1-\frac1d$ a separate estimate involving the minimal eigenvalue captures the decreasing branch of the exact distance curve. The dimension-independent bound $T-\frac12T^2$ also follows as a consequence.

%====================================================%
\section{Preliminaries}\label{sec:preliminaries}
%====================================================%

For a finite probability vector $\bdlambda$, we use the
determinant representation of subentropy~\cite[Eq.~(14)]{Jozsa2015},
with the reciprocal change of integration variable:
\begin{equation}\label{eq:determinant-integral}
 \Q(\bdlambda)=\int_0^\infty \Pa{\frac{1}{1+t}-\frac1{\prod_i(1+\lambda_i t)}}
 \frac{\dif t}{t^2}.
\end{equation}
For a finite-dimensional quantum state $\rho$, set
$\Q(\rho)=\Q(\bdlambda(\rho))$. Zero coordinates leave the
integral unchanged.

If the $Z_i$ are independent random variables with the unit exponential
law and $X_{\bsp}=\sum_i p_iZ_i$, then, with Euler's constant $\gamma$,
\begin{equation}\label{eq:gaussian}
 \Q(\bsp)=1-\gamma + \bbE[\eta(X_{\bsp})],\qquad \eta(x):=-x\log x.
\end{equation}
Indeed, for $x>0$,
\begin{eqnarray*}
 -\eta(x)-x+1 =\int_0^\infty   \Pa{\ee^{-tx}-\ee^{-t}+(x-1)t\ee^{-t}}t^{-2}\dif t.
\end{eqnarray*}
The identity follows by differentiating twice in $x$ and matching the
value and first derivative at $x=1$. Both sides extend to $x=0$.
The integrand is nonnegative by convexity in $x$.
Taking expectations for $X_{\bsp}$ and for one unit exponential, and
subtracting, gives \eqref{eq:gaussian}. In a fixed finite
simplex the expectation is continuous by dominated convergence,
because $0\leqslant X_{\bsp}\leqslant \sum_i Z_i$ and the latter sum has
finite second moment.

For the vector $\bsu_d$ uniform on $d$ coordinates, $dX_{\bsu_d}$
has the $\operatorname{Gamma}(d,1)$ law. Its logarithmic moment is
$-\bbE[\eta(dX_{\bsu_d})]=d(H_d-\gamma)$, where
$H_d=\sum_{j=1}^d1/j$. This follows by differentiating
$\bbE[(dX_{\bsu_d})^s]=\Gamma(d+s)/\Gamma(d)$ at $s=1$.
Thus $\Q(\bsu_d)=1+\log d-H_d$. For $b\in[0,1]$, we also write
\begin{equation}\label{eq:diffuse-atom-fin}
 \Q^\circ(b):=1-\gamma + \bbE[\eta(bZ+1-b)]  = \int_0^\infty  \Pa{\frac{1}{1+t}-\frac{\ee^{-(1-b)t}}{1+bt}}   \frac{\dif t}{t^2},
\end{equation}
where $Z$ is a unit exponential random variable; it represents the
large-dimension limit of $\Q(b,(1-b)/K,\ldots,(1-b)/K)$ as $K\to\infty$,
with $\Q^\circ(1)=0$ and $\Q^\circ(0)=1-\gamma$.

We use natural logarithms throughout. For a state $\rho$ let
$\S_L(\rho)=1-\Tr{\rho^2}$. For a probability vector $\bdlambda$ put
$\S_L(\bdlambda)=1-\sum_i\lambda_i^2$. For a multipartite state $\rho$ on
$\bigotimes_{i=1}^N\cH_i$ with single-party marginals $\rho_i$, we
write
\begin{eqnarray*}
 \Delta_{\Q}(\rho)&=&\sum_{i=1}^N\Q(\rho_i)-\Q(\rho),\qquad \Delta_L(\rho)=\sum_{i=1}^N \S_L(\rho_i)- \S_L(\rho),\\
 \Delta_L\Pa{\bigotimes_{i=1}^N\rho_i} &=&\sum_{i=1}^N \S_L(\rho_i)-1+\prod_{i=1}^N(1-\S_L(\rho_i)).
\end{eqnarray*}
For $N=2$, $\Delta_L(\rho_A\ot\rho_B)=\S_L(\rho_A)\S_L(\rho_B)$.
The same notation applies to a probability table $\bsP=(p_{ij})$,
with row and column marginals $\bdalpha=(\sum_j p_{ij})_i$ and
$\bdbeta=(\sum_i p_{ij})_j$ in place of $\rho_A$ and $\rho_B$.
We regard $\bsP$ as the probability vector of its cells. Set
$$
\cR=\Q-\frac13\S_L,\qquad \F=\Delta_{\Q}-\frac13\Delta_L.
$$
If a marginal of a bipartite table is supported on one index, the other
marginal and the table have the same nonzero entries, so both gaps
vanish.

%===========================================%
\section{Main results}\label{sec:results}
%===========================================%

\begin{lem}[Diagonal core]\label{lem:diagonal-core}
For every joint probability table $\bsP=(p_{ij})\in\Delta_{d-1}$
with $d=mn$ and marginals $\bsr=(\sum_jp_{ij})_{i=1}^m$ and
$\bsc=(\sum_ip_{ij})_{j=1}^n$, one has
\begin{eqnarray*}
 \Delta_{\sfQ}(\bsP)=\sfQ(\bsr)+\sfQ(\bsc)-\sfQ(\bsP) \geqslant
 \frac13\Pa{1-\norm{\bsr}_2^2-\norm{\bsc}_2^2    +\norm{p}_2^2}=\frac13\Delta_L(\bsP).
\end{eqnarray*}
Equality holds precisely when one marginal is pure or, after
deleting zero rows and columns, $\bsP$ is the uniform $2\times2$ table.
\end{lem}

\begin{thrm}\label{th:infinite-sharp}
Let $\cH_1,\ldots,\cH_N$ be finite-dimensional Hilbert spaces, $2\leqslant N<\infty$,
and let $\rho$ be a state on $\bigotimes_{i=1}^N\cH_i$.
Then
\begin{equation}\label{eq:infinite-sharp}
 \sum_{i=1}^N\sfQ(\rho_i)-\sfQ(\rho) \geqslant\frac13\Br{\sum_{i=1}^N\sfS_L(\rho_i)- \sfS_L(\rho)}.
\end{equation}
The coefficient $1/3$ is optimal whenever at least two local spaces have
dimension at least two. Equality holds exactly for product states
either with at most one mixed factor or with exactly two mixed
factors, both maximally mixed on two-dimensional supports, and
all other factors pure.
\end{thrm}

\begin{prop}\label{prop:single-state-sharp}
For every finite-dimensional state $\rho$,
\begin{equation}\label{eq:single-state-sharp}
 (2\log2-1)\sfS_L(\rho)\leqslant\sfQ(\rho) \leqslant\sfQ^\circ\Pa{\sqrt{1-\sfS_L(\rho)}} \leqslant\frac12\sfS_L(\rho).
\end{equation}
In the lower bound, equality holds exactly when $\rho$ is pure or maximally
mixed on a two-dimensional support, and the coefficient $2\log2-1$ is
optimal in every ambient Hilbert space of dimension at least two. Both
upper inequalities are strict whenever $\rho$ is mixed; the coefficient
$1/2$ is optimal in every ambient Hilbert space of dimension at least two,
and $\sfQ^\circ(\sqrt{1-s})$ is the sharp large-dimension supremum of
$\sfQ(\rho)$ at every fixed $s=\sfS_L(\rho)\in(0,1)$.
\end{prop}

\begin{prop}\label{prop:linear-marginal-product}
Let $\cH_1,\ldots,\cH_N$ be finite-dimensional Hilbert spaces,
$2\leqslant N<\infty$, and let $\rho$ be a state on $\bigotimes_{i=1}^N\cH_i$.
Then
\begin{equation}\label{eq:linear-marginal-product}
 \sum_{i=1}^N \sfS_L(\rho_i)-\sfS_L(\rho) \geqslant\frac12\Delta_L\Pa{\bigotimes_{i=1}^N\rho_i}
 =\frac12\Br{\sum_{i=1}^N \sfS_L(\rho_i)-1  +\prod_{i=1}^N(1-\sfS_L(\rho_i))}.
\end{equation}
Equality holds exactly when at most one marginal $\rho_i$ is mixed.
The coefficient $1/2$ is optimal whenever at least two local spaces have
dimension at least two.
\end{prop}

Combining Theorem~\ref{th:infinite-sharp} with
Proposition~\ref{prop:linear-marginal-product} gives
$$
\Delta_{\Q}(\rho)\geqslant\frac13\Delta_L(\rho)  \geqslant\frac16\Delta_L\Pa{\bigotimes_{i=1}^N\rho_i}.
$$
The bound obtained by this chain has coefficient $1/6$.
The next theorem gives the sharp coefficient $1/3$. Its bipartite
case uses the independent probability-table estimate proved in
Appendix~\ref{app:sharp-marginal-product}, with a finite interval
certificate for the remaining parameter regions.

\begin{thrm}\label{th:subentropy-marginal-product}
Let $\cH_1,\ldots,\cH_N$ be finite-dimensional Hilbert spaces,
$2\leqslant N<\infty$, and let $\rho$ be a state on $\bigotimes_{i=1}^N\cH_i$.
Then
\begin{equation}\label{eq:subentropy-marginal-product}
 \sum_{i=1}^N\sfQ(\rho_i)-\sfQ(\rho) \geqslant\frac13\Delta_L\Pa{\bigotimes_{i=1}^N\rho_i}
 =\frac13\Br{\sum_{i=1}^N \sfS_L(\rho_i)-1  +\prod_{i=1}^N(1-\sfS_L(\rho_i))}.
\end{equation}
The coefficient $1/3$ is optimal whenever at least two local spaces have
dimension at least two. Equality holds for the exact same product families
as in Theorem~\ref{th:infinite-sharp}.
\end{thrm}

The joint purity determines which multipartite bound is larger:
$$
\Delta_L(\rho)-\Delta_L\Pa{\bigotimes_{i=1}^N\rho_i}=\Tr{\rho^2}-\prod_{i=1}^N\Tr{\rho_i^2}.
$$
The two bounds agree on product states. For $N=2$, on the diagonal state
with table $(1+t,1-t,1-t,1+t)/4$ ($0<|t|\leqslant1$), both marginal
purities are $1/2$ and the joint purity is $(1+t^2)/4$, so the gap term is
larger. For the table $(1/2,1/4,1/4,0)$, the joint purity is $3/8$ while
the product of marginal purities is $25/64$, so the marginal-product term
is larger. Note also a sharp contrast between single-state and gap
comparisons: although Proposition~\ref{prop:single-state-sharp} shows that
$\Q(\rho)$ and $\S_L(\rho)$ are globally equivalent within a factor
$1/(4\log2-2)<1.30$, the gap ratio $\Delta_{\Q}/\Delta_L$ is unbounded
above; on $\rho_\varepsilon=\diag(1-2\varepsilon,\varepsilon,\varepsilon,0)$
we compute in Section~\ref{sec:single-marginal-cont-fin} that
$\Delta_{\Q}(\rho_\varepsilon)/\Delta_L(\rho_\varepsilon)
=\frac12\log(1/\varepsilon)-1+O(\varepsilon\log(1/\varepsilon))\to\infty$.

\begin{thrm}[Sharp Audenaert-type continuity]\label{th:finite-continuity}
Let $\rho,\sigma$ be states on $\bbC^d$ ($d\geqslant2$) with trace distance
$T=\frac12\norm{\rho-\sigma}_1\in[0,1]$. For $t\in[0,1]$ put
$\bsc_t=(1-t,t/(d-1),\ldots,t/(d-1))$. The maximum at fixed distance $T$ is
\begin{eqnarray}\label{eq:finite-continuity}
 \max_{\frac12\norm{\rho-\sigma}_1=T}|\sfQ(\rho)-\sfQ(\sigma)|=\sfQ(\bsc_T).
\end{eqnarray}
The maximum over distances at most $T$ is $\sfQ(\bsc_0)$ for $\bsc_0=\bsc_{\min\Set{T,1-\frac1d}}$.
This equals $\log d-H_d+1$ when $T\geqslant1-\frac1d$. Across all dimensions,
$$
|\sfQ(\rho)-\sfQ(\sigma)|\leqslant\sfQ^\circ(1-T)    \leqslant\min\Set{T-\tfrac12T^2,1-\gamma}.
$$
More sharply, if $\sfQ(\sigma)\geqslant\sfQ(\rho)$,
$t=\frac12\norm{\bdlambda^\downarrow(\rho)-\bdlambda^\downarrow(\sigma)}_1
 \leqslant T$, and
$r_1=\min\Set{\lambda_{\max}(\rho),\lambda_{\max}(\sigma)}$, then
\begin{eqnarray}\label{eq:finite-continuity-refined}
 \sfQ(\sigma)-\sfQ(\rho)\leqslant  \sfQ\Pa{1-t,\frac{t}{d-1},\ldots,\frac{t}{d-1}} -\frac23t(1-t-r_1).
\end{eqnarray}
At $T=0$, every identical pair attains the maximum. At every $T>0$,
the maximizing pairs in \eqref{eq:finite-continuity} are precisely, up to
exchange,
$$\rho=\proj{\psi},\qquad  \sigma=(1-T)\proj{\psi}+\frac{T}{d-1}(\I_d-\proj{\psi}),$$
where $\ket{\psi}$ is a unit vector.
\end{thrm}

%=============================================%
\section{Proof strategy}\label{sec:proof-strategy}
%=============================================%

\subsection{The joint linear-entropy gap}

For a probability table $\bsP$, put 
$$
\F(\bsP)=\Delta_{\Q}(\bsP)-\frac13\Delta_L(\bsP).
$$
If $\bsa=(a_j)$ and $\bsb=(b_j)$ are its two smallest positive rows,
of masses $A$ and $B$, and $\bsP^{\rm m}$ is the table after merging them,
then
\begin{eqnarray*}
\F(\bsP) - \F(\bsP^{\rm m})  \geqslant  \Pa{243\log 2-\frac{70741}{420}}  \sum_{j\ne\ell}a_jb_\ell.
\end{eqnarray*}
The loss of $\Delta_L$ is exactly twice the sum in this formula.
The determinant integral controls the loss of $\Delta_{\Q}$ through
four-cell rectangles. After applying the row and column inequalities,
the remaining $2\times2$ problem is a one-variable estimate at fixed
marginals. The second derivative in the correlation variable is at
least $1/30$ for $\F$. The tangent at the product table is nonnegative
throughout the feasible interval because its two endpoint values are
nonnegative. The ten steps in Section~\ref{sec:diagonal-proof}
derive the constants and identify the zero-loss mergers.

For a bipartite quantum state, dephasing in the marginal eigenbases
gives a diagonal state $\tau$ with the same marginals. Consequently
$$
 \F(\rho)=\F(\tau)+\cR(\tau)-\cR(\rho), \qquad \cR=\Q-\tfrac13 \S_L.
$$
The scalar comparison controls the first term. Schur--Horn and the
spectral transfer bound in Proposition~\ref{prop:mixing}
control the second. At equality the spectra of $\tau$ and $\rho$
coincide. Their purity difference is the sum of the squared moduli
of all off-diagonal entries of $\rho$ in the dephasing basis, so equality
forces $\rho=\tau$. Summing over successive bipartitions gives the
multipartite inequality. Section~\ref{sec:quantum-main-fin}
checks the equality cases at each cut and the sharpness example.

\subsection{The gap for the product of the marginals}

For $\bsP$ with marginals $\bsr,\bsc$, write $R=\sum_i r_i^2$,
$C=\sum_j c_j^2$ and $\pi_{\bsP}=\sum_{ij}p_{ij}^2$. Then
$$
 \Delta_L(\bsP)- \S_L(\bsr)\S_L(\bsc)=\pi_{\bsP}- R C.
$$
A direct comparison is needed when $\pi_{\bsP}<RC$. Appendix~\ref{app:sharp-marginal-product}
treats the minimum of $\Delta_{\Q}(\bsP)- \frac13\S_L(\bsr)\S_L(\bsc)$. It derives the geometry of a minimizing table, proves analytic
bounds in regions adjoining the boundary, and gives the interval certificate covering the other regions. Dephasing then proves the
bipartite quantum bound. In the multipartite induction the linear entropy of the remaining subsystem group enters two affine
bounds. They intersect when that group's linear entropy equals the linear entropy of the product of its marginals. The comparison
and the equality conditions are given in Section~\ref{sec:other-proofs-fin}.

The same supplement gives the single-state comparison, using a
two-coordinate variation at the minimum, and the continuity theorem,
using a table built from the overlap of two ordered spectra. In the
continuity proof the low-distance range is controlled by the scalar
gap, and the high-distance range by a bound on the smallest eigenvalue.

%=============================%
\section{Conclusion}
%=============================%

We established a sharp comparison between the subentropy and linear-entropy subadditivity gaps for arbitrary finite-dimensional multipartite states. The main inequality bounds the subentropy gap from below by one third of the larger of the joint linear-entropy gap and the marginal-product linear-entropy gap. The constant $1/3$ is optimal whenever at least two local dimensions are at least two, and the two bounds share exactly the same equality families: product states with at most one mixed factor, or with exactly two maximally mixed two-dimensional factors and all remaining factors pure.

We further obtained three complementary results: a sharp two-sided single-state sandwich for subentropy in terms of linear entropy, with constants $2\log2-1$ and $1/2$; a sharp multipartite linear-entropy comparison with coefficient $1/2$; and a sharp Audenaert-type continuity theorem that determines the maximal subentropy difference at every fixed trace distance, together with the dimension-free bound $T-\frac12T^2$.

Methodologically, the comparison is reduced to joint probability tables, where row merging, convexity at fixed marginals, endpoint estimates, and quantitative Schur concavity control the two gaps. Dephasing in marginal eigenbases and a strict spectral transfer estimate lift the table comparison to quantum states and preserve the equality classification. These results show that, although subentropy and linear entropy are globally equivalent at the single-state level, their subadditivity deficits are quantitatively comparable and share the same saturation structure, while separating what is determined by the marginals from what depends on the joint purity.

\subsubsection*{Acknowledgments}

The authors used AI tools, namely ChatGPT Pro (GPT-6 Pro), Codex (gpt-6.1-sol) and Claude
Opus 5.5 (via Claude Code), for literature search, computation and double-checking, including
formal verification of the results in Lean. All proofs can be checked by hand, and the computations
can be reproduced from the verification package. All outputs were manually verified, and
the authors take full responsibility for the work.

%------------------------------------------------------------%
\newpage
\appendix
\appendixpage
\addappheadtotoc
%------------------------------------------------------------%

%===========================================================%
\section{Detailed proofs of the main results}\label{app:detailed-fin}
%===========================================================%

The proof of Lemma~\ref{lem:diagonal-core} starts with the determinant
integral and ends with its equality classification. The spectral transfer
argument then converts the table comparison into a quantum inequality.
The later subsections give the single-state, marginal-purity, and
continuity arguments. Appendix~\ref{app:sharp-marginal-product}
supplies the independent bipartite marginal-product comparison.

\subsection{Proof of Lemma~\ref{lem:diagonal-core}}\label{sec:diagonal-proof}

\paragraph{Step 1. Express the two gaps and remove zero rows and columns.} L
et $\bsr$ and $\bsc$ be the row and column marginals of $\bsP$.
The quantity to be bounded is
$$
 \F(\bsP)=\Q(\bsr)+\Q(\bsc)-\Q(\bsP)  - \frac13\Pa{1-\norm{\bsr}_2^2-\norm{\bsc}_2^2+\norm{\bsP}_2^2}.
$$
The determinant representation \eqref{eq:determinant-integral} will be
used to compare the gaps under a merge, and the exponential representation
\eqref{eq:gaussian} will be used to treat the $2\times2$ case.
Deleting zero rows and columns preserves both gaps. If one positive
row or column remains, the other marginal and the table have the same
nonzero entries. Since $\Q(1)=\S_L(1)=0$, both gaps vanish. In the following
steps both marginals have at least two positive entries.
\paragraph{Step 2. Compute the change under one merge.}
For a nonnegative table $\bsT$ with row masses $r_i$, column masses
$c_j$ and total mass $|\bsT|$, put
\begin{equation}\label{eq:G}
 G_{\bsT}(t)=\frac1{1+t|\bsT|} -\frac1{\prod_i(1+tr_i)}-\frac1{\prod_j(1+tc_j)}  +\frac1{\prod_{i,j}(1+tT_{ij})}.
\end{equation}
If $|\bsT|=1$, \eqref{eq:determinant-integral} gives
$$
\Delta_{\Q}(\bsT)=\int_0^\infty \frac{G_{\bsT}(t)}{t^2}\dif t.
$$
The constant and linear terms cancel in \eqref{eq:G}, since the row,
column and cell masses all sum to $|\bsT|$. Thus $G_{\bsT}(t)=O(t^2)$ at zero.
The four reciprocal terms are bounded at infinity, so the integral against $t^{-2}\dif t$ converges
absolutely.

We use the row and column determinant comparison of
Patil~\cite[Proposition~4]{Patil2026} in the following explicit form.

Merge two rows of an arbitrary nonnegative table, with entries
$a_j,b_j$ and totals $A=\sum_j a_j$, $B=\sum_jb_j$.
Let
$$
 R_{\rm rest}=\prod_{\text{other rows }i}(1+tr_i),\qquad
 P_{\rm rest}=\prod_{\text{cells outside the two rows}}(1+tp_{ij}).
$$
Grouping the cells in each other row shows
$P_{\rm rest}\geqslant R_{\rm rest}$, because
$\prod_j(1+tp_{ij})\geqslant1+t\sum_jp_{ij}$.
Put
\begin{eqnarray*}
 D&:=&\frac1{1+t(A+B)}-\frac1{(1+tA)(1+tB)}\geqslant0,\\
 \mathscr S&:=&\frac1{\prod_j(1+t(a_j+b_j))}
               -\frac1{\prod_j(1+ta_j)(1+tb_j)}\geqslant0.
\end{eqnarray*}
The inequalities follow by expanding the positive denominator products.
Let $\bsP_{{\rm sel}}$ denote the selected two-row table before merging.
The total mass and every column mass are the same on both sides
of the merge, so the first and third terms of \eqref{eq:G} cancel.
The change contributed by the second term is $D/R_{\rm rest}$.
The change contributed by the last term is $-\mathscr S/P_{\rm rest}$,
because the reciprocal of the cell product increases after the merge.
Thus subtraction before and after gives
\begin{eqnarray}
 G_{\rm before}-G_{\rm after} &=&\frac D{R_{\rm rest}}-\frac{\sS}{P_{\rm rest}}=\frac{D-\sS}{R_{\rm rest}}
    +\left(\frac1{R_{\rm rest}}-\frac1{P_{\rm rest}}\right)\sS\notag\\
 &\geqslant&\frac{G_{P_{\rm sel}}(t)}{R_{\rm rest}},
 \label{eq:merge}
\end{eqnarray}
where $D-\sS=G_{\bsP_{\rm sel}}(t)$.
Transposition gives the same identity for column merging.
\paragraph{Step 3. Establish the bound on a four-entry table.}
For a rectangle $\bsT=\Pa{\begin{smallmatrix}a&b\\c&d\end{smallmatrix}}$
with nonnegative entries, we establish
\begin{equation}\label{eq:rectangle-bound}
 G_{\bsT}(t)\geqslant (ad+bc)t^2 h_*(t|\bsT|),\qquad  h_*(s)=\frac{2+10s+22s^2+29s^3+14s^4+7s^5}{2(1+s)^8}.
\end{equation}
At $t=0$ both sides vanish. For $t>0$ we can absorb $t$ into the cell
masses and work at parameter $1$.

Using the scaled entries, set
$$x=a+d,\qquad y=b+c,\qquad p=ad,\qquad q=bc,\qquad
 S=x+y,\qquad X=1+S,\qquad K=p+q.$$
Then $0\leqslant p\leqslant x^2/4$ and
$0\leqslant q\leqslant y^2/4$. Let $R=(1+a+b)(1+c+d)$ and
$C=(1+a+c)(1+b+d)$. Write $W=RC$, $V=R+C$, and
$J=(1+a)(1+b)(1+c)(1+d)$. These three quantities are functions of $x,y,p,q$ given by
\begin{eqnarray*}
 W&=&X^2+X(xy+2K)+K^2+Kxy+qx^2+py^2-4pq,\\
 V&=&2X+xy+2K,\qquad J=(1+x+p)(1+y+q).
\end{eqnarray*}
For instance, writing $U=(a+b)(c+d)$ and $T=(a+c)(b+d)$ gives
$U+T=xy+2K$ and $UT=(p-q)^2+(p+q)xy+qx^2+py^2$.
The numerator of the kernel over $XWJ$ is $N=WJ-XVJ+XW$.
By \eqref{eq:rectangle-bound}, it suffices to show
\begin{equation}\label{eq:rectangle-polynomial}
 H=2NX^7-K(2+10S+22S^2+29S^3+14S^4+7S^5)WJ\geqslant0.
\end{equation}

Write $H=\sum_{\alpha,\beta=0}^4 H_{\alpha\beta}(x,y)
p^\alpha q^\beta$. Substituting $p=x^2\lambda/4$ and
$q=y^2\mu/4$, with $\lambda,\mu\in[0,1]$, gives Bernstein
coefficients of degree $(4,4)$
\begin{equation}\label{eq:rectangle-bernstein}
 B_{ij}(x,y)=\sum_{\alpha=0}^i\sum_{\beta=0}^j
 H_{\alpha\beta}(x,y)\,
 \frac{x^{2\alpha}y^{2\beta}}{4^{\alpha+\beta}}\,
 \frac{\binom i\alpha\binom j\beta}
      {\binom4\alpha\binom4\beta}
 \qquad (0\leqslant i,j\leqslant4).
\end{equation}
Indeed, expand $\lambda^\alpha$ by multiplying by
$(\lambda+(1-\lambda))^{4-\alpha}$ and do the same for
$\mu^\beta$. The resulting Bernstein basis functions are nonnegative, and
$$H=\sum_{i,j=0}^4 B_{ij}(x,y)\binom4i\lambda^i(1-\lambda)^{4-i}
\binom4j\mu^j(1-\mu)^{4-j}.$$ It suffices to prove that
$B_{ij}(x,y)\geqslant0$.

Swapping the columns interchanges $(x,p)$ and $(y,q)$, so it suffices
to check $x=y+e$ with $e,y\geqslant0$. The coefficient expansion of
\eqref{eq:rectangle-polynomial}--\eqref{eq:rectangle-bernstein} has the form
\begin{equation}\label{eq:rectangle-remainder}
 B_{ij}(y+e,y)=A_{ij}(e,y)+
 \mathbf1_{\{i=j=4\}}
 \left(\frac{y^3}{16}P_0(y)+\frac{ey^2}{16}P_1(y)
                        +\frac{e^2y}{32}P_2(y)\right).
\end{equation}
The following coefficient calculation verifies the signs of $A_{ij}$
and the three displayed remainders. The counts of positive coefficients
of $A_{ij}$, with rows indexed by $i$ and columns by $j$, are
$$
 \begin{pmatrix}
 0&74&85&85&85\\
 96&108&120&120&120\\
 111&124&124&124&124\\
 113&127&127&127&127\\
 113&127&127&127&88
 \end{pmatrix}.
$$
For explicit coefficient extraction, put
$h_{\alpha\beta uv}=[x^uy^v]H_{\alpha\beta}(x,y)$. The coefficient of
$e^my^n$ in $B_{ij}(y+e,y)$ is
\begin{equation}\label{eq:rectangle-coefficient}
 \sum_{\substack{\alpha\leqslant i,\ \beta\leqslant j,\ u,v\geqslant0\\
         u+v+2\alpha+2\beta=m+n}}
 h_{\alpha\beta uv}
 \frac{\binom i\alpha\binom j\beta\binom{u+2\alpha}{m}}
 {4^{\alpha+\beta}\binom4\alpha\binom4\beta},
\end{equation}
where the binomial coefficient vanishes if $m>u+2\alpha$. In this
calculation the degree in $e,y$ is at most $16$. For each $i,j$, the
smallest coefficient of $A_{ij}$ over $m,n\geqslant0$, $m+n\leqslant16$
is zero. These checks can be done in integers by clearing the same
positive denominator $4^8\cdot12^2$ in each sum. The case $(i,j)=(0,0)$
gives the zero polynomial.
The summands in parentheses in \eqref{eq:rectangle-remainder}
are the full coefficients of $e^0,e^1,e^2$ for $(i,j)=(4,4)$.
The coefficients $[y^j]P_k$ are listed below.
\begin{center}
$\begin{array}{r|r|r|r}
j&P_0&P_1&P_2\\\hline
0&16&24&48\\
1&166&332&844\\
2&516&1290&4744\\
3&671&2013&13353\\
4&-270&-945&18300\\
5&-2864&-11456&-4685\\
6&-4736&-21312&-47546\\
7&-1401&-7005&-9202\\
8&5582&30701&155810\\
9&8420&50520&271026\\
10&5228&33982&198686\\
11&1488&10416&66064\\
12&144&1080&7360
\end{array}$
\end{center}

One split at $y=5/8$ proves positivity of all three polynomials.
For $P(y)=\sum_{j=0}^{12}p_jy^j$ use $y=5z/[8(1+z)]$ on
$0\leqslant y<5/8$ and $y=(5+z)/8$ on $y\geqslant5/8$, with
$z\geqslant0$. The coefficients of $z^k$ in the two cleared
polynomials $8^{12}(1+z)^{12}P(5z/[8(1+z)])$ and
$8^{12}P((5+z)/8)$ are respectively
$$
 \sum_{j=0}^k p_j5^j8^{12-j}\binom{12-j}{k-j},
 \qquad
 \sum_{j=k}^{12}p_j5^{j-k}8^{12-j}\binom jk .
$$
Using the displayed $p_j$, the smallest coefficients in the first
list for $P_0,P_1,P_2$ are $16\cdot8^{12}$, $24\cdot8^{12}$,
and $48\cdot8^{12}$. In the second list the minima are
$144$, $1080$, and $7360$, respectively. Both transformations
therefore give $P_k(y)>0$ on their domains.

Equations \eqref{eq:rectangle-remainder} and
\eqref{eq:rectangle-coefficient} prove that every Bernstein
coefficient is nonnegative, hence $H\geqslant0$. If $x=0$ or $y=0$,
the corresponding product $p$ or $q$ vanishes and the same
substitution applies with the corresponding Bernstein variable
set to zero. Restoring the factor $t$ gives the rectangle weight
$(ad+bc)t^2$ in \eqref{eq:rectangle-bound}. The coefficient
calculation can also be reproduced with
\nolinkurl{verify_rectangle_kernel.py} in the accompanying verification
directory.
\paragraph{Step 4. Evaluate the two scalar integrals.}
Direct differentiation gives
$$
 \psi(s):=-h_*'(s)-h_*(s)
 =\frac{4+14s+13s^2+38s^3-22s^4-7s^6}{2(1+s)^9}.
$$
The nonzero numerator coefficients have exactly one sign change.
The numerator is positive at zero and has a negative leading coefficient. Descartes' rule, together with continuity, therefore
shows that there is exactly one positive root $\xi$, with
$\psi>0$ before $\xi$ and $\psi<0$ after $\xi$.

The merge estimate uses the two integrals
$$
 J=\int_0^\infty \frac{s\psi(s)}{1+s/2}\dif s,
 \qquad \beta=\frac32\int_0^\infty \frac{h_*(s)}{1+s/2}\dif s.
$$
Their integrands have the exact partial fraction forms
\begin{eqnarray*}
 \frac{s\psi(s)}{1+s/2}
 &=&-\frac{2152}{s+2}+\sum_{j=1}^9\frac{a_j}{(s+1)^j},\\
 \frac32\frac{h_*(s)}{1+s/2}
 &=&-\frac{243}{s+2}+\sum_{j=1}^8\frac{b_j}{(s+1)^j}.
\end{eqnarray*}
Their coefficients are
\begin{center}
$\begin{array}{c|r|r}
 j&a_j&b_j\\\hline
 1&2152&243\\
 2&-2152&-243\\
 3&2145&243\\
 4&-2096&-465/2\\
 5&1927&201\\
 6&-1534&-273/2\\
 7&930&60\\
 8&-360&-12\\
 9&64&0
\end{array}$
\end{center}
Multiplication by a common denominator verifies both identities. In each case the two simple-pole
coefficients cancel at infinity. Integrate them together using
one common upper cutoff. The two results are
$2152\log2$ and $243\log2$. For $j\geqslant2$,
$\int_0^\infty(s+1)^{-j}\dif s=1/(j-1)$. Hence
$$
 J=2152\log2-\frac{626491}{420},\qquad
 \beta=243\log2-\frac{23487}{140}.
$$
The positive atanh series gives the strict rational lower bound
$$
 \log2=2\sum_{j=0}^\infty
 \frac1{(2j+1)3^{2j+1}}
 >2\sum_{j=0}^4\frac1{(2j+1)3^{2j+1}}
 =\frac{4297606}{6200145}.
$$
Substituting it in the preceding formulas gives exactly
$$
 J>\frac{125389}{24800580}>0,\qquad
 \eta:=\beta-\frac23=243\log2-\frac{70741}{420}>\frac{361}{102060}>0.
$$
The script \nolinkurl{verify_smallest_row_analytic.py} in the same
directory checks these rational identities and signs.
\paragraph{Step 5. Bound the weighted integral uniformly.}
For $0<v\leqslant u\leqslant2/3$, put $k=2/u-2\geqslant1$ and define
\[
 I(u,v)=\int_0^\infty e^{-(u-v)t}(1+ut/2)^{-k}h_*(vt)\dif t.
\]
We first compare it with $I(u,u)$ and then find a lower bound independent
of $u$. The derivative in $v$ is taken for $0<v<u$.

Let $q=u/v\geqslant1$. Differentiating under the integral and
substituting $s=vt$ gives
$$
 \partial_v I(u,v)=-\frac1{v^2}\int_0^\infty
 s\psi(s)e^{-(q-1)s}(1+qs/2)^{-k}\dif s.
$$
Relative to the integrand defining $J$, the extra
factor is
$$W(s)=e^{-(q-1)s}(1+qs/2)^{-k}(1+s/2).$$
It is positive and nonincreasing, since
$$
 \frac{W'(s)}{W(s)}=-(q-1)-\frac{kq}{2+qs}+\frac1{2+s}
 \leqslant0.
$$
Indeed $k,q\geqslant1$ and $q/(2+qs)\geqslant1/(2+s)$.
Put $f(s)=s\psi(s)/(1+s/2)$. On both sides of $\xi$,
$f(s)(W(s)-W(\xi))\geqslant0$. Integrating this pointwise
inequality yields
$$\int_0^\infty f(s)W(s)\dif s\geqslant W(\xi)J>0.$$
Thus $\partial_vI(u,v)<0$ and $I(u,v)\geqslant I(u,u)$.
On each compact positive parameter interval, $h_*(s)=O(s^{-3})$,
$h_*'(s)=O(s^{-4})$ and $k\geqslant1$ give integrable bounds for these
differentiations, including at $v=u$.

For the endpoint integral use $u=2/(k+2)$, $s=ut$, and then
$z=2/(s+2)$. The exact result is
$$I(u,u)=(k+2)\int_0^1z^{k+1}g(z)\dif z,$$
where
$$g(z)=\frac{112-448z+788z^2-752z^3+382z^4-81z^5}{(2-z)^8}.$$
Its derivative is $g'(z)=zp(z)/(2-z)^9$, with
\begin{eqnarray*}
 p(z)=16(1-z)^4+280z(1-z)^3+40z^2(1-z)^2+22z^3(1-z)+3z^4.
\end{eqnarray*}
Consequently $g$ is increasing on $[0,1]$.
The probability density $(k+2)z^{k+1}$ has distribution function
$z^{k+2}$, which decreases pointwise with $k$. Its expectation of
an increasing function therefore increases with real $k$.
Integration by parts makes this monotonicity explicit.
$$
 (k+2)\int_0^1z^{k+1}g(z)\dif z
 =g(1)-\int_0^1z^{k+2}g'(z)\dif z.
$$
The minimum for $k\geqslant1$ is attained at $k=1$. Hence
$$
 I(u,u)\geqslant\frac32\int_0^\infty
       \frac{h_*(s)}{1+s/2}\dif s
 =\beta.
$$
\paragraph{Step 6. Merge the two smallest rows.}\phantomsection\label{step:smallest-row-merge}
Let $\bsP$ have at least three positive rows. Merge the two smallest,
with entries $a_j,b_j$ and masses $A,B$, to obtain $\bsP^{\rm m}$. Write
$u=A+B>0$. Every outside row has mass
$r_i\geqslant\max(A,B)\geqslant u/2$. There is at least one
such row, so $u\leqslant2/3$ and the bound for $I(u,v)$ applies.
The function $x\mapsto\log(1+tx)/x$ is nonincreasing for
$x>0$, for each fixed $t\geqslant0$. Therefore
\begin{eqnarray*}
 \log R_{\rm rest}(t)=\sum_{\rm outside}\log(1+tr_i)\leqslant\sum_{\rm outside}\frac{2r_i}{u}\log(1+ut/2)=\frac{2(1-u)}u\log(1+ut/2).
\end{eqnarray*}
With $k=2(1-u)/u\geqslant1$, this proves
\begin{equation}\label{eq:smallest-row-outside}
 R_{\rm rest}(t)\leqslant(1+ut/2)^k.
\end{equation}
Write $G_{a,b}$ for the kernel of the selected two-row table.
Integrating \eqref{eq:merge} gives
\begin{equation}\label{eq:smallest-row-merge-integral}
 \Delta_{\Q}(\bsP)-\Delta_{\Q}(\bsP^{\rm m})
 \geqslant\int_0^\infty \frac{G_{a,b}(t)}{t^2R_{\rm rest}(t)}\dif t.
\end{equation}
Inside the selected two-row table merge columns in any fixed
order. At step $j$ the local rectangle is
$$
 T_j=\begin{pmatrix}
 \sum_{\ell<j}a_\ell&a_j\\
 \sum_{\ell<j}b_\ell&b_j
 \end{pmatrix}.
$$
Write $v_j=|T_j|\leqslant u$ and
$$K_j=\left(\sum_{\ell<j}a_\ell\right)b_j
        +a_j\left(\sum_{\ell<j}b_\ell\right).$$
For $j=2,\ldots,n$, each cross-column product occurs once, so
\begin{equation}\label{eq:smallest-row-K-sum}
 \sum_jK_j=AB-\sum_ja_jb_j=:K_{a,b}.
\end{equation}
Thus $K_{a,b}=\sum_{j\ne\ell}a_jb_\ell$ is the total cross-column
mass of the selected rows. For two positive rows it vanishes exactly
when both are concentrated in one common column.
The column version of the same merge identity gives
$G_{a,b}(t)\geqslant\sum_jG_{T_j}(t)/C_j(t)$.
Here $C_j(t)=\prod_{\ell>j}(1+t(a_\ell+b_\ell))$.
The final one-column table has $G=0$, which closes this telescope.
The masses in $C_j$ sum to $u-v_j$, and consequently
$C_j(t)\leqslant e^{(u-v_j)t}$. \eqref{eq:rectangle-bound}
therefore yields
$$
 G_{a,b}(t)\geqslant t^2\sum_jK_j
 e^{-(u-v_j)t}h_*(v_jt).
$$
All summands are nonnegative. Substituting this and
\eqref{eq:smallest-row-outside} in
\eqref{eq:smallest-row-merge-integral}, and using the lower bound for $I(u,v_j)$ gives
$$
 \Delta_{\Q}(\bsP)-\Delta_{\Q}(\bsP^{\rm m}) \geqslant\sum_jK_jI(u,v_j)\geqslant \beta K_{a,b}.
$$
A term with $v_j=0$ has $K_j=0$ and is omitted. The rational kernel
identities also apply at zero entries because all denominators are positive.
Merging the rows increases the row marginal purity by $2AB$ and
the joint purity by $2\sum_j a_jb_j$. The column marginal purity
is fixed. Thus $\Delta_L(\bsP)-\Delta_L(\bsP^{\rm m})=2K_{a,b}$.
Subtract $2K_{a,b}/3$ to obtain
\begin{equation}\label{eq:smallest-row-quantitative}
 \F(\bsP)-\F(\bsP^{\rm m})\geqslant \eta K_{a,b},\qquad \eta>0.
\end{equation}
Transposing the table gives the same estimate for a column merge.
\paragraph{Step 7. Parametrize the binary table and bound the second derivative.}\phantomsection\label{step:binary-curvature}
Let $a$ and $b$ be the row and column biases. Write the four entries as
\begin{equation}\label{eq:Walsh}
 (x,y,z,w)=\frac14(1+a+b+c,\,1+a-b-c,\,
                             1-a+b-c,\,1-a-b+c),
\end{equation}
and put
$$
 \Q_2(a)=\Q\Pa{\frac{1+a}2,\frac{1-a}2},\qquad \F(a,b,c)=\Q_2(a)+\Q_2(b)-\Q(x,y,z,w)  -\frac{1-a^2-b^2+c^2}{12}.
$$
Indeed, the marginal squared norms are $\frac12(1+a^2)$ and
$\frac12(1+b^2)$, and the joint squared norm is $\frac14(1+a^2+b^2+c^2)$.
Hence the last term in $\F$ is $\frac13\Delta_L$.
Row and column permutations and
transposition allow us to assume $0\leqslant b\leqslant a\leqslant1$.
The complete feasible interval for the remaining coordinate is
\begin{equation}\label{eq:c-interval}
 c_-=a+b-1\leqslant c\leqslant1-a+b=c_+.
\end{equation}
The table is the product of its marginals at $c=ab$, and
\begin{equation}\label{eq:fiber-distances}
 ab-c_-=(1-a)(1-b),\qquad c_+-ab=(1-a)(1+b).
\end{equation}

For a positive probability vector $\bsp=(x,y,z,w)$, set
$\bsv=(1,-1,-1,1)/4$ and let the four $Z_i$ be independent
$\operatorname{Exp}(1)$ variables. The exponential representation
\eqref{eq:gaussian}, differentiated along this trace-zero direction,
gives
$$
\F_{cc}=\bbE\Br{\frac{\Inner{\bsv}{\bsZ}^2}{\Inner{\bsv}{\bsZ}}}-\frac16.
$$
Denote the expectation by $\cH(\bsp)$. It is convex in $\bsp$.
The four permutations
$$
(1),\quad(12)(34),\quad(13)(24),\quad(14)(23)
$$
are transitive and send $\bsv$ to $\bsv$ or $-\bsv$. Exchangeability of the
exponentials makes $\cH$ invariant under these permutations.
Their average sends every $\bsp$ to $\bsu_4$, so
$\cH(\bsp)\geqslant\cH(\bsu_4)$.

For completeness put $S=\sum_iZ_i$ and $V=Z_1-Z_2-Z_3+Z_4$.
To compute $\bbE[V^2e^{-tS}]$, each of the four square terms
contributes $2/(1+t)^6$. The mixed terms together contribute
$-4/(1+t)^6$ because the sum of the four signs in $V$ is zero.
Hence
$$
 \bbE[V^2e^{-tS}]=\frac4{(1+t)^6},\qquad \cH(\bsu_4)=\frac14\int_0^\infty \frac4{(1+t)^6}\dif t=\frac15.
$$
Tonelli applies to the nonnegative integrand used to represent $1/S$.
Consequently
\begin{equation}\label{eq:binary-curvature}
 \F_{cc}\geqslant\frac1{30}.
\end{equation}
On the positive simplex the differentiations are locally dominated.
If $p_i\geqslant\varepsilon$, then
$\Inner{\bsp}{\bsZ}\geqslant\varepsilon S$ and $|\Inner{\bsv}{\bsZ}|\leqslant S/4$,
so the second derivative is dominated by $S/(16\varepsilon)$.
The first derivative is dominated by an integrable multiple of
$S(1+|\log S|)$.

If $a=1$, feasibility forces $c=b$ and both gaps vanish. Assume $a<1$.
The product point then has four positive entries. Put
$H_0=F(a,b,ab)$ and $g_0=\F_c(a,b,ab)$. Integrating
\eqref{eq:binary-curvature} along the fiber yields
\begin{equation}\label{eq:binary-tangent}
\F(a,b,c)\geqslant H_0+g_0(c-ab)+\frac{(c-ab)^2}{60}.
\end{equation}
Continuity extends this inequality to both endpoints of
\eqref{eq:c-interval}. In the exponential formula, a common integrable
bound is $1/\mathrm e+S\log_+S$ for $|\eta(\Inner{\bsp}{\bsZ})|$.
\paragraph{Step 8. Compute the value and slope at the product table.}\phantomsection\label{step:binary-product}
Put $A=a^2$, $B=b^2$, $\kappa=ab$, $s=A+B$, and
$D(u)=(1-Au^2)(1-Bu^2)$. Define
\[
 T=\int_0^1(1-u^2)\left(\frac{1}{D(u)}-1\right)\dif u,\qquad
 J=\int_0^1\frac{(1-u^2)^2}{D(u)}\dif u.
\]

For a strictly positive probability vector $\bdlambda$ with distinct entries $\lambda_i$, put
$\mathcal D_i=\prod_{j\ne i}(\lambda_i-\lambda_j)$. Partial fractions
in the determinant integral give
$$
 \Q(\bdlambda)=-\sum_i\frac{\lambda_i^d\log\lambda_i}{\mathcal D_i}.
$$
Indeed, define $A_i=\lambda_i^{d-1}/\mathcal D_i$. The decomposition
$$
 \frac1{\prod_i(1+t\lambda_i)}=\sum_i\frac{A_i}{1+t\lambda_i}
$$
implies $\sum_iA_i=\sum_iA_i\lambda_i=1$.
In the integral for $\Q$, use
$(1+tz)^{-1}=1-tz+t^2z^2/(1+tz)$ for each factor in the partial
fraction sum. The constant and linear terms cancel, so integration
up to $R$ gives
$\log(1+R)-\sum_iA_i\lambda_i\log(1+R\lambda_i)$.
Its limit is $-\sum_i A_i\lambda_i\log\lambda_i$, since
$\sum_i A_i\lambda_i=1$.

First suppose $0<b<a<1$ and put $A=a^2$, $B=b^2$. At the product point the four eigenvalues are
$$
 \lambda_1=\frac{(1+a)(1+b)}4,\quad
 \lambda_2=\frac{(1+a)(1-b)}4,\quad
 \lambda_3=\frac{(1-a)(1+b)}4,\quad
 \lambda_4=\frac{(1-a)(1-b)}4.
$$
The six differences are
\begin{eqnarray*}
 \lambda_1-\lambda_2&=&b(1+a)/2,\qquad
 \lambda_1-\lambda_3=a(1+b)/2,\qquad
 \lambda_1-\lambda_4=(a+b)/2,\\
 \lambda_2-\lambda_3&=&(a-b)/2,\qquad
 \lambda_2-\lambda_4=a(1-b)/2,\qquad
 \lambda_3-\lambda_4=b(1-a)/2.
\end{eqnarray*}
Multiplying the three signed differences belonging to each index yields
\begin{equation}\label{eq:product-denominators}
 (\mathcal D_1,\ldots,\mathcal D_4)=\frac{ab}{2}
 \bigl(\lambda_1(a+b),-\lambda_2(a-b),
                  \lambda_3(a-b),-\lambda_4(a+b)\bigr).
\end{equation}
Substitute these into the spectral formula and group
$\log\lambda_{\sigma\tau}=\log(1+\sigma a)+\log(1+\tau b)-\log4$.
Collect the logarithms using the two binary identities
$\Q_2(a)=-\sum_{\sigma=\pm1}\sigma(1+\sigma a)^2
\log((1+\sigma a)/2)/(4a)$ and the same identity with $b$.
In $\Delta_{\Q}$ the $\log 4$ terms cancel. The coefficient of
$\log(1+a)$ is $-(1-A)^2(1-B)/(16a(A-B))$, with the opposite
coefficient for $\log(1-a)$. The coefficient of $\log(1+b)$ is
$(1-A)(1-B)^2/(16b(A-B))$, with the opposite coefficient for
$\log(1-b)$. Thus the product gap is
\begin{eqnarray}\label{eq:product-log-value}
 \Delta_{\Q}(a,b,ab)
 &=&\frac{(1-A)(1-B)}{16(A-B)}
 \left[\frac{1-B}{b}\log\frac{1+b}{1-b}
       -\frac{1-A}{a}\log\frac{1+a}{1-a}\right]\notag\\
 &=&\frac{(1-A)(1-B)}8
            \int_0^1\frac{1-u^2}{D(u)}\dif u.
\end{eqnarray}
To verify the second equality, define
$I_0(z)=\int_0^1(1-zu^2)^{-1}\dif u$ and
$R_0(z)=(1-z)I_0(z)$. Then
$I_0(a^2)=\operatorname{artanh}(a)/a$ and
$$
 \frac{R_0(B)-R_0(A)}{A-B}
       =\int_0^1\frac{1-u^2}{D(u)}\dif u.
$$
At the product point,
$\Delta_L=(1-A)(1-B)/4$. Subtracting one third of this quantity
from \eqref{eq:product-log-value} gives the first identity below
because $\int_0^1(1-u^2)\dif u=2/3$.

For the derivative, first differentiate the spectral formula
with respect to $c$. Define
$$
 \varepsilon=\frac14(1,-1,-1,1),\qquad
 C_i=\frac{\lambda_i^4}{\mathcal D_i}.
$$
Then
\begin{eqnarray*}
 C_i'&=&C_i\left[\frac{4\varepsilon_i}{\lambda_i}
       -\sum_{j\ne i}\frac{\varepsilon_i-\varepsilon_j}
                            {\lambda_i-\lambda_j}\right],\\
 \Q_{4,c}&=&-\sum_iC_i'\log\lambda_i
             -\sum_i\frac{\varepsilon_i\lambda_i^3}{\mathcal D_i}.
\end{eqnarray*}
The nonlogarithmic sum is
\[
 -\sum_i\frac{\varepsilon_i\lambda_i^3}{\mathcal D_i}
 =-\frac1{2ab}\left(
 \frac{\lambda_1^2-\lambda_4^2}{a+b}
 +\frac{\lambda_2^2-\lambda_3^2}{a-b}\right)=-\frac1{4ab}.
\]
Indeed, the respective differences of the two eigenvalues in the
numerators are $(a+b)/2$ and $(a-b)/2$, and all four eigenvalues
sum to one. For the logarithmic part, let
$D_i^*=\sum_{j\ne i}(\varepsilon_i-\varepsilon_j)/(\lambda_i-\lambda_j)$.
The six differences above give
\[
\begin{array}{c|l}
i & D_i^*\\ \hline
1 & [b(1+a)]^{-1}+[a(1+b)]^{-1}\\
2 & [b(1+a)]^{-1}-[a(1-b)]^{-1}\\
3 & [a(1+b)]^{-1}-[b(1-a)]^{-1}\\
4 & -[a(1-b)]^{-1}-[b(1-a)]^{-1}
\end{array}
\]
In the spectral formula $\sum_i C_i=1$, as follows by matching
the first-order term of the reciprocal product at zero. Hence
$\sum_i C_i'=0$. Splitting $\log\lambda_i$ into the row and column
logarithms, the coefficients of the two log ratios are
$-C'_1-C'_2$ and $-C'_1-C'_3$, respectively. From the table
and $C_i'=C_i(4\varepsilon_i/\lambda_i-D_i^*)$ these are
\[
 -\frac{b(1-A)^2}{8A(A-B)}
 \quad\hbox{and}\quad
 \frac{a(1-B)^2}{8B(A-B)}.
\]
Therefore
\begin{eqnarray}\label{eq:product-log-derivative}
 \left.\Q_{4,c}\right|_{c=ab}
 &=&-\frac1{4ab}+\frac1{8(A-B)}
 \left[\frac{a(1-B)^2}{B}\log\frac{1+b}{1-b}
       -\frac{b(1-A)^2}{A}\log\frac{1+a}{1-a}\right]\notag\\
 &=&-\frac{ab}{4}\int_0^1\frac{(1-u^2)^2}{D(u)}\dif u.
\end{eqnarray}
The last identity follows from the pointwise rational identity
$$
 \frac{A(1-B)^2/(1-Bu^2)-B(1-A)^2/(1-Au^2)}{A-B}
       =1-\frac{AB(1-u^2)^2}{D(u)}.
$$
Since $F_c=-\Q_{4,c}-c/6$, the two identities are
\begin{equation}\label{eq:product-value-gradient}
 H_0=\frac{(1-A)(1-B)}8T,\qquad
 g_0=\frac{\kappa}{12}(3J-2).
\end{equation} The compact integrals and the
correlation derivative are continuous for $a,b<1$. Thus the identities
extend to $a=b$ and $b=0$ by taking limits of the compact integral
expressions.
\paragraph{Step 9. Compare the tangent at the two endpoints.}
Let $\ell(c)=H_0+g_0(c-ab)$. It suffices to check its two endpoint
values. Equations \eqref{eq:fiber-distances} and
\eqref{eq:product-value-gradient} give
\begin{eqnarray*}
 \ell(c_-)&=&\frac{(1-a)(1-b)}{12}E_-,\qquad
 E_-=\frac32(1+a)(1+b)T-\kappa(3J-2),\\
 \ell(c_+)&=&\frac{(1-a)(1+b)}{12}E_+,\qquad
 E_+=\frac32(1+a)(1-b)T+\kappa(3J-2).
\end{eqnarray*}
The two prefactors are positive when $a<1$. The exact integral identities
$$
 T=\int_0^1\frac{u^2(1-u^2)(s-\kappa^2u^2)}{D(u)}\dif u,\qquad
 3J-2=3\int_0^1
        \frac{u^2(1-u^2)(s-1-\kappa^2u^2)}{D(u)}\dif u
$$
give
\begin{eqnarray}
 E_-&=&\frac32\int_0^1\frac{u^2(1-u^2)}{D(u)}
       \bigl[(1+a+b-\kappa)(s-\kappa^2u^2)+2\kappa\bigr]\dif u,
       \label{eq:tangent-minus}\\
 E_+&=&\frac32\int_0^1\frac{u^2(1-u^2)}{D(u)}
       \bigl[(1+a-b+\kappa)(s-\kappa^2u^2)-2\kappa\bigr]\dif u.
       \label{eq:tangent-plus}
\end{eqnarray}
The first integrand is nonnegative. For the second, both factors
being bounded below are nonnegative, and
$$
 1+a-b+\kappa\geqslant1+\kappa,\qquad
 s-\kappa^2u^2\geqslant2\kappa-\kappa^2.
$$
Hence its bracket is at least
$$
 (1+\kappa)(2\kappa-\kappa^2)-2\kappa   =\kappa^2(1-\kappa)\geqslant0.
$$
It follows that $\ell(c)\geqslant0$ throughout the interval.
By \eqref{eq:binary-tangent},
$$
 \F(a,b,c)\geqslant\ell(c)+\frac{(c-ab)^2}{60}\geqslant0.
$$
If $ab>0$, the brackets at both endpoints are strictly positive.
If $b=0<a$, both brackets equal $(1+a)a^2>0$.
When $a=b=0$, we have $\F(0,0,c)\geqslant c^2/60$ and
$\F(0,0,0)=H_0=0$ by \eqref{eq:product-value-gradient}.
Thus, with two positive row masses and two positive column masses,
equality holds exactly at the uniform table.
\paragraph{Step 10. Reduce the original table to the binary case.}
With the zero rows and columns removed, assume both marginals are mixed.
If there are more than two rows, apply \eqref{eq:smallest-row-quantitative}
to the two smallest positive rows. Repeat until there are two rows.
Every column mass is preserved during these merges. Now transpose,
merge the two smallest positive columns until two columns remain,
and transpose back. At each merge $\F$ cannot increase. The resulting
table $\bsU$ has two positive row masses and two positive column masses,
so the preceding step gives $\F(\bsP)\geqslant \F(\bsU)\geqslant0$.

If $\F(\bsP)=0$, the final table $\bsU$ is uniform and every merge has
$K_{a,b}=0$ by \eqref{eq:smallest-row-quantitative}. The product
$a_jb_\ell$ then vanishes whenever $j\ne\ell$. Because both selected
rows have positive mass, choosing a positive entry in each row shows
that both are supported in one common column. If a row merge occurs,
the resulting row is supported in a single column. Any later zero-loss
row merge that involves it preserves that support. A later column merge
maps its only support column to one column of the new table. It would
therefore yield a row of $\bsU$ with a zero cell, which is impossible.
If the only merges are column merges, the transposed argument gives
the same contradiction. Thus the table already had two positive rows
and columns, and was uniform after discarding zero rows and columns.
Conversely, the binary calculation gives equality for the uniform
table, and Step 1 gives equality for a pure marginal. This proves
Lemma~\ref{lem:diagonal-core}, including the equality classification.\qed

\subsection{Spectral mixing}

We next quantify the change in $\Q$ when two spectral coordinates
are moved toward one another while their sum is preserved. A reciprocal expectation controls the derivative along such a
transfer and gives Schur concavity of $\mathcal R$.

\begin{lem}\label{lem:exponential}
Let $Z_1,\ldots,Z_d$ be independent exponentials of mean one. For any
probability vector $\bdlambda$ and distinct $i,j$, setting
$w=\lambda_i+\lambda_j$,
\begin{equation}\label{eq:moment}
 \bbE\left[\frac{Z_iZ_j}{\sum_k\lambda_kZ_k}\right]
 \geqslant\frac23+\frac2{15}(\lambda_i-\lambda_j)^2
          +\frac{(1-w)^2}{3(1+w)}\geqslant\frac23.
\end{equation}
\end{lem}

\begin{proof}
Put $a=\lambda_i$, $b=\lambda_j$, $\theta=1-w$, $x=Z_i$, $y=Z_j$,
$X=(x+y)/2$, and $Y=ax+by+B$, where $B=\sum_{k\notin\{i,j\}}\lambda_kZ_k$
is independent of $(x,y)$ and has mean $\theta$. (At a boundary vector the
left side of \eqref{eq:moment} may be $+\infty$, in which case the bound is
trivial; assume it is finite.) Conditioning on $(x,y)$ and applying
Jensen's inequality to the convex reciprocal replaces $B$ by $\theta$.
Write $S=wX+\theta$ and $L=(a-b)(x-y)/2$. Since $S\pm L>0$ almost surely,
averaging the two orders of $(x,y)$ gives
$$\bbE\!\left[\frac{xy}{Y}\right]
 \geqslant\bbE\!\left[\frac{xyS}{S^2-L^2}\right]
 \geqslant\bbE\!\left[\frac{xy}{S}\right]
         +\bbE\!\left[\frac{xyL^2}{S^3}\right].$$
The variable $V=x/(x+y)$ is uniform on $[0,1]$ and independent of $X$,
which has the gamma law with shape two and rate two ($\bbE X=1$,
$\bbE X^2=3/2$, $\bbE X^3=3$). Thus
$$\bbE\!\left[\frac{xyL^2}{S^3}\right]
 =4(a-b)^2\bbE\bigl[V(1-V)(2V-1)^2\bigr]\,\bbE\!\left[\frac{X^4}{S^3}\right]
 \geqslant\frac2{15}(a-b)^2,$$
since $\bbE[V(1-V)(2V-1)^2]=1/30$ and Jensen's inequality for the convex
perspective $x^4/s^3$ gives $\bbE[X^4/S^3]\geqslant(\bbE X)^4/(\bbE S)^3=1$.
For the first term, write
$1/S=1/X-(S-X)/X^2+(S-X)^2/(X^2S)$ with $S-X=\theta(1-X)$. Independence of
$xy/X^2=4V(1-V)$ from $X$ cancels the linear term ($\bbE[1-X]=0$) and gives
$\bbE[xy/S]-\frac23=\frac23\theta^2\bbE[(1-X)^2/S]$. Cauchy--Schwarz yields
$$\bbE\!\left[\frac{(1-X)^2}{S}\right]
 \geqslant\frac{(\bbE[(1-X)^2])^2}{\bbE[(1-X)^2S]}
 =\frac{(1/2)^2}{(1+w)/2}=\frac1{2(1+w)},$$
completing the proof of \eqref{eq:moment}.
\end{proof}

Schur concavity of subentropy was established
in~\cite[Section~VI]{Mintert2004} and follows also from
\cite[Corollary~2]{Datta2014}. The following proposition establishes the
quantitative Schur-concavity estimate used in the quantum reduction,
including an explicit spectral margin that is asymptotically exact near
the uniform rank-two spectrum ($1/3$) and along a sequence with vanishing
leading eigenvalues ($1/2$).

\begin{prop}\label{prop:mixing}
The function $\sfQ-\frac13\sfS_L$ is Schur concave on every probability
simplex. More sharply, if $\bdlambda(\sigma)\prec\bdlambda(\rho)$ and
$s=\lambda_1^\downarrow(\rho)+\lambda_2^\downarrow(\rho)\in(0,1]$
is the sum of the two largest eigenvalues (padding a rank-one
spectrum with zero), then
\begin{equation}\label{eq:mixing}
 \sfQ(\sigma)-\sfQ(\rho) \geqslant\Pa{\frac13+\frac{(1-s)^2}{6(1+s)}}(\Tr{\rho^2}-\Tr{\sigma^2})
 \geqslant\frac13(\Tr{\rho^2}-\Tr{\sigma^2}),
\end{equation}
The first inequality and the comparison of the leftmost and rightmost
expressions are strict whenever the ordered spectra differ. Equality
in the second inequality holds exactly when $s=1$ or the ordered spectra
coincide.
The universal coefficient $1/3$ is optimal.
\end{prop}

\begin{proof}
First suppose that all eigenvalues are positive and set
$Y=\sum_k\lambda_kZ_k$. Differentiation of \eqref{eq:gaussian} gives
$$
 \partial_{\lambda_i}\Q-\partial_{\lambda_j}\Q =-\bbE[(Z_i-Z_j)\log Y].
$$
An arbitrary smooth extension away from the trace-one hyperplane may
be used here, since only differences of partial derivatives occur.
For independent exponentials, integration by parts yields
$$
 \bbE[(Z_i-Z_j)f] =\bbE[Z_iZ_j(\partial_{Z_i}f-\partial_{Z_j}f)],
$$
Taking $f=\log Y$ gives
\begin{equation}\label{eq:gradient}
 \partial_{\lambda_i}\Q-\partial_{\lambda_j}\Q  =-(\lambda_i-\lambda_j)  \bbE\Br{\frac{Z_iZ_j}{Y}}.
\end{equation}
Integration by parts is justified by truncation followed by dominated
convergence: positive eigenvalues bound $Y$ below by a positive multiple
of $\sum_k Z_k$, and the exponential density controls the tails.

Set $C(w)=\frac13+\frac{(1-w)^2}{6(1+w)}$, which decreases on $[0,1]$
($C'(w)=-(1-w)(3+w)/[6(1+w)^2]\leqslant0$). Along an averaging transfer
$\bsp(t)=\bsp-t\bse_i+t\bse_j$ with $d=p_i-p_j>0$ and $0\leqslant t\leqslant d/2$, the
pair sum $w=p_i+p_j$ is constant, $\frac{\dif}{\dif t}\Q(\bsp(t))
=(d-2t)\bbE[Z_iZ_j/Y_t]$, and $-\frac{\dif}{\dif t}\norm{\bsp(t)}_2^2=2(d-2t)$.
Integrating \eqref{eq:moment} over $[0,t]$ (handling boundary zeros by
mixing with the uniform vector with weight $\varepsilon\downarrow0$ and
scaling $t$ by $1-\varepsilon$) gives
\begin{equation}\label{eq:transfer-gain-fin}
 \Q(\bsp(t))-\Q(\bsp)\geqslant  C(w)\Pa{\norm{\bsp}_2^2-\norm{\bsp(t)}_2^2}  +\frac{d^4-(d-2t)^4}{60}.
\end{equation}
Realize $\bdlambda(\sigma)\prec\bdlambda(\rho)$ by a finite sequence of
averaging transfers (any overshoot $t\in(d/2,d]$ is an averaging transfer
followed by a transposition, and equal-weight pairs need no transfer). Every
intermediate spectrum $\bsr$ along the chain obeys $\bsr\prec\bdlambda(\rho)$, so
every selected pair sum satisfies $w\leqslant s$ and hence
$C(w)\geqslant C(s)$. Telescoping \eqref{eq:transfer-gain-fin} proves
\eqref{eq:mixing}, with strict inequality whenever at least one nontrivial
transfer occurs (by the strictly positive quartic term), i.e., whenever
$\bdlambda(\sigma)\ne\bdlambda(\rho)$.

For optimality of the universal coefficient $1/3$, take a two-point
spectrum $(1/2+\varepsilon,1/2-\varepsilon)$ and mix it to $(1/2,1/2)$.
Equation \eqref{eq:gradient} gives
$\Q(1/2,1/2)-\Q(1/2+\varepsilon,1/2-\varepsilon)=\frac23\varepsilon^2+o(\varepsilon^2)$,
while the purity decrease is $2\varepsilon^2$, so their ratio tends to $1/3$.

For the high-rank limit, let $\bsu_n$ be uniform on $n$ coordinates.
The spectrum $\bsu_n$ majorizes $\bsu_{2n}$ after zero padding, and
$$
 \Q(\bsu_{2n})-\Q(\bsu_n)
 =\sum_{j=n+1}^{2n}\left(\log\frac{j}{j-1}-\frac1j\right),
 \qquad
 \norm{\bsu_n}_2^2-\norm{\bsu_{2n}}_2^2
 =\sum_{j=n+1}^{2n}\frac1{j(j-1)}.
$$
By the Taylor formula for $-\log(1-1/j)$,
$j(j-1)(\log(j/(j-1))-1/j)\to1/2$. The ratio of the two
sums is a weighted average of these terms, so it tends to $1/2$
as the top-two mass $2/n$ tends to zero.
\end{proof}

\subsection{The quantum and multipartite comparison}\label{sec:quantum-main-fin}

For illustration, we now give the full proof of the subentropy estimate used above in the bipartite case, a special one in Theorem~\ref{th:infinite-sharp}:
\begin{eqnarray}\label{eq:native-Q-purity}
 \Delta_{\Q}(\rho_{AB})\geqslant \frac13\Delta_L(\rho_{AB}),\qquad d=mn.
\end{eqnarray}
This derivation does not assume bipartite subadditivity or its equality
characterization. If one local dimension is one, the state is a product with a pure
one-dimensional factor. Padding invariance of $\Q$ then makes both
sides vanish.

For $m,n\geqslant2$, we first work with a pinched state
$$
\rho'=\sum^{m,n}_{i,j=1}\Pi^A_i\ot\Pi^B_j\rho_{AB}\Pi^A_i\ot\Pi^B_j=\sum^{m,n}_{i,j=1}p_{ij}\out{ij}{ij},\qquad p_{ij}=\Innerm{ij}{\rho}{ij},\sum^{m,n}_{i,j=1}p_{ij}=1,
$$
with row and column sums $r_i=\sum^n_{j=1}p_{ij},c_j=\sum^m_{i=1}p_{ij}$ such that $\rho'_A=\diag(\bsr)$ and $\rho'_B=\diag(\bsc)$.
Write $d=mn$. The symbols $\bsr=(r_1,\ldots,r_m)$ and $\bsc=(c_1,\ldots,c_n)$ denote the marginal probability vectors, and $\bsP=(p_{ij})$ denotes the probability table of all entries $p_{ij}$. With these notations, we rewrite
\begin{eqnarray*}
\Delta_{\Q}(\rho_{AB})& = &\Q(\rho_A) + \Q(\rho_B) - \Q(\rho_{AB}) = \Q(\rho'_A) + \Q(\rho'_B) - \Q(\rho_{AB}) \\
&=&  [\Q(\rho'_{AB}) - \Q(\rho_{AB})] +  [\Q(\rho'_A) + \Q(\rho'_B) -  \Q(\rho'_{AB})] \\
&=&  [\Q(\rho') - \Q(\rho)] + [\Q(\bsr) + \Q(\bsc) -  \Q(\bsP)] \\
&=& [\Q(\rho') - \Q(\rho)] + \Delta_{\Q}(\bsP)  
\end{eqnarray*}
and
\begin{eqnarray*}
\Delta_{L}(\rho_{AB})& = &\S_L(\rho_A) + \S_L(\rho_B) - \S_L(\rho_{AB}) = \S_L(\rho'_A) + \S_L(\rho'_B) - \S_L(\rho_{AB}) \\
&=&  [\S_L(\rho'_{AB}) - \S_L(\rho_{AB})] + [\S_L(\rho'_A) + \S_L(\rho'_B) -  \S_L(\rho'_{AB})] \\
&=&  [\S_L(\rho') - \S_L(\rho)] +  [\S_L(\bsr) + \S_L(\bsc) -  \S_L(\bsP)] \\
&=&  [\S_L(\rho') - \S_L(\rho)] + \Delta_{L}(\bsP).
\end{eqnarray*}
In order to show \eqref{eq:native-Q-purity}, it suffices to show that
\begin{eqnarray}
\Delta_{\Q}(\bsP)&\geqslant&\frac13 \Delta_{L}(\bsP),\label{eq:term1}\\
\Q(\rho') - \Q(\rho)&\geqslant& \frac13[\S_L(\rho') - \S_L(\rho)].\label{eq:term2}
\end{eqnarray}
In fact, \eqref{eq:term1} follows immediately from Lemma~\ref{lem:diagonal-core} and \eqref{eq:term2} follows from Proposition~\ref{prop:mixing}.

\begin{proof}[Proof of Theorem~\ref{th:infinite-sharp}]

\begin{itemize}
\item \textbf{Step 1. Dephase in the marginal eigenbases.}
For a bipartite state, choose eigenbases of its two marginals and dephase
in their product basis. The resulting diagonal state $\tau$ has the same
marginals. Hence
$$
\F(\rho)=\F(\tau)+\cR(\tau)-\cR(\rho).
$$
The classical term is nonnegative by Lemma~\ref{lem:diagonal-core}.
By the Schur--Horn theorem~\cite[Chapter~9]{Marshall2011}, the diagonal
of $\rho$ is majorized by its spectrum. Proposition~\ref{prop:mixing}
therefore gives $\mathcal R(\tau)\geqslant\mathcal R(\rho)$.
\item \textbf{Step 2. Classify bipartite equality.}
At equality strictness in Proposition~\ref{prop:mixing} gives equality
of the two spectra up to permutation. Thus $\Tr{\tau^2}=\Tr{\rho^2}$.
In the dephasing basis the difference of these purities is the sum of
the squared moduli of the off-diagonal entries of $\rho$. All these
entries vanish and $\rho=\tau$. Lemma~\ref{lem:diagonal-core} then
gives either a pure marginal or a product of two maximally mixed
rank-two states. A pure
marginal forces a product state: if $\rho_A=\proj{a}$ and
$\bsE=\proj{a}\otimes\I$, then $\Tr{(\I-\bsE)\rho}=0$ implies
$\norm{(\I-\bsE)\rho^{1/2}}_2^2=0$. Hence $\rho=\bsE\rho \bsE=\proj{a}\otimes\rho_B$.
\item \textbf{Step 3. Sum the bipartite gaps.}
For $N$ parties, put $B_i=A_i\cdots A_N$ and apply the
bipartite comparison to the reduced state $\rho_{B_i}$ across
$A_i\mid B_{i+1}$, for $1\leqslant i<N$.
The deficits telescope:
$$
\sum_{i=1}^{N-1} \Br{\cR(\rho_{A_i})+\cR(\rho_{B_{i+1}})  - \cR(\rho_{B_i})} =\sum_{i=1}^N\cR(\rho_{A_i})-\cR(\rho).
$$
\item \textbf{Step 4. Classify multipartite equality.}
Equality requires equality at
every cut, and each bipartite equality state is a product across
that cut. Induction thus forces a full product state. It remains
to determine which factors of this product may be mixed. If it has
at least two mixed factors, the first such factor and its
mixed suffix must both be maximally mixed on two-dimensional
supports. The support of
a tensor product is the tensor product of the supports; a suffix
of rank two consequently contains exactly one mixed factor,
also maximally mixed on a two-dimensional support. All other
factors are pure. If there
is at most one mixed factor, every cut has a pure side.
These states all give equality, completing the classification.
\item \textbf{Step 5. Verify sharpness.}
Finally, $\Q(\I_d/d)=\log d-H_d+1$. For two maximally mixed
rank-two factors, subtraction gives
$$\Delta_{\Q}=2(\log2-1/2)-(\log4-13/12)=1/12,$$
while $\Delta_L=2(1/2)-3/4=1/4$.
Pure spectator factors leave the nonzero spectra unchanged.
Thus the coefficient $1/3$ is optimal in all spaces admitting
two rank-two local supports.\qedhere
\end{itemize}
\end{proof}

\subsection{Single-state, marginal-purity, and continuity bounds}
\label{sec:single-marginal-cont-fin}\label{sec:other-proofs-fin}

Fix the coordinates outside a selected pair and write
$$
 \bdlambda(v)=(\alpha(1+v),\alpha(1-v),\lambda_3,\ldots,\lambda_d),
 \qquad |v|\leqslant1,\quad\alpha>0.
$$
Here the remaining coordinates are nonnegative and
$2\alpha+\sum_{j\geqslant3}\lambda_j=1$.
Let the $Z_i$ be independent $\operatorname{Exp}(1)$ variables and put
$$
 X=\alpha(Z_1+Z_2)+\sum_{j\geqslant3}\lambda_jZ_j,\qquad
 Y=\alpha(Z_1-Z_2).
$$
Then $X>0$, $|Y|\leqslant X$, $\bbE[X]=1$, and $(X,Y)$ has
the same law as $(X,-Y)$. The exponential representation and the identity
\begin{eqnarray*}
 \frac12\big((X+vY)\log(X+vY)+(X-vY)\log(X-vY)\big)-X\log X=X\sum_{k\geqslant1}\frac{(vY/X)^{2k}}{2k(2k-1)}
\end{eqnarray*}
give the expansion
\begin{eqnarray}\label{eq:pair-moment-series}
 \Q(\bdlambda(0))-\Q(\bdlambda(v)) &=&\sum_{k\geqslant1}A_kv^{2k},\\
 A_k&=&\frac1{2k(2k-1)}
             \bbE\left[\frac{Y^{2k}}{X^{2k-1}}\right]>0.
 \label{eq:pair-moment-coeff}
\end{eqnarray}

The pointwise identity follows by integrating the series of
$1/(1-z^2)$ twice, starting at $z=0$. The coefficients sum to
$\log2$, so the series extends to $z=\pm1$ and is bounded by
$X\log2$. In particular the expectation gives a series uniform
for $v^2\in[0,1]$. The strict coefficient signs follow from
$Y\ne0$ almost surely.

\begin{proof}[Proof of Proposition~\ref{prop:single-state-sharp}]
Put $c=2\log2-1$ and, for a probability vector $\bdlambda$, define
$$
\cG(\bdlambda)=\Q(\bdlambda)-c\S_L(\bdlambda).
$$
Fix a finite probability simplex. Keep all coordinates but two
fixed and write the selected pair as
$\alpha(1+v),\alpha(1-v)$, with $\alpha>0$ and $|v|\leqslant1$.
Put $D=v^2$. The expansion \eqref{eq:pair-moment-series} holds on
the entire interval $D\in[0,1]$. Since
$$
\S_L(\bdlambda(v))=\S_L(\bdlambda(0))-2\alpha^2D,$$
we obtain
\begin{equation}\label{eq:single-state-pair-concavity}
 \cG(\bdlambda(v)) =\cG(\bdlambda(0))+(2c\alpha^2-A_1)D-\sum_{k\geqslant2}A_kD^k.
\end{equation}
This function is strictly concave in $D$, since $A_2>0$.

Continuity gives a global minimizer of $\cG$ on the finite
simplex. If it had two unequal positive coordinates, their parameter
would satisfy $0<D<1$, and strict concavity would give a value
strictly greater than the smaller endpoint value in
\eqref{eq:single-state-pair-concavity}. This contradicts minimality.
Every minimizer is therefore uniform on its positive support.

Let $\bsu_k$ be uniform on $k$ coordinates. For $k\geqslant2$,
the uniform subentropy formula gives
\begin{eqnarray*}
 \Q(\bsu_k) &=&\log k-H_k+1 =\sum_{j=2}^k\left(\log\frac j{j-1}-\frac1j\right),\\
 \S_L(\bsu_k)&=&1-\frac1k =\sum_{j=2}^k\frac1{j(j-1)}.
\end{eqnarray*}
For $x\geqslant1$, put $b(x)=x(x+1)\log(1+1/x)-x$. Then
$b'(x)=(2x+1)\log(1+1/x)-2>0$, since strict convexity of $1/t$ and the
midpoint integral bound give $\log(1+1/x)=\int_x^{x+1}\dif t/t>2/(2x+1)$.
It follows that
$$
j(j-1)\Pa{\log\frac j{j-1}-\frac1j} =b(j-1)\geqslant b(1)=c,
$$
strictly for $j>2$. Summation yields $\Q(\bsu_k)\geqslant c\S_L(\bsu_k)$,
with equality only for $k=2$ among mixed uniform vectors. The pure vector
$\bsu_1$ also has $\cG(\bsu_1)=0$, so the zeros of $\cG$ are
exactly the pure and uniform binary vectors. Since
$\Q(\I_2/2)/\S_L(\I_2/2)=2\log2-1$, the lower coefficient is optimal in
every dimension at least two.

For the upper inequalities, put $b=\sqrt{\Tr{\rho^2}}=\sqrt{1-\S_L(\rho)}\in(0,1]$.
For each $t>0$, the function $f_t(v)=t\sqrt v-\log(1+t\sqrt v)$ on
$[0,\infty)$ satisfies $f_t(0)=0$, $f_t'(v)=t^2/[2(1+t\sqrt v)]$, and
$f_t''(v)=-t^3/[4\sqrt v(1+t\sqrt v)^2]<0$ for $v>0$. Strict concavity and
$f_t(0)=0$ imply subadditivity $f_t(x)+f_t(y)\geqslant f_t(x+y)$, strictly
when $x,y>0$. Summing over the eigenvalues $(\lambda_i)$ of $\rho$ yields
$$\sum_i\bigl(t\lambda_i-\log(1+t\lambda_i)\bigr)
 =\sum_i f_t(\lambda_i^2)\geqslant f_t(b^2)=tb-\log(1+tb),$$
whence $\prod_i(1+t\lambda_i)^{-1}\geqslant\ee^{-(1-b)t}/(1+bt)$.
Integrating in \eqref{eq:determinant-integral} gives
$\Q(\rho)\leqslant\Q^\circ(b)$, strictly whenever $\rho$ is mixed.

Next, differentiating \eqref{eq:diffuse-atom-fin} on $(0,1)$ (dominated on
compact subintervals by an integrable multiple of $(1+Z)(1+Z+|\log Z|)$)
and integrating by parts against the exponential density yields, for
$x=(1-b)/b>0$,
\begin{equation}\label{eq:Qcirc-deriv-fin}
 -\frac{\dif}{\dif b}\Q^\circ(b)
 =\bbE\bigl[(Z-1)\log(bZ+1-b)\bigr]
 =\bbE\!\left[\frac{Z}{x+Z}\right]
 =\int_0^\infty\frac{\ee^{-xu}}{(1+u)^2}\,\dif u
 <\frac1{1+x}=b,
\end{equation}
where the strict inequality is Jensen's inequality for the strictly concave
function $z\mapsto z/(x+z)$. Integrating \eqref{eq:Qcirc-deriv-fin} from
$b$ to $1$ gives $\Q^\circ(b)<\frac12(1-b^2)=\frac12\S_L(\rho)$ for all
$b\in(0,1)$ (and $\Q^\circ(0)=1-\gamma<1/2$), with equality at $b=1$.
For sharpness of the upper envelope at any $s=1-b^2\in(0,1)$ as dimension
grows, take $K$ with $(K+1)b^2\geqslant1$, set
$a_K=(1+\sqrt{K((K+1)b^2-1)})/(K+1)$, and form
$\bsp_K=(a_K,(1-a_K)/K,\ldots,(1-a_K)/K)$. The equation
$a_K^2+(1-a_K)^2/K=b^2$ gives $\S_L(\bsp_K)=s$ exactly and
$a_K\to b$. For fixed $t>0$ its Laplace transform is
$$
 (1+a_Kt)^{-1}(1+(1-a_K)t/K)^{-K} \longrightarrow (1+bt)^{-1}\ee^{-(1-b)t}.
$$
For any probability vector the transform is bounded above by $(1+t)^{-1}$
by multiplication of the positive factors, and bounded below by $\ee^{-t}$
by $1+x\leqslant\ee^x$. Thus the integrable function
$[(1+t)^{-1}-\ee^{-t}]/t^2$ dominates the subentropy integrands, giving
$\Q(\bsp_K)\to\Q^\circ(b)$. Finally, the spectral formula for a binary law yields
$$
 \Q(1-\varepsilon,\varepsilon) =\frac{-(1-\varepsilon)^2\log(1-\varepsilon)  +\varepsilon^2\log\varepsilon}{1-2\varepsilon} =\varepsilon+O(\varepsilon^2|\log\varepsilon|).
$$
Its linear entropy is $2\varepsilon(1-\varepsilon)$, so the ratio tends
to $1/2$. Zero padding preserves this example in every dimension
at least two.
\end{proof}

\begin{proof}[Proof of Proposition~\ref{prop:linear-marginal-product}]
For $1\leqslant i\leqslant N$ put $u_i=\sqrt{\Tr{\rho_{A_i}^2}}\in(0,1]$,
and set $w=\sqrt{\Tr{\rho^2}}\in(0,1]$. We first prove
\begin{equation}\label{eq:schatten-two-marginal-comparison}
 \sum_{i=1}^N u_i-(N-1)\leqslant w.
\end{equation}
Choose marginal eigenbases $(e^{(i)}_{k_i})$, write
$e_{\bsk}=\bigotimes_{i=1}^N e^{(i)}_{k_i}$, and define the $N$-way
probability table $p_{\bsk}=\Innerm{e_{\bsk}}{\rho}{e_{\bsk}}$,
whose $i$-th marginal $\bsr^{(i)}$ is the eigenvalue list of $\rho_i$.
Then $\sum_{k_i}(r^{(i)}_{k_i})^2=u_i^2$ and $\sum_{\bsk}p_{\bsk}^2\leqslant w^2$.
Set $\alpha^{(i)}_{k_i}=r^{(i)}_{k_i}/u_i\in[0,1]$, so
$\sum_{k_i}(\alpha^{(i)}_{k_i})^2=1$. Inducting from $x+y-1\leqslant xy$
on $[0,1]$ gives
$\sum_{i=1}^N\alpha^{(i)}_{k_i}-(N-1)\leqslant\prod_{i=1}^N\alpha^{(i)}_{k_i}$.
Multiplying by $p_{\bsk}$, summing over $\bsk$, and applying Cauchy--Schwarz
gives
$$
\sum_{i=1}^N u_i-(N-1) =\sum_{\bsk}p_{\bsk}\!\left(\sum_{i=1}^N\alpha^{(i)}_{k_i}-(N-1)\right)
 \leqslant\sum_{\bsk}p_{\bsk}\prod_{i=1}^N\alpha^{(i)}_{k_i}
 \leqslant\left(\sum_{\bsk}p_{\bsk}^2\right)^{1/2}\leqslant w,
$$
proving \eqref{eq:schatten-two-marginal-comparison}.

Set $e_i=1-u_i\in[0,1)$, $S=\sum_{i=1}^N e_i$,
$E_2=\sum_{1\leqslant i<j\leqslant r}e_ie_j$, and
$\Pi=\prod_{i=1}^N(1-e_i)$. Twice the difference of the two sides of
\eqref{eq:linear-marginal-product} is
$D:=(N-1)-\sum_{i=1}^N u_i^2+2w^2-\prod_{i=1}^N u_i^2$.
If $S\leqslant1$, then $w\geqslant1-S\geqslant0$ and Bonferroni's
inequality gives $0\leqslant\Pi\leqslant1-S+E_2$; squaring yields
$$
D\geqslant\bigl[(1-S)^2+2E_2\bigr]-\bigl[(1-S)^2+2(1-S)E_2+E_2^2\bigr]   =E_2(2S-E_2)\geqslant0,
$$
since $2E_2\leqslant S^2\leqslant S$. Whenever at least two marginals are
mixed, $E_2>0$ and $D>0$ strictly.

If $S>1$, set $G(u_1,\ldots,u_N)=(N-1)-\sum_{i=1}^N u_i^2-\prod_{i=1}^N u_i^2$.
Then $D\geqslant G(u_1,\ldots,u_N)$ and $\partial_{u_i}G<0$ on $(0,1]^N$.
Since $S-1<S=\sum_{i=1}^N(1-u_i)$, we can increase $u_i$ to
$\tilde u_i\in[u_i,1]$ with $\sum_{i=1}^N\tilde u_i=N-1$. Because every
$\tilde u_i\geqslant u_i>0$, at least two $\tilde u_i$ are strictly below
$1$, so the $S=1$ case gives
$D\geqslant G(u_1,\ldots,u_N)\geqslant G(\tilde u_1,\ldots,\tilde u_N)>0$.
If at most one marginal is mixed, $\rho$ is a product with at most one
mixed factor and both sides of \eqref{eq:linear-marginal-product} vanish.

For optimality, embed
$\rho_\varepsilon=\diag(1-2\varepsilon,\varepsilon,\varepsilon,0)$
($0<\varepsilon<1/2$) on two local two-dimensional supports and tensor with
pure states on the remaining factors. Then
$\Delta_L(\rho_\varepsilon)=2\varepsilon^2$, while
$\Delta_L(\bigotimes_i\rho_i)=4\varepsilon^2(1-\varepsilon)^2$, whose
ratio tends to $1/2$ as $\varepsilon\downarrow0$. On this same bipartite
family, expanding the confluent spectral formula for
$(1-2\varepsilon,\varepsilon,\varepsilon)$ and the binary formula for
$(1-\varepsilon,\varepsilon)$ yields
$$\Delta_{\Q}(\rho_\varepsilon)
 =\varepsilon^2\bigl(\log(1/\varepsilon)-2\bigr)
  +\varepsilon^3\bigl(6\log(1/\varepsilon)-9\bigr)
  +O\bigl(\varepsilon^4\log(1/\varepsilon)\bigr),$$
and hence
$\Delta_{\Q}(\rho_\varepsilon)/\Delta_L(\rho_\varepsilon)
 =\frac12\log(1/\varepsilon)-1+O(\varepsilon\log(1/\varepsilon))\to\infty$,
confirming the remark after Theorem~\ref{th:subentropy-marginal-product}.
\end{proof}

\begin{proof}[Proof of Theorem~\ref{th:subentropy-marginal-product}]
\begin{itemize}
\item \textbf{Step 1. Prove the bipartite bound by dephasing.}
For $N=2$, dephase in the product of the marginal eigenbases to obtain
$\tau$. Both marginals are preserved, and Proposition~\ref{prop:mixing}
gives
$\Q(\tau)-\Q(\rho)\geqslant\frac13(\Tr{\rho^2}-\Tr{\tau^2})\geqslant0$,
with equality iff $\rho=\tau$. Appendix~\ref{app:sharp-marginal-product}
proves the bipartite classical bound and its equality classification,
yielding \eqref{eq:subentropy-marginal-product} and its equality cases for
$N=2$.
\item \textbf{Step 2. Combine the two suffix bounds.}
Let $N\geqslant3$ and assume the theorem holds for $N-1$ parties.
Write $B=A_2\cdots A_N$, $s_i=\S_L(\rho_i)\in[0,1)$, $x=\S_L(\rho_B)\in[0,1)$,
$S=\sum_{i=2}^Ns_i$, and $P=\prod_{i=2}^N(1-s_i)>0$. Decompose
$\Delta_{\Q}(\rho)=\Delta_{\Q}(\rho_{A_1\mid B})+\Delta_{\Q}(\rho_B)$.
The bipartite case on $A_1\mid B$ gives
$\Delta_{\Q}(\rho_{A_1\mid B})\geqslant\frac13 s_1 x$, while combining
Theorem~\ref{th:infinite-sharp} and the induction hypothesis on $\rho_B$
gives $\Delta_{\Q}(\rho_B)\geqslant\frac13\max\set{S-x,\,S-1+P}$. Adding the
two bounds gives
$$
3\Delta_{\Q}(\rho)\geqslant\max\Set{S-(1-s_1)x,\;s_1x+S-1+P}.
$$
\item \textbf{Step 3. Compare the two affine bounds at their intersection.}
The first affine branch is nonincreasing in $x$ since $s_1<1$,
and the second is nondecreasing since $s_1\geqslant0$. In detail,
subtracting the second branch from the first gives $1-P-x$.
Thus the first is the maximum up to $x=1-P$ and the second
is the maximum thereafter. They intersect at $x_*=1-P$. For $x\leqslant x_*$ use the first bound, and for $x\geqslant x_*$ use the second. At $x_*$,
both bounds equal $s_1(1-P)+S-1+P=\Delta_L(\bigotimes_{i=1}^N\rho_i)$,
proving \eqref{eq:subentropy-marginal-product}.
\item \textbf{Step 4. Classify equality and verify sharpness.}
For equality, if $s_1=0$ then $\rho=\rho_{A_1}\otimes\rho_B$ and equality
reduces to the induction hypothesis on $\rho_B$. If $s_1>0$, then
$0<s_1<1$, so the two slopes have strictly opposite signs; equality forces
$x=1-P$, bipartite equality across $A_1\mid B$, and equality in both
suffix bounds for $\rho_B$. If $x=0$, then $P=1$, so every suffix marginal
is pure and $\rho$ is a product with only $\rho_{A_1}$ mixed. If $x>0$,
bipartite equality forces $\rho=(\bsP_{A_1}/2)\otimes(\bsP_B/2)$, while equality
in Theorem~\ref{th:infinite-sharp} forces $\rho_B$ to be a full product of
rank two, hence having one uniform rank-two factor and pure remaining
factors. Conversely, every product state in the equality class of
Theorem~\ref{th:infinite-sharp} attains equality. Embedding
$(\bsP_A/2)\otimes(\bsP_B/2)$ with pure spectators proves optimality of $1/3$.\qedhere
\end{itemize}
\end{proof}

\begin{proof}[Proof of Theorem~\ref{th:finite-continuity}]
\begin{itemize}
\item \textbf{Step 1. Compare the trace distance and the spectral distance.}
Without loss of generality assume $\Q(\sigma)\geqslant\Q(\rho)$, and let
$\bsp=\bdlambda^\downarrow(\rho)$ and $\bsq=\bdlambda^\downarrow(\sigma)$ be
their nonincreasing eigenvalue vectors in $\Delta_{d-1}$. By Mirsky's
eigenvalue inequality~\cite[Chapter~9]{Marshall2011},
$\tau:=\frac12\norm{\bsp-\bsq}_1\leqslant T$. If $\tau=0$, then $\bsp=\bsq$,
so the subentropy difference is zero and
\eqref{eq:finite-continuity-refined} holds with both sides zero.
\item \textbf{Step 2. Construct the probability table from the shared masses.}
Suppose $\tau>0$. Set $r_i=\min(p_i,q_i)$, $u_i=p_i-r_i$, and
$v_i=q_i-r_i$ for $1\leqslant i\leqslant d$, so $r_1\geqslant\cdots\geqslant r_d$,
$\sum_{i=1}^d r_i=1-\tau$, and $\sum_{i=1}^d u_i=\sum_{i=1}^d v_i=\tau$.
Because $\bsp\neq \bsq$, at least one index $i_0$ has $p_{i_0}>q_{i_0}$, so
$v_{i_0}=0$; let $w=(w_1,\ldots,w_{d-1})$ list the remaining $d-1$ entries
of $\bsv$. Define the $d\times d$ probability table $\bsP$ with first row
$(r_1,\frac{\tau}{d-1},\ldots,\frac{\tau}{d-1})$ and $i$-th row
$(r_i,0,\ldots,0)$ for $2\leqslant i\leqslant d$. Its entries, after discarding the unused zero positions, form
$\bsP_{\rm vec}=(r_1,\ldots,r_d,\frac{\tau}{d-1},\ldots,\frac{\tau}{d-1})$,
its row marginal is $\bsr+\tau \bse_1=(r_1+\tau,r_2,\ldots,r_d)$, and its column
marginal is $\bsc_\tau:=(1-\tau,\frac{\tau}{d-1},\ldots,\frac{\tau}{d-1})$.
Observe three relations:
\begin{enumerate}[(1)]
\item Splitting each $q_i=r_i+v_i$ into $(r_i,v_i)$ via $2\times2$ doubly
stochastic matrices and then averaging $w$ to
$(\frac{\tau}{d-1},\ldots,\frac{\tau}{d-1})$ yields the majorization chain
$\bsq\oplus0_{d-1}\succ(r,w)\succ \bsP_{\rm vec}$; by zero-padding invariance
and Proposition~\ref{prop:mixing}, $\Q(\bsq)\leqslant\Q(\bsP_{\rm vec})$.
\item For every $1\leqslant k\leqslant d$,
$\sum_{i=1}^k p_i=\sum_{i=1}^k r_i+\sum_{i=1}^k u_i
 \leqslant\sum_{i=1}^k r_i+\tau=\sum_{i=1}^k(\bsr+\tau \bse_1)_i$,
so $\bsp\prec \bsr+\tau \bse_1$ and $\Q(\bsp)\geqslant\Q(\bsr+\tau \bse_1)$.
\item Direct computation of squared norms gives
$\Delta_L(\bsP)=2\tau\sum_{i=2}^d r_i=2\tau(1-\tau-r_1)$.
\end{enumerate}
\item \textbf{Step 3. Apply the diagonal estimate at small spectral distance.}
Applying Lemma~\ref{lem:diagonal-core} to $\bsP$ gives
$$
\Q(\bsq)-\Q(\bsp)\leqslant\Q(\bsP_{\rm vec})-\Q(\bsr+\tau \bse_1) \leqslant\Q(\bsc_\tau)-\frac23\tau(1-\tau-r_1),
$$
proving \eqref{eq:finite-continuity-refined}. Note that $r_1\geqslant1/d$
forces $\tau\leqslant1-1/d$. Since $t\mapsto\Q(\bsc_t)$ is strictly
increasing on $[0,1-1/d]$ (by strict Schur concavity in
Proposition~\ref{prop:mixing}) with maximum $\Q(\bsu_d)=\log d-H_d+1$,
and $\tau\leqslant T$, this gives the upper bound
$\Q(\bsc_{\min\{T,1-1/d\}})$ for distances at most $T$. Moreover,
$$
(1+\tau z/(d-1))^{-(d-1)}>\ee^{-\tau z}
$$ 
for all $z>0$ when $\tau>0$. Thus \eqref{eq:determinant-integral} gives
$\Q(\bsc_\tau)<\Q^\circ(1-\tau)$. Because $\Q^\circ(1-t)$ is strictly
increasing on $[0,1]$ (by \eqref{eq:Qcirc-deriv-fin}) and satisfies
$\Q^\circ(1-t)\leqslant\min\{t-t^2/2,1-\gamma\}$ (from integrating
\eqref{eq:Qcirc-deriv-fin} over $[1-t,1]$), monotonicity in $\tau\leqslant T$
yields 
$$
|\Q(\rho)-\Q(\sigma)|\leqslant\Q^\circ(1-T)\leqslant\min\set{T-T^2/2,1-\gamma}.
$$
\item \textbf{Step 4. Use the least eigenvalue at large trace distance.}
For $T>1-1/d$, let $\bsE$ be the projection onto the strictly positive
spectral subspace of $\rho-\sigma$. It is nonzero and
$$
T=\Tr{\bsE\rho}-\Tr{\bsE\sigma}\leqslant1-\Tr{\bsE\sigma}\leqslant1-q_d.
$$
Thus $q_d\leqslant1-T<1/d$. Averaging the first $d-1$ entries of $q$ gives
a probability vector majorized by $q$. Its first $d-1$ entries equal
$(1-q_d)/(d-1)\geqslant T/(d-1)$, so it majorizes the decreasing
rearrangement of $\bsc_T$. Schur concavity and nonnegativity then give
$\Q(\sigma)-\Q(\rho)\leqslant\Q(\bsq)\leqslant\Q(\bsc_T)$. Equality forces
$\rho=\proj{\psi}$ to be pure and, by strict Schur concavity, $q$ to
be a permutation of $\bsc_T$. In particular, $q_d=1-T$ and
$\Tr{\bsE\rho}=1$, $\Tr{\bsE\sigma}=q_d$. If $T<1$, the latter identity
implies that $E$ has rank one, since $\sigma\geqslant q_d\I_d$ and
$q_d>0$. Its range is therefore the eigenspace of $\sigma$ at $q_d$
and contains $\ket{\psi}$. If $T=1$, $\Tr{\bsE\sigma}=0$ implies that
the range of $\bsE$ lies in the kernel of $\sigma$, which is
one-dimensional and again contains $\ket{\psi}$.
\item \textbf{Step 5. Classify equality in the small-distance range.}
For $0<T\leqslant1-1/d$, strict monotonicity of $t\mapsto\Q(\bsc_t)$
forces $\tau=T$ at equality. The correction $\frac23\tau(1-\tau-r_1)$
must vanish,
so $r_1=1-T$ and $r_2=\cdots=r_d=0$. Thus $r+\tau e_1=(1,0,\ldots,0)$
and equality gives $\Q(\bsp)=0$, so $\rho=\proj{\psi}$ is pure. Strict
Schur concavity in $\Q(\bsq)=\Q(\bsP_{\rm vec})=\Q(\bsc_T)$ gives $\bsq=\bsc_T$.
At $T=1-1/d$ this gives $\sigma=\I_d/d$. At $T<1-1/d$ the top
eigenvalue $1-T$ is simple. Since
$T\geqslant1-\Innerm{\psi}{\sigma}{\psi}\geqslant1-(1-T)=T$,
the vector $\ket{\psi}$ belongs to this top eigenspace.
\item \textbf{Step 6. Realize the maximum at every distance.}
Conversely, for every $T\in(0,1]$ the stated pair has
$\rho-\sigma$ with eigenvalues $T,-T/(d-1),\ldots,-T/(d-1)$ and
subentropy difference $\Q(\bsc_T)$. This proves the fixed-distance formula
and its equality classification, including $T=0$. Taking such a pair at
distance $\min\set{T,1-\frac1d}$ realizes the maximum over distances at most $T$.\qedhere
\end{itemize}
\end{proof}

\section{Proof of the sharp marginal-product bound}
\label{app:sharp-marginal-product}

For row and column marginals $\bsr,\bsc$, write
$$
 R=\sum_{i<\infty}r_i^2,\qquad C=\sum_{j<\infty}c_j^2,\qquad  \pi_{\bsP}=\sum_{i,j<\infty}p_{ij}^2,
$$
where diffuse mass contributes zero to every squared sum. The largest
ordinary marginal atoms are $a,b$, and the second largest are
$a_2=r_2,b_2=c_2$, with zero padding when necessary. Throughout this appendix set
$$
 \begin{aligned}
 H(\bsP)&=\Delta_{\Q}(\bsP)-\tfrac13(1-R)(1-C),\\
 \Phi(\bsr,\bsc)&=\Q(\bsr)+\Q(\bsc)-\tfrac13(1-R)(1-C),\\
 \cJ(\bsp)&=1-\gamma-\Q(\bsp),\qquad (s)_+=\max(s,0).
 \end{aligned}
$$
For a generalized spectrum, $\Q$ in these expressions means the
diffuse completion $\Q^\circ$ defined next.

The two linear-entropy expressions differ by
$\Delta_L(\bsP)-\S_L(\bsr)\S_L(\bsc)=\pi_{\bsP}-RC$. Consequently
$$
H(\bsP)=\Pa{\Delta_{\Q}(\bsP)-\frac13\Delta_L(\bsP)}   +\frac{\pi_{\bsP}-RC}{3}.
$$
For finite tables with $\pi_{\bsP}\geqslant RC$, Lemma~\ref{lem:diagonal-core}
already gives $H(\bsP)\geqslant0$. When $\pi_{\bsP}<RC$, the deficit in that lemma
must also cover $(RC-\pi_{\bsP})/3$. We prove the full comparison by
reducing a nonpositive minimum of $H$ to finitely many marginal
parameters. The reduction includes countable tables and diffuse limits
and gives the equality statement used in the companion trace-class paper.

\begin{prop}[Sharp marginal-product comparison]\label{prop:mp-comparison}
Every finite, countable, or generalized probability table satisfies
$H(\bsP)\geqslant0$. Equality holds exactly when one marginal is pure,
or, after deleting zero rows and columns and permuting labels, when
$\bsP$ is the uniform $2\times2$ table. In particular the coefficient
$1/3$ is optimal.
\end{prop}

The proof proceeds in four stages. First, compactness produces a minimizing
generalized table. Second, first variations constrain the rows and columns
of a minimizer with mixed marginals. Third, block estimates bound the
objective in the binary and boundary regions and reduce the remaining
cases to interval tests. Fourth, the coverage argument
in Section~\ref{sec:mp-coverage} combines these tests with the analytic
estimates to exclude a negative minimum and classify equality.
Table~\ref{tab:mp-coverage} summarizes the parameter domains, the
estimates used on each domain and the finite coverage checks.
The accompanying \texttt{verification} directory contains the complete
source files for these checks, the terminal counts and a
\nolinkurl{README.md} giving the build and reproduction instructions
and source file digests.

\subsection{Generalized tables and compactness}\label{generalized-tables-and-stationarity}

\begin{lem}[Diffuse completion and compactness]\label{lem:mp-compactness}
The generalized tables obtained by allowing one diffuse row and one
diffuse column attain the infimum of $H$ over finite tables. Their
subentropies and purities are continuous under sorted truncation, and
every finite set of retained data is approximable by genuine finite
tables.
\end{lem}

\begin{proof}
For a subprobability vector $\bsp$, interpret the missing mass as the
limit of atoms of vanishing size and define

\begin{equation}\label{eq:mp-11}
L_{\bsp}^\circ(z)=\ee^{-z(1-\sum_i p_i)}\prod_i(1+zp_i)^{-1},
 \qquad  \Q^\circ(\bsp)=\int_0^\infty  \frac{(1+z)^{-1}-L_{\bsp}^\circ(z)}{z^2}\,\dif z.
\end{equation}
With independent unit exponentials $Z_i$, put
$X=1-\sum_i p_i+\sum_i p_iZ_i$. Then $\bbE [X]=1$, and Jensen's
inequality gives $\bbE[X^2]\leqslant2$ by treating the deterministic
term as a weighted constant. Its Laplace transform is $L_{\bsp}^\circ$.
The scalar integral identity in Section~\ref{sec:preliminaries}
therefore gives $\Q^\circ(\bsp)=1-\gamma+ \bbE[\eta(X)]$.

Since 
$$
 \log L_{\bsp}^\circ(z)=-z+\sum_i(zp_i-\log(1+zp_i)),
$$
the transforms are between $\ee^{-z}$ and $(1+z)^{-1}$. The upper bound follows for
finite $p$ by the product inequality $\prod_i(1+zp_i)\ee^{z(1-\sum_i p_i)}\geqslant1+z$ and
passes to the limit. Thus the integrand is dominated by the integrable function
$((1+z)^{-1}-\ee^{-z})/z^2$. If a collection of retained atoms is replaced by diffuse mass
and the sum of squares of those atoms is at most $\sigma$, the change in $\Q^\circ$ is
nonnegative and at most

\begin{equation}\label{eq:mp-12}
E(\sigma)=\frac{\sigma}{2}\log(1+\sigma^{-1/2})
            +\frac{\sigma}{2}.
\end{equation}

Indeed $0\leqslant zx-\log(1+zx)\leqslant z^2x^2/2$ for $z,x\geqslant0$. The decrease of $L_{\bsp}^\circ$ is at
most $(1+z)^{-1}\min(1,z^2\sigma/2)$. Integrate below and above $z=\sigma^{-1/2}$ to
obtain \eqref{eq:mp-12}.

For a table sorted by both decreasing marginal masses, keep the first $m$ rows and $m$
columns and replace the discarded rows and columns by diffuse mass. The retained data
are the $m^2$ cells and the first $m$ masses of each marginal. In each of the three
spectra, any omitted atom has mass at most $1/(m+1)$. Its squared-sum error is at most
$1/(m+1)$, and its subentropy error is at most $E(1/(m+1))$.

Each set of finite retained data arises from a nonnegative $(m+1)\times(m+1)$ block
table of mass one. The last row and column contain the aggregate masses outside the
retained rows or columns, and are interpreted as diffuse. Conversely, any such data can
be approximated by actual tables: replace the last row and column by $K$ equally
weighted rows or columns, distributing the last corner uniformly over the $K^2$ cells,
and let $K\to\infty$. The noncore joint and marginal atoms have sizes $O(1/K)$, so
\eqref{eq:mp-12} gives convergence of their subentropies and the corresponding squared
sums converge as well. Therefore a negative $H$ on one of these block tables would imply
a finite classical counterexample.

For compactness, represent the generalized table by a nonnegative array $\bsP$ on $(\bbN\cup\set{\infty})^2$, where a cell with an $\infty$ index counts as diffuse in the joint
spectrum. An $\infty$ index is diffuse on the corresponding marginal. Take a sequence of
ordinary finite tables with decreasing marginal masses along which $H$ tends to its
infimum. By diagonal subsequence extraction, each marginal atom and each cell with two
finite indices converges. In the limit put 
$$
 p_{i\infty}=r_i-\sum_{j<\infty}p_{ij},\qquad
 p_{\infty j}=c_j-\sum_{i<\infty}p_{ij},\qquad
 p_{\infty\infty}=1-\sum_{i<\infty}r_i-\sum_{j<\infty}c_j+
                   \sum_{i,j<\infty}p_{ij}.
$$
The first two values are nonnegative by passing to the limit in finite sums. The last
is the limit of the lower-right mass beyond successively larger finite rectangles, after
taking the sequence limit at each fixed rectangle. It too is nonnegative. Estimate
\eqref{eq:mp-12} and the squared-sum error imply convergence of all three subentropies
and purities, hence convergence of $H$. Conversely, truncation followed by the
diffuse approximation just described approximates each generalized table in these
data by finite tables. Thus the extracted limit minimizes $H$ over the generalized
tables, and its value is the infimum over finite tables.

For later use, the finite approximants can satisfy the same upper bounds
on the largest marginal atoms, the marginal purities, or their sum as
the limit, provided those bounds are positive. If an approximant
exceeds a bound, split every row in proportions $1-\varepsilon,\varepsilon$
and every column in the same proportions, with $0\leqslant\varepsilon\leqslant1/2$.
Each largest marginal atom is multiplied by $1-\varepsilon$ and each
marginal purity by $(1-\varepsilon)^2+\varepsilon^2$.
Since any excess tends to zero, $\varepsilon$ can be chosen to tend to zero.
After padding and keeping the original labels for the main copies, the
split and unsplit joint lists differ in $\ell^1$ by
$2[1-(1-\varepsilon)^2]$. The exponential representation gives
continuity uniformly over these lists. Indeed $\eta$ obeys
$|\eta(x)|\leqslant1/\ee+x^2$, the corresponding exponential sums have
fourth moments bounded by $24$ by Jensen, and their coupled difference
has second moment equal to the squared $\ell^2$ distance of the lists.
Consequently this split preserves the limiting $H$.
\end{proof}

\begin{prop}[Pure and entirely diffuse marginals]\label{prop:mp-endpoints}
A pure marginal gives $H=0$. If one marginal is entirely diffuse
and the other is mixed, then $H>0$. Thus a nonpositive generalized
minimum with mixed marginals has a positive ordinary atom on each side.
\end{prop}

\begin{proof}
A pure marginal gives $H=0$, since the other marginal has the same
spectrum as the joint table. Consider a global minimum with $H\leqslant0$
and both marginals mixed. If all row mass were diffuse, the joint spectrum would be entirely
diffuse and $H=\Q^\circ(\bsc)-\S_L(\bsc)/3$. The diffuse approximation and
Proposition~\ref{prop:single-state-sharp} give
$$
H\geqslant(2\log 2-4/3)\S_L(\bsc)>0.
$$
Transposition treats an entirely diffuse column marginal.
Thus each marginal has a positive ordinary atom at such a minimum.
\end{proof}

\subsection{Spectrum calculus}

\begin{lem}[Spectrum calculus]\label{lem:mp-calculus}
For a mixed generalized spectrum $p$ of ordinary mass $1-\delta$, put
$$
 I_{\bsp}=\int_0^\infty L_{\bsp}^\circ(z)\,\dif z=\bbE[1/X_{\bsp}],\qquad
 g_{\bsp}(x)=\int_0^\infty L_{\bsp}^\circ(z)\frac{x}{1+zx}\,\dif z,
 \qquad g_{\bsp}(0)=0.
$$
The function $g_{\bsp}$ is increasing and concave, and $g_{\bsp}(x)/x$ decreases
on $x>0$. With diffuse compensation, the derivative of $\cJ$
in the ordinary coordinates is $\sum_i h_i g_{\bsp}(p_i)$. Equivalently,
$$
 \rD\sfQ^\circ(\bsp)[\bsh]=-\sum_i h_i g_{\bsp}(p_i).
$$
For an ordinary probability spectrum, $\bsh$ is a finitely supported
trace-zero direction. Moreover,
$$
 \sum_i g_{\bsp}(p_i)=1-\delta I_{\bsp},\qquad I_{\bsp}\geqslant1.
$$
If $x>0$ is the largest atom, then
$$
 \frac{g_{\bsp}(x)}x\leqslant\frac{1-\delta I_{\bsp}}{1-\delta}\leqslant1.
$$
The tangent and secant comparisons proved below include diffuse mass.
\end{lem}

\begin{proof}
Differentiate the determinant representation, writing
$X_{\bsp}=1+\sum_i p_i(Z_i-1)$. The logarithmic derivative of $L_{\bsp}^\circ$ in an ordinary
coordinate is $z-z/(1+zp_i)$. Cancellation of the trace term gives
the stated derivative. A mixed spectrum has either two positive
reciprocal factors or a positive diffuse exponential; these persist
on a short segment and dominate at infinity. Trace cancellation
controls the origin. Zero coordinates have derivative zero. The
integral formula proves the monotonicity and concavity assertions.
Since
$$
 -\frac{\dif}{\dif z}L_{\bsp}^\circ(z)
 =L_{\bsp}^\circ(z)\left(\delta+\sum_i\frac{p_i}{1+zp_i}\right),
$$
integration gives the gradient sum. Jensen gives
$I_{\bsp}=\bbE[1/X_{\bsp}]\geqslant1$. Decreasing $g_{\bsp}(y)/y$ implies
$\sum_i g_{\bsp}(p_i)\geqslant(1-\delta)g_{\bsp}(x)/x$.
The denominator is positive whenever an ordinary atom is present.

At an arbitrary table with its largest column of mass $b=1-\epsilon$, move mass
$\epsilon$ from the one-column table with row marginal $\bsr$ into the other columns. The
concavity of $\Q$ and its directional derivative in the determinant integral give

\begin{equation}\label{eq:mp-8}
\Q(\bsP)-\Q(\bsr)\leqslant\epsilon g_{\bsr}(a).
\end{equation}

Indeed, transfers out of cells of mass $r_i$ have marginal weights $g_{\bsr}(r_i)$, which are
at most $g_{\bsr}(a)$. This remains valid with diffuse mass, with $L_{\bsp}$ replaced by
$L_{\bsp}^\circ$ and zero gradients assigned to diffuse cells.
Pure row marginals follow by continuity.

For any two positive atom masses $x,y$ of a marginal spectrum $p$, or a positive atom $x$ and $y=0$, one has

\begin{equation}\label{eq:mp-13}
\frac{g_{\bsp}(x)-g_{\bsp}(y)}{x-y} =\int_0^\infty\frac{L_{\bsp}^\circ(z)}{(1+zx)(1+zy)}\,\dif z \ \geqslant\ \frac23 \qquad (x\neq y).
\end{equation}

To verify this, put $s=x+y$. The other atom factors and the diffuse factor are at least
$\ee^{-(1-s)z}$ in the Laplace transform. The two factors in $x,y$ occur twice in the
integrand and their product is bounded above by $(1+sz/2)^4$. The resulting lower bound
on the integral is $\bbE[1/(1-s+sZ)]$, where $Z$ is gamma with shape four and rate two.
This is convex in $s\in[0,1]$. Its left derivative at one equals $\bbE[1/Z^2-1/Z]=0$,
since both inverse moments are $2/3$. For domination near one, the difference quotients
are bounded by a constant times $(1+Z)/Z^2$, which is integrable. A convex function with
zero left derivative at the right endpoint is minimized there. This proves
\eqref{eq:mp-13}, with strict inequality if $s<1$ because the expectation is strictly
convex. If $y=0,s=1$, the comparison of the two products is strict for $z>0$, so the
slope for $y=0$ is always strictly larger than $2/3$.
We write $K_{\bsp}(x,y)$ for the integral in \eqref{eq:mp-13},
also at $x=y$ by continuity.

\end{proof}

\begin{lem}[Transform envelopes and normalized kernel]\label{lem:mp-transforms}
Let $\bsp$ have purity $u=\sum_i p_i^2$. For $z\geqslant0$,
$$
 \frac{\ee^{-(1-\sqrt u)z}}{1+\sqrt u\,z}
 \leqslant L_{\bsp}^\circ(z)\leqslant(1+uz)^{-1/u},
$$
with exponential limits at $u=0$. If $u<1$, then
\begin{equation}\label{eq:mp-72}
 I_{\bsp}=\int_0^\infty L_{\bsp}^\circ(z)\,\dif z  \leqslant\frac1{1-u}.
\end{equation}
For $z,y>0$ define the normalized kernel
$$
 b(z,y):=\frac{zy-\log(1+zy)}{z^2y^2}
       =\int_0^1\frac{v\,\dif v}{1+zyv},\qquad b(z,0)=\frac12.
$$
Its continuous value at $z=0$ is also $1/2$. It decreases in each
argument and is a positive mixture of reciprocals in either argument.
Consequently
 \begin{eqnarray*}
 D_{\bsp}(z)&:=&\log L_{\bsp}^\circ(z)+z=z^2\sum_i p_i^2 b(z,p_i),\\
 D_{\bsP}(z)&=&z^2\sum_{i,j<\infty}p_{ij}^2 b(z,p_{ij}).
 \end{eqnarray*}
\end{lem}

\begin{proof}
Give an atom size $p_i$ probability $p_i$ and give size zero the diffuse
probability $\delta$. Its mean is $u$. The function
$y\mapsto\log(1+zy)/y=\int_0^z(1+vy)^{-1}\,\dif v$ is convex,
including its value $z$ at zero. Size-biased Jensen therefore gives
the upper transform. Integration uses
$\int_0^\infty(1+uz)^{-1/u}\,\dif z=1/(1-u)$ for $0<u<1$;
the exponential limit at zero has integral one. For the pure endpoint
$u=1$ the integral diverges.

For the lower transform, the scalar function
$v\mapsto z\sqrt v-\log(1+z\sqrt v)$ is nonnegative, increasing,
concave, and zero at zero, as in the proof of
Proposition~\ref{prop:single-state-sharp}. Concavity gives
$\sum_i[zp_i-\log(1+zp_i)]\geqslant z\sqrt u-\log(1+z\sqrt u)$.
Adding $-z$ and exponentiating proves the claim. Monotone convergence handles countably many atoms. The same argument
applies to a tail after normalizing its positive mass. The integral identity
for $b$ and its monotonicity follow by direct integration; summing
$zp_i-\log(1+zp_i)$ gives the last formula.
\end{proof}

\subsection{Stationary minima}

\begin{prop}[Structure of a nonpositive minimum]\label{prop:mp-stationarity}
At a generalized global minimum with $H\leqslant0$ and mixed
marginals, both ordinary marginal supports are finite. Diffuse row
mass attaches only to a unique largest column and diffuse column
mass only to a unique largest row. The doubly diffuse cell vanishes.
If diffuse row mass is present, no finite row is supported entirely
in its attachment column; the transposed assertion also holds.
\end{prop}

\begin{proof}
At a global minimum with $H\leqslant0$ and mixed marginals, put \[
 A_i=g_{\bsr}(r_i)-\tfrac23 \S_L(\bsc)r_i,\qquad
 B_j=g_{\bsc}(c_j)-\tfrac23 \S_L(\bsr)c_j
\] on finite indices, with $A_\infty=B_\infty=0$. The derivative representative at a
cell of two finite indices is $g_{\bsP}(p_{ij})-A_i-B_j$. At a cell with one diffuse index it
is $-A_i-B_j$, and at the doubly diffuse cell it is zero. Lemma~\ref{lem:mp-calculus} justifies these formulas, including
zero atoms and diffuse attachments, by transfers between cells.

All positive cells must have one common derivative $\kappa$, by transfers in both
directions, and every zero cell has derivative at least $\kappa$. These tests do not
require an interior optimum. If a cell has a finite index, labels can always be
reordered after a perturbation. By \eqref{eq:mp-13} each $A_i>0$ for $r_i>0$ and each
$B_j>0$ for $c_j>0$. Also $A_i>A_k$ when $r_i>r_k>0$, and likewise for $B$: in
\eqref{eq:mp-13} strictness holds if $r_i+r_k<1$, and otherwise the two product factors
are strictly below their balanced upper bound when the two masses differ. Consequently the derivatives of the cells $(i,\infty)$ and
$(\infty,j)$ give $\kappa<0$. The cell $(\infty,\infty)$ has derivative zero and hence
is unoccupied. If $p_{i\infty}>0$, then $A_i=-\kappa$ and every $A_k\leqslant-\kappa$, so
$r_i$ is largest. The assertion for $p_{\infty j}>0$ follows from the same reasoning
with $B$.

Suppose there were infinitely many positive row atoms. Then $A_i\to0$. Indeed
$g_r(r_i)/r_i\leqslant\int L_r^\circ(z)\dif z<\infty$. For all sufficiently small positive rows
the diffuse-column cell cannot be occupied, since $-A_i>\kappa$. In each of their
positive cells one has $g_P(p_{ij})=A_i+B_j+\kappa>0$. The only columns this permits for
sufficiently small rows are those with $B_j+\kappa=0$. To see this, first restrict to
the finite set $B_j\geqslant-\kappa/2$ since $B_j\to0$. Then use $B_j+\kappa\leqslant0$ and take
$A_i$ smaller than each strictly positive difference $-B_j-\kappa$ in that finite set. A
column with $B_j+\kappa=0$ must be occupied in each such row, since the derivative at a
zero cell would be $-A_i-B_j<\kappa$. In these columns every entry of the row has the
same value, as $g_P$ is strictly increasing. Thus all sufficiently small rows have a
common conditional profile, uniform on the indicated finite set of columns.

For completeness, at a minimum of $H$, two rows with the same conditional distribution
on finite columns cannot have unequal positive masses. For either row pair, vary its
masses as $a(1\pm v)$ while holding its normalized conditional profile fixed, so the
column marginal stays fixed. The even expansion of $\Q$ in \eqref{eq:pair-moment-series}
has coefficients \[
 \frac{ \bbE[Y^{2k}/X^{2k-1}]}{2k(2k-1)}
\] for the decrease from the balanced point, with $X$ the exponential sum at $v=0$ and
$Y$ the coefficient of $v$. In this variation the joint sum has the same law for
$v$ and $-v$ because exchanging the selected rows permutes their independent exponentials. In a generalized spectrum include the diffuse mass as a constant.
The bound $|Y|\leqslant X$, together with $\bbE [X]=1$, still gives uniform convergence of the
expansion on $0\leqslant v^2\leqslant1$.

The joint coefficients are at most the row-marginal coefficients. To construct a
coupling, give each finite joint cell an independent exponential variable. Select one
column in each row according to its conditional profile. For the chosen row pair use the
same selection in both rows. The two selected exponentials are nevertheless independent
because their conditional joint law given the selected column is the product of two
mean-one exponential laws, independent of the column. In any other row use independent
selections. If it selects a diffuse column, supply a fresh independent exponential and
average it out. The row diffuse mass is constant already. Each joint sum and
corresponding direction is the conditional mean of the marginal sum and direction given
the finite-cell exponentials. Apply conditional Jensen to the convex perspective
$Y^{2k}/X^{2k-1}$ to get the coefficient comparison.

The strict convexity of $s\mapsto s^4$ makes the coefficient comparison strict for $k=2$
whenever the chosen row conditional profile has two positive entries, or any other
positive nondiffuse row has a choice of different cells or a diffuse component. For the
pair, the ratios $Y/X$ after two different choices differ almost surely because they
depend on independent continuous cell variables. If the pair has a single choice and
another row varies, $Y$ is nonzero almost surely and $X$ varies under that row's
selection or its fresh exponential. In the only remaining case each positive nondiffuse
row is supported on one finite column, so the effective joint spectrum equals the row
spectrum for the entire variation and all coefficients agree.

It follows that $H$ along this variation is either strictly concave in $v^2$ or, in the
last case, strictly increasing affine in $v^2$ because $S_L(c)>0$. A minimum cannot
occur at $0<v^2<1$. In particular any two positive rows with the common small-row
profile must have equal mass. This contradicts the existence of infinitely many such
rows with masses tending to zero. The same proof with rows and columns exchanged shows
that there are finitely many positive column atoms.

If the row marginal has diffuse mass, it is incident to a unique largest column at a
minimum. To see this, let $k$ be the multiplicity of the largest column mass. At fixed
marginals the joint subentropy is strictly concave in the vector of finite cell masses.
This follows from $\Q^\circ=1-\gamma + \bbE[\eta(X)]$ with the diffuse mass included as a
constant in $X$: two different core vectors give different affine exponential sums
almost surely. Therefore the core at a maximum is invariant under permuting equal-mass
columns. The diffuse-row entries in these columns are their marginal masses minus their
core column sums, so they too are equal. Since the diffuse row is present and its mass
can only be incident to largest columns, all $k$ of these columns have positive
diffuse-row entries.

If $k\geqslant2$, transfer a small mass between two of these diffuse-row entries while
preserving their sum. The row and joint spectra are fixed. The second derivative of $H$
at the balanced column pair is $-2K_{\bsc}(c_1,c_1)+\tfrac43 \S_L(\bsr)<0$, where $K_{\bsc}$ is the
integral in \eqref{eq:mp-13} evaluated at $x=y=c_1$. The proof of \eqref{eq:mp-13} gives
$K_{\bsc}(c_1,c_1)\geqslant2/3$, and $\S_L(r)<1$ because the row marginal has a positive ordinary
atom. This contradicts the two-sided minimum, so $k=1$. The same argument applies to
diffuse column mass.

Diffuse row mass also excludes every finite row supported entirely on the first column.
Let $L_{\bsr}^\circ,L_{\bsP}^\circ$ denote the two transforms at the minimum and $k_{\bsc}=\tfrac23
\S_L(\bsc)$. Move $\epsilon$ from $(\infty,1)$ to $(i,1)$ at a new finite row label $i$. The
two transforms are multiplied by the same factor $\ee^{z\epsilon}/(1+z\epsilon)$. The row
squared sum increases by $\epsilon^2$ and the column spectrum stays fixed. Dividing the
change in $H$ by $\epsilon^2/2$ and taking $\epsilon\downarrow0$ gives

\begin{equation}\label{eq:mp-14}
k_{\bsc} \ \geqslant\ \int_0^\infty(L_{\bsr}^\circ(z)-L_{\bsP}^\circ(z))\,\dif z .
\end{equation}

For dominated convergence use $0\leqslant z\epsilon-\log(1+z\epsilon)
\leqslant z^2\epsilon^2/2$ and retain at least half of the positive diffuse mass as an
exponential factor. Grouping the joint masses by rows gives $L_{\bsP}^\circ\leqslant L_{\bsr}^\circ$. If
a finite row of mass $y>0$ is supported entirely on the first column, the two-sided
transfer between its occupied cell and $(\infty,1)$ gives $g_{\bsP}(y)-g_{\bsr}(y)+k_{\bsc} y=0$. This
requires
$$
 k_{\bsc}=\int_0^\infty\frac{L_{\bsr}^\circ(z)-L_{\bsP}^\circ(z)}{1+zy}\,\dif z .
$$ 
If the transforms differ, the last integral is strictly smaller than the one in
\eqref{eq:mp-14}. If they agree, the last integral is zero, whereas $k_{\bsc}>0$ because the
column marginal is mixed. Both cases give a contradiction. Exchange rows and columns for
the corresponding exclusion when the column marginal has diffuse mass.
\end{proof}

\subsection{Largest entry and block purities}\label{largest-entry-and-block-purities}

\begin{prop}[Fixed-marginal geometry]\label{prop:mp-fixed-marginal}
A joint-subentropy maximizer with fixed finite marginals has
entrywise sorted rows and columns when its marginals are sorted. The largest
entry $x=p_{11}$ lies in $[(a+b-1)_+,ab]$. The reverse rectangle,
prefix, and purity inequalities below hold. They also hold at a
generalized nonpositive global minimum with mixed marginals.
\end{prop}

\begin{proof}
At a finite maximizer of joint subentropy for fixed marginals, sort both marginals
decreasingly. Put $a=r_1,\ b=c_1,\ x=p_{11}$, $s=1-a,\ t=1-b,\ d=1-a-b+x$. We shall show that $x$ is
the largest joint entry and $d$ is the mass outside the first row and first column. To
justify the ordering and the sign used below, minimize $1-\gamma-\Q(\bsP)$ on the
transportation polytope. Delete zero marginal entries so the product table is a strictly
positive feasible point. The derivative of this objective on trace-zero directions is
represented by $g_{\bsP}(p_{ij})$, by Lemma~\ref{lem:mp-calculus}. The convex optimality conditions
give row and column potentials $\lambda_i,\mu_j$ with $g_{\bsP}(p_{ij})=\lambda_i+\mu_j$ on
positive cells and $\lambda_i+\mu_j\leqslant0$ on zero cells. Thus each entry is the value of
the clipped inverse of $g_{\bsP}$ at $\lambda_i+\mu_j$. A larger row potential gives
entrywise larger row entries and a strictly larger total if the smaller row has positive
mass. This orders the potentials as claimed, and the same argument orders the column
potentials.

On positive arguments the logarithm of the inverse is concave. Indeed its derivative at $g_{\bsP}(y)$ equals $1/(y g'_{\bsP}(y))$, and
$$
 \frac{\dif}{\dif y}(y g'_{\bsP}(y))
 =\int_0^\infty L_{\bsP}(z)\frac{1-zy}{(1+zy)^3}\,\dif z
 =-\int_0^\infty L'_{\bsP}(z)\frac{z}{(1+zy)^2}\,\dif z>0.
$$
The boundary terms in integration by parts vanish for $y>0$. Differentiation is
justified on a compact positive $y$-interval by the determinant bound
$L_{\bsP}(z)\leqslant(1+z)^{-1}$. The logarithmic derivative comparison on a rectangle with all
four entries positive gives $p_{11}p_{ij}\leqslant p_{1j}p_{i1}$. When $p_{ij}=0$ the same
inequality follows from nonnegativity. Consequently \[
 p_{1j}p_{i1}-x p_{ij}=e_{ij}\geqslant0\qquad(i,j>1).
\] Summing these inequalities and using the marginal constraints gives

\begin{equation}\label{eq:mp-26}
\sum_{i,j>1} e_{ij}=\delta=ab-x,\qquad
 p_{i1}=\frac{x r_i+\sum_{j>1}e_{ij}}{a},\qquad
 p_{1j}=\frac{x c_j+\sum_{i>1}e_{ij}}{b}.
\end{equation}

The domain of $x$ is $[(a+b-1)_+,ab]$. If $x=ab$, \eqref{eq:mp-26} forces the joint
table to be its product table. This exact endpoint relation strengthens the purity
comparison near the independent coupling.

Since $g_P$ is concave and increasing, its inverse extended by zero
on the negative half-line is convex. Summing the reverse rectangle
inequality over a cut with marginal masses $a',b'$ and prefix mass $x'$ gives
$x'(1-a'-b'+x')\leqslant(a'-x')(b'-x')$, or $x'\leqslant a'b'$. This justifies the prefix sign used
in summation by parts.

At a finite fixed-marginal maximizer, the convexity of the clipped inverse of $g_{\bsP}$ implies, for all $i,j>1$,

\begin{equation}\label{eq:mp-36}
p_{ij}\geqslant(p_{i1}+p_{1j}-p_{11})_+.
\end{equation}

Indeed, stationarity expresses each entry as the convex nondecreasing function
$f(\lambda_i+\mu_j)$, with $f(s)=0$ for $s\leqslant0$ and $f(s)=g_{\bsP}^{-1}(s)$ on its positive
domain. Both sets of potentials are ordered with their marginal masses. For a convex
function the increment over a fixed nonnegative interval increases with its left
endpoint. Apply this with interval length $\lambda_1-\lambda_i$ and left endpoints
$\lambda_i+\mu_1$ and $\lambda_i+\mu_j$. Together with nonnegativity this gives
\eqref{eq:mp-36}. In particular, with the notation of \eqref{eq:mp-26}, the nonnegative
defect also obeys $e_{ij}\leqslant(x-p_{i1})(x-p_{1j})$.

At a nonpositive generalized minimum with mixed marginals the same clipped-inverse formula holds on the finite
cells, with potentials supplied by the first variation. Its finite rows and columns are
ordered because the potentials increase with the corresponding marginal masses. With a
diffuse label placed after all finite labels, any rectangle whose southeast cell is
diffuse has that cell zero. The reverse rectangle inequality holds on the full array.
The same rectangle and prefix arguments apply there,
with squared norms taken over ordinary atoms and block masses
including diffuse attachments.

First,
\begin{equation}\label{eq:mp-71}
\pi_{\bsP}\leqslant \sum_{i,j<\infty}p_{ij}r_ic_j\leqslant RC .
\end{equation}
The proof applies directly with diffuse mass. Use one last label on each side for its
diffuse mass and form the full array of masses $\bar P$. The reverse rectangle
inequality from \eqref{eq:mp-26} holds on finite cells. In an ordered rectangle whose
southeast cell has a diffuse index, that cell is zero, because diffuse row mass attaches
only to the first column and diffuse column mass only to the first row. Hence the
reverse inequality holds on the full array. Every prefix sum of $\bar P-\bar
r\otimes\bar c$ is nonpositive. For the first summation by parts, test this difference
against the array of finite cell masses extended by zero on diffuse cells. Its mixed
adjacent difference $p_{ij}-p_{i+1,j}-p_{i,j+1}+p_{i+1,j+1}$ is nonnegative, by
convexity of the clipped inverse. At a diffuse boundary the same difference is
nonnegative by monotonicity of the finite row or column. The inner product is
$\pi_{\bsP}-\sum_{i,j<\infty}p_{ij}r_ic_j$. For the second summation by parts, test against $r_i
c_j$ on finite cells and zero on diffuse cells. Its adjacent mixed differences are
products of nonnegative differences of the sorted marginal atom masses, also at the
diffuse boundary. This gives the second inequality in \eqref{eq:mp-71}. For an ordinary
finite optimizer the same proof works without the diffuse labels.
\end{proof}

\begin{prop}[Additional stationary purity constraint]\label{prop:mp-stationary-purity}
At a generalized global minimum with $H\leqslant0$ and mixed
marginals, let $x>0$ be the largest joint atom. For the secants
$K_{\bsp}(x,y)$ defined in \eqref{eq:mp-13}, one has
\begin{equation}\label{eq:mp-74}
(a-R)\big[K_{\bsr}(a,a_2)-\tfrac23(1-C)\big]
 +(b-C)\big[K_{\bsc}(b,b_2)-\tfrac23(1-R)\big]
 \leqslant (x-\pi_{\bsP})\,g_P(x)/x
 \leqslant x-\pi_{\bsP} .
\end{equation}
\end{prop}

\begin{proof}
Write $E_{\bsp}=\sum_\nu p_\nu g_{\bsp}(p_\nu)$, summing only the actual atoms. At a generalized minimum with $H\leqslant0$ and mixed marginals, denote the top cell by $x$. Set $h=-\kappa$, where
$\kappa$ is the common derivative at the occupied cells. Averaging
their first variations, including the diffuse attachments, gives
$$
 h= E_{\bsr}+E_{\bsc}-E_{\bsP}-\frac23\big((1-C)R+(1-R)C\big).
$$
Indeed, average the cell derivative with respect to the full array of masses, using a
zero value for a gradient at a diffuse joint cell. The row and column gradient terms
average to the corresponding marginal sums. Subtract this identity from
$h=A_1+B_1-g_{\bsP}(x)$ to obtain 
$$
\begin{aligned}
 g_{\bsP}(x)-E_{\bsP}={}&g_{\bsr}(a)-E_{\bsr}+g_{\bsc}(b)-E_{\bsc}\\
 &-\frac23\big((1-C)(a-R)+(1-R)(b-C)\big).
\end{aligned}
$$
For any spectrum with largest atom $m$, purity $u$ and diffuse
mass $\delta$, write $E_{\bsp}=\sum_i p_i g_{\bsp}(p_i)$. The exact decomposition is
$$
 g_{\bsp}(m)-E_{\bsp}
 =\sum_i p_i(m-p_i)K_{\bsp}(m,p_i)+\delta m K_{\bsp}(m,0).
$$
The last term uses $g_{\bsp}(0)=0$ and retains the diffuse probability
$\delta$ at atom size zero. The total weight is
$m(1-\delta)-u+\delta m=m-u$. In a marginal the term at its chosen
largest atom is zero, so the secants are bounded below by the value of the integral in
\eqref{eq:mp-13} at that marginal's two largest atoms. In the joint spectrum they are
bounded above by $g_{\bsP}(x)/x$, since the other reciprocal factor can be bounded above by
one. Substituting these bounds into the identity for $g_{\bsP}(x)-E_{\bsP}$
gives the first inequality in \eqref{eq:mp-74}.
The last inequality is the largest-gradient estimate in
Lemma~\ref{lem:mp-calculus}. Thus global marginal stationarity
supplies a pivot-dependent upper bound on joint purity in addition
to \eqref{eq:mp-71}.
\end{proof}

\begin{lem}[Block purities and caps]\label{lem:mp-blocks}
Assume the geometry of Proposition~\ref{prop:mp-fixed-marginal}.
Put $s=1-a$, $t=1-b$, $\delta=ab-x$,
$u=\sqrt{R-a^2}$, $v=\sqrt{C-b^2}$, $a_2=r_2$, $b_2=c_2$,
and $d=1-a-b+x$. The three blocks outside $(1,1)$ have the lower
purities and upper atom caps below. The bounds use ordinary squared
sums and full block masses, so they include the permitted diffuse
attachments.
\end{lem}

\begin{proof}
Write $R=\sum r_i^2,\ C=\sum c_j^2$, $u=\sqrt{R-a^2}$, $v=\sqrt{C-b^2}$, and $a_2=r_2,\
b_2=c_2$. Without retaining the second masses explicitly one may use $a_2\leqslant\min(a,s,u)$
and $b_2\leqslant\min(b,t,v)$. Let $V=(p_{i1})_{i>1}$, $U=(p_{1j})_{j>1}$, and
$W=(p_{ij})_{i,j>1}$. Lower bounds for their squared sums are respectively

\begin{equation}\label{eq:mp-27}
\begin{aligned}
 \ell_V&=\max(0,u-d,xu/a)^2,\\
 \ell_U&=\max(0,v-d,xv/b)^2,\\
 \ell_W&=\max\!\left(0,u+v-s-t+d,\frac{xuv}{ab}-\frac{\delta}{x}\right)^2.
\end{aligned}
\end{equation}

The last expression omits the fraction when $x=0$. For $V,U$, use the norm triangle
inequality when the marginal tails are split into the strip and the row or column sums
of the inner block, and use \eqref{eq:mp-26} for the other bound. For $W$, the same
splitting implies that the norm of its row-sum vector is at least $u-(s-d)$ and its column-sum norm is at least $v-(t-d)$. If $d>0$, apply the
two-party version of \eqref{eq:schatten-two-marginal-comparison}
to $W/d$. After multiplying by $d$, it gives
$\|W\|_2\geqslant\|\operatorname{row}(W)\|_2+
\|\operatorname{col}(W)\|_2-d$.
The same homogeneous inequality holds at $d=0$ because the block is
then zero. Substituting the two lower marginal norms gives
$u+v-s-t+d$. Also \eqref{eq:mp-26} gives $p_{ij}\geqslant x r_i
c_j/(ab)-e_{ij}/x$; the norm of $(e_{ij})$ is at most $\delta$. This proves
\eqref{eq:mp-27}.

Upper bounds on the largest atom of each block are

\begin{equation}\label{eq:mp-28}
\begin{aligned}
 m_V&=\min\{x,a_2,s-d,(xa_2+\delta)/a\},\\
 m_U&=\min\{x,b_2,t-d,(xb_2+\delta)/b\},\\
 m_W&=\min\{x,a_2,b_2,d,m_Vm_U/x\}.
\end{aligned}
\end{equation}

These come from the marginal sums and \eqref{eq:mp-26}, with the last upper bound
following from $x p_{ij}\leqslant p_{i1}p_{1j}$. Omit its fraction if $x=0$. Infeasible
parameters can be discarded by comparing each $\ell$ to its block mass times the
corresponding $m$.

In this calculation use $r_\infty=c_\infty=0$ as atom sizes in
the products and sums, while retaining the full table masses $p_{ij}$
on diffuse cells. Let $q_{ij}=r_i c_j$. Expansion of
$\sum_{ij}p_{ij}(a-r_i)(b-c_j)$ gives
$$
 \Inner{\bsP}{\bsq} \geqslant E:=
 RC-(a-R)(b-C)+d(a-a_2)(b-b_2).
$$
The extra term comes from the $d$ mass outside the first row and column.
Cauchy--Schwarz, applied on the whole array or on all but its first cell, yields

\begin{equation}\label{eq:mp-29}
\pi_{\bsP}
 \geqslant \underline{\pi}_{\bsP}:=\max\!\left\{(\sqrt R+\sqrt C-1)_+^2,\,
    E_+^2/(RC),\,
    x^2+(E-xab)_+^2/(RC-a^2b^2)\right\}.
\end{equation}

The last lower bound uses zero for the quotient when its denominator is zero, as then
$\bsq$ vanishes from the complement. It is necessary that $\underline{\pi}_{\bsP}\leqslant x$.

Lemma~\ref{lem:mp-transforms} gives the following lower bound on
the joint logarithmic surplus:

\begin{equation}\label{eq:mp-30}
\begin{aligned}
 \log L_{\bsP}(z)+z\geqslant{}& zx-\log(1+zx)\\
 &+z^2\max\{\underline{\pi}_P-x^2,\ell_U+\ell_V+\ell_W\}\,b(z,x)\\
 &+z^2\sum_{J\in\{U,V,W\}}\ell_J\big(b(z,m_J)-b(z,x)\big).
\end{aligned}
\end{equation}

Indeed, apply $b(z,p_{ij})\geqslant b(z,m_J)$ in each block and separate off the common
coefficient $b(z,x)$. Equations \eqref{eq:mp-26}--\eqref{eq:mp-30} use only the stated
largest marginal masses and purities.

At a fixed-marginal optimizer, $x\in[(a+b-1)_+,ab]$. The three quantities in
\eqref{eq:mp-27} bound the block purities. In addition,

\begin{equation}\label{eq:mp-50}
\|V\|^2\geqslant u^2-2a_2d,\qquad
 \|U\|^2\geqslant v^2-2b_2d,\qquad
 \|W\|^2\geqslant \left(\frac{xuv}{ab}\right)^2-\frac{2a_2b_2\delta}{ab}.
\end{equation}

For the first two, expand the squared sums of the marginal tails as those of their
strips plus the corresponding sums of $W$. The cross term is bounded by $a_2d$ or
$b_2d$, using that strip entries do not exceed the marginal atom caps. Equivalently,
expanding the squared sum of $V=r_{\rm tail}-w_r$ gives $u^2-2\langle r_{\rm
tail},w_r\rangle+\|w_r\|^2$. For the last bound, use $p_{ij}\geqslant x r_i c_j/(ab)-e_{ij}/x$
from \eqref{eq:mp-26} and square with the lower estimate $y^2\geqslant z^2-2z(z-y)_+$ for
nonnegative $y,z$. Sum and use $\sum e_{ij}=\delta$; if $x=0$ the bound is trivial. A
box can upper bound $a_2,b_2$ by $\min(a,s,u)$ and $\min(b,t,v)$. This needs no extra
coordinates.

Dropping the nonnegative correction in \eqref{eq:mp-29} gives the useful corollary

\begin{equation}\label{eq:mp-32}
\pi_{\bsP}\geqslant \frac{(RC-(a-R)(b-C))_+^2}{RC}.
\end{equation}
\end{proof}

\begin{lem}[Canonical determinant envelopes]\label{lem:mp-canonical-tail}
For fixed mass and purity, a canonical spectrum minimizes the
determinant: equal larger atoms and at most one smaller atom, with a
diffuse endpoint at zero purity. A fixed additional largest atom is
handled by applying the same comparison to its remainder. The
resulting derivative bounds and branch concavity are as follows.
\end{lem}

\begin{proof}
A spectrum bounded by $a>0$ has squared sum at most

\begin{equation}\label{eq:mp-35}
k a^2+(1-ka)^2,\qquad k=\lfloor1/a\rfloor.
\end{equation}

The spectrum with $k$ copies of $a$ and the remaining mass in one atom majorizes the
given spectrum because its largest $\ell$ atoms always sum to $\min(1,\ell a)$. This
also proves \eqref{eq:mp-35} for diffuse limits by finite approximation.

For a precise upper bound on a marginal Laplace transform, fix the largest two atoms and
apply the following canonical tail comparison to the rest. At fixed total one and fixed squared
sum $\alpha$, $\sum_i\log(1+zp_i)$ is minimized by $m$ equal atoms and at most one
smaller atom, with $m=\lfloor1/\alpha\rfloor$. To see this in finite dimension, minimize
on the compact intersection of the probability simplex with the prescribed purity. On
the positive support of a nonuniform minimizer, the sum and purity constraints have
independent gradients. The stationarity equation for $\log(1+zp_i)$ has at most two
solutions because its derivative is strictly convex and is equated to an affine function
of $p_i$. If the smaller of the two values occurred twice, an opposite variation of the
two equal small atoms would be tangent to both constraints. The Lagrangian second
derivative along it is negative: the derivative slope between the smaller and larger
values is strictly greater than the second derivative of $\log(1+z p_i)$ at the smaller
value. This contradicts the second-order necessary condition. The resulting shape and
the prescribed purity determine the atoms, including the uniform endpoint when $\alpha$
is reciprocal to an integer. Normalize before applying this to any nonzero tail mass.
A zero-mass tail is omitted and assigned normalized purity zero.
The normalized larger atom is \[
 \frac{1+\sqrt{((m+1)\alpha-1)/m}}{m+1},
\] and the smaller is the remaining mass. The diffuse endpoint follows by continuity of the transforms.

Fix the largest atom $a$, the mass $s=1-a$ of its tail, and the sum $u^2$ of the squares
of the tail atoms. If its largest tail atom is $t$, the tail determinant extremizer with
this additional constraint consists of $t$, $m$ equal atoms $x$, and one smaller atom
$y$. Here \[
 mx+y=s-t,\qquad mx^2+y^2=u^2-t^2,\qquad 0\leqslant y\leqslant x\leqslant t.
\] Indeed, apply the canonical comparison just proved to
the part after $t$. Its canonical largest atom is the smallest possible largest atom at
that purity, by \eqref{eq:mp-35}. It therefore respects the cap $t$ whenever that cap is
feasible. Let $t_{\min}$ be the smallest feasible largest tail
atom at the specified mass and purity. The possible values of
$t$ lie in $[t_{\min},\min(a,u)]$ since $a$ is the largest marginal
atom and $t^2\leqslant u^2$. Conversely, in this interval the
maximum attainable squared sum at cap $t$, given by
\eqref{eq:mp-35} scaled to mass $s$, includes one atom of mass $t$.
The residual target $u^2-t^2$ lies between zero and the cap
maximum for the remaining mass $s-t$. Its canonical realization
therefore has largest atom at most $t$. If $u=0$ the tail is
entirely diffuse and $t=0$. A remainder of positive mass and
zero purity is likewise interpreted as diffuse mass.

On every smooth branch of this comparison spectrum, containing no coincident positive
remainder masses $x=y$, let $\Q_*(t)$ denote its marginal subentropy. Then

\begin{equation}\label{eq:mp-54}
\Q_*'(t)=(t-x)(t-y)
 \int_0^\infty\frac{zL_{r_*}(z)}{(1+zt)(1+zx)(1+zy)}\,\dif z
 \geqslant \frac{(t-x)(t-y)}{(1+2s/3)^3}.
\end{equation}

To obtain the identity, differentiate the two moment constraints. They give
$mx'=(y-t)/(x-y)$ and $y'=(t-x)/(x-y)$. The derivative of the log determinant of the
tail is $(t-x)$ times the difference of the two secant slopes of $p\mapsto z/(1+zp)$ on
$[x,t]$ and $[y,x]$. That difference equals $z^3(t-y)/[(1+zt)(1+zx)(1+zy)]$. Substitute
in the determinant formula for $\Q$. The derivative integrals converge locally along
each positive branch by the explicit positive atom factors, and the expansion at zero
gives a bounded integrand there. For the inequality, use $L_{r_*}(z)\geqslant e^{-z}$ and
bound the denominator above by $(1+zs/3)^3$ since $t+x+y\leqslant s$. Apply Jensen with the
gamma density $ze^{-z}$, whose mean is two. Continuity of the transforms at the uniform
branch endpoints and at the diffuse remainder gives the monotonic comparison for the
entire feasible interval.

For $a,u,\alpha\in[0,1]$, let the first two row masses be
$a,(1-a)u$ and the remaining normalized purity be $\alpha$.
Replace the row remainder by the canonical remainder of purity
$\alpha$ from the canonical comparison above. This bounds marginal subentropy below without changing
purity. Put $s=(1-a)(1-u)$. If $s=0$, the remainder is absent and
the subentropy is constant in $\alpha$. For $s>0$ and
$\alpha\in(1/(k+1),1/k)$, let the normalized remainder consist
of $k$ equal atoms $y$ and a smaller atom $w=1-ky$.
Differentiation within this branch gives

\begin{equation}\label{eq:mp-79}
\frac{\dif}{\dif\alpha}\Q(r_{\rm can})
      =-\frac{s^2}{2}K_{r_{\rm can}}(sy,sw).
\end{equation}

In fact $\dif\alpha/\dif y=2k(y-w)$ and the derivative of $\Q$ in $y$ is $-ks[g_{r_{\rm
can}}(sy)-g_{r_{\rm can}}(sw)]$. The transform $L_{r_{\rm can}}$ increases with $\alpha$
by majorization. The denominator $(1+zsy)(1+zsw)$ in $K$ decreases because both $y+w$
and $yw$ are nonincreasing in $y$. Thus $\Q(r_{\rm can})$ is
concave in $\alpha$ on each branch, including its endpoints by
continuity. At a reciprocal-purity breakpoint the derivative
formulas are one-sided, with $y,w$ taken from the respective branch.
\end{proof}

\begin{lem}[Power-series loss bound]\label{lem:mp-series}
If $m=ab$ bounds the largest joint atom and $J_0\geqslant0$ bounds
its purity below, then the following four-term estimate holds.
Each kernel value admits the rational four-interval lower rule below.
\begin{equation}\label{eq:mp-33}
1-\gamma-\sfQ(\bsP) \ \geqslant\ \sum_{k=1}^{4}\frac{(2k-2)!}{k!}       \left[J_0 b(2k-1,m)\right]^k.
\end{equation}
\end{lem}

\begin{proof}
Let $J_0$ be any
nonnegative lower bound on the joint purity. Write $m=ab$ and use the normalized kernel $b(z,m)$ from
Lemma~\ref{lem:mp-transforms}. Since $D(z)=\log L_{\bsP}(z)+z\geqslant z^2 J_0 b(z,m)$, retain four terms in $\ee^{D(z)}-1$.

Indeed $b(z,m)^k$ is convex in $z$ because $b$ is nonnegative, decreasing and convex,
and the $k$-th power is increasing and convex on its nonnegative range. Integrate
$D^k/k!$ against $\ee^{-z}\dif z/z^2$ and apply Jensen under the gamma density
$\ee^{-z}z^{2k-2}/(2k-2)!$, of mean $2k-1$. For a rational bound on $b$, partition $[0,1]$
into four equal intervals. On interval $j$, the mass under $y\,\dif y$ is $(2j-1)/32$ and
the mean of $y$ is $(3j^2-3j+1)/(6(2j-1))$. A Jensen lower bound on each interval gives
the rational sum used by the program.
\end{proof}

\subsection{Common convexity and quadrature estimates}\label{quadrature-and-determinant-comparisons}

\begin{lem}[Positive Laplace integrands]\label{lem:mp-positive-integrands}
For every generalized spectrum $\bsp$, the functions
\begin{equation}\label{eq:mp-39}
f_{\bsp}(z)=\frac{(1+z)^{-1}-L_{\bsp}^\circ(z)}{z^2},\qquad j_{\bsp}(z)=\frac{L_{\bsp}^\circ(z)-\ee^{-z}}{z^2}
\end{equation}
are Laplace transforms of finite nonnegative measures. They are therefore
completely monotone; midpoint sums are lower bounds, trapezoid sums
are upper bounds, and the stated tail bounds are uniform.
\end{lem}

\begin{proof}
Let $X_{\bsp}$ be its weighted exponential sum with the diffuse mass
included as a constant. Choose one of the exponentials with probability equal to its
weight, independently of their values, and use a fresh unit exponential when the diffuse
part is chosen. If $Y$ is the selected value, it has the unit exponential law and its
conditional mean given the retained exponentials is $X_{\bsp}$. For $U,V\geqslant0$, Taylor's
formula gives 
$$
 \frac{\ee^{-zV}-\ee^{-zU}+z\ee^{-zU}(V-U)}{z^2} = (V-U)^2\int_0^1(1-v)e^{-z((1-v)U+vV)}\,\dif v.
$$
Use $(U,V)=(X_{\bsp},Y)$ for $f_{\bsp}$ and $(U,V)=(1,X_{\bsp})$ for $j_{\bsp}$. Taking expectations
cancels the linear term and represents each function as the claimed transform. The
representing mass is finite by the finite second moments. The integral tails above $T>0$
are at most $1/(2T^2)$ for either function, by $L_{\bsp}^\circ(z)\leqslant(1+z)^{-1}\leqslant z^{-1}$.
These facts give midpoint lower sums and trapezoid upper sums with a uniform tail bound,
including exact tail envelopes with many equal atoms.
\end{proof}

\begin{lem}[Marginal endpoint comparison]\label{lem:mp-majorization}
Increase one or two largest marginal atoms and rescale the remaining
mass. Replace the rescaled tail by the canonical spectrum at its
normalized purity, and then increase that canonical tail's normalized
purity. For $0\leqslant c_0\leqslant1/3$, the resulting comparison
spectra give the lower bound \eqref{eq:mp-51} for the marginal part
of the deficit.
\end{lem}

\begin{proof}
There is also a simple way to avoid bounding opposite endpoints of a marginal Laplace
integrand separately. If $a$ is the largest atom, increasing it to $A\geqslant a$ and
rescaling the rest to mass $1-A$ majorizes the original spectrum. For a finite spectrum
the largest $k$-atom cumulative sum increases by $(A-a)(1-v_k)$, where $v_k$ is the
fraction of the original tail mass in its first $k-1$ atoms. The generalized conclusion
follows by approximating the diffuse mass. When retaining two largest atoms, one can
similarly increase both and rescale the rest if the two new masses sum to at most one.
They are still the two largest masses since the remaining atoms only shrink. For $k\geqslant2$
the increase in the largest $k$-atom cumulative sum is the sum of the two increases
times one minus the normalized tail fraction in the first $k-2$ atoms. For $k=1$, the
largest atom is at least as large as before. If the two upper endpoints sum to more than
one, the binary spectrum $(A,1-A)$, where $A$ is the first upper
endpoint, majorizes the original spectrum by the cumulative-sum test.
This leaves no remainder. When a positive tail is rescaled, its
normalized purity stays fixed. Lemma~\ref{lem:mp-canonical-tail}
then permits replacing the rescaled tail by the canonical determinant
extremizer at that purity. Within the canonical family, increasing the
normalized purity gives a majorizing spectrum. Within a branch its equal
larger atoms increase, and neighboring branches meet at uniform spectra.

The marginal part of the deficit can be bounded using the same comparison spectrum for
both its subentropy and its linear entropy. For any $0\leqslant k\leqslant1/3$, the function
$\bsp\mapsto \Q(\bsp)-k\S_L(\bsp)$ is Schur-concave. This follows from
Proposition~\ref{prop:mixing}. Hence, for $0\leqslant c_0\leqslant1/3$, $\Q(\bsr)+\Q(\bsc)-c_0
\S_L(\bsr)\S_L(\bsc)$ decreases when either marginal is replaced by a spectrum majorizing it.
Holding the other marginal fixed gives the preceding Schur-concave function with
coefficient $k=c_0 \S_L$ of that other marginal.

One may first increase a retained head atom or pair of head atoms and rescale the rest
as just described. Replacing the new tail by its determinant extremizer at fixed
tail purity decreases $\Q$ and keeps $\S_L$ fixed. Increasing that canonical tail's
purity is another majorization, so it finally yields a comparison spectrum $\bsr^\sharp$
determined by upper endpoints for the largest marginal masses and tail purity. The same
holds for $\bsc^\sharp$. It follows that

\begin{equation}\label{eq:mp-51}
\Q(\bsr)+\Q(\bsc)-c_0\S_L(\bsr)\S_L(\bsc)\geqslant  \Q(\bsr^\sharp)+\Q(\bsc^\sharp)-c_0 \S_L(\bsr^\sharp)\S_L(\bsc^\sharp).
\end{equation}

Only the marginal part is compared in \eqref{eq:mp-51}. The joint element in the deficit
is still bounded for the original marginal box.
\end{proof}

\begin{lem}[Gauss and Lobatto enclosures]\label{lem:mp-gauss}
For a completely monotone integrand on a finite interval, the
four-point Gauss--Legendre rule is a lower bound and the four-point
Gauss--Lobatto rule is an upper bound. The same signs hold after
the change of variable $w=z^3/3$ for an integrand $z^2k(z)$ with
$k$ completely monotone.
\end{lem}

\begin{proof}
If a completely monotone integrand is multiplied by $z^2$, use
$w=z^3/3$, so $\dif w=z^2\dif z$. Composition with $z=(3w)^{1/3}$
preserves complete monotonicity. In a chain-rule term of order $k$,
an outer derivative of order $j$ and inner derivative orders summing
to $k$ have total sign $(-1)^j(-1)^{k-j}=(-1)^k$.

Use
four-point Gauss--Legendre quadrature for lower bounds and four-point Gauss--Lobatto for
upper bounds. The signs follow from the Hermite interpolation remainder: for Gauss the
remainder has the sign of the eighth derivative times the product of four squared node
differences. For Lobatto the corresponding factor contains both endpoint differences and
the two squared interior differences, and the relevant derivative is the sixth. The
integration rules integrate these interpolation polynomials exactly. For an interval beginning at zero after the change of variable,
apply the quadrature inequalities first to
$w\mapsto k((3(w+\varepsilon))^{1/3})$ with $\varepsilon>0$.
This function has the required derivatives throughout the interval.
Here $k$ is continuous at zero and bounded on finite intervals.
Dominated convergence as $\varepsilon\downarrow0$ passes the
integrals to the limit, and continuity passes each weighted node
value to the limit, including the Lobatto endpoint.

On $[-1,1]$ the Gauss nodes are
$\pm\sqrt{(3\pm2\sqrt{6/5})/7}$, with respective inner and outer
weights $(18+\sqrt{30})/36$ and $(18-\sqrt{30})/36$.
The Lobatto nodes are $-1,-1/\sqrt5,1/\sqrt5,1$, with weights
$1/6,5/6,5/6,1/6$. Affine scaling gives each interval rule.
\end{proof}

\begin{lem}[Rational constants]\label{lem:mp-constants}
The constants used below satisfy
\[
 1-\gamma>\frac{21}{50},\qquad
 1-\gamma>\frac{2113}{5000},\qquad
 0.42278433<1-\gamma<0.42278434,
\]
and
\[
 3.141592653589793<\pi<3.141592653589794,
 \qquad \frac6{\pi^2}>\frac35.
\]
In the $\pi$ enclosure the verifier uses the binary64 endpoints
represented by the displayed decimal strings; these also enclose
$\pi$.
\end{lem}

\begin{proof}
A uniform spectrum of rank $200$ gives
$1-\gamma\geqslant1+\log200-H_{200}>21/50$.
Use $\log200=7\log2+\log(25/16)$ and the first ten positive terms
of $2\sum_{j\geqslant0}z^{2j+1}/(2j+1)$ at $z=1/3,9/41$.
The rank-$4096$ comparison, with the first fourteen positive terms
for $\log2$, similarly gives $2113/5000$.

For the tight interval take $K=8192$ and use
$$
 H_K-\log K-\frac1{2K}<\gamma
 <H_K-\log K-\frac1{2K}+\frac1{12K^2}.
$$
These bounds follow by summing the trapezoidal errors for $1/x$ on
$[k,k+1]$. With $s=(2k+1)^{-1}$, the error is
$2\sum_{j\geqslant1}(2j/(2j+1))s^{2j+1}$, bounded above termwise
by $(1/k^2-1/(k+1)^2)/12$. Enclose $\log K=13\log2$ by the
positive series at $1/3$. For the upper bound, rounding each harmonic
summand down to a multiple of $10^{-30}$ and using $32$ logarithm
terms with remainder at most
$2(1/3)^{65}/[65(1-1/9)]$ suffices. The lower bound follows from
the same harmonic and logarithm enclosures.

Machin's formula
$\pi=16\arctan(1/5)-4\arctan(1/239)$, with $25$ terms and
alternating-series remainders, gives the decimal and binary64
endpoint comparisons by rational arithmetic. The upper enclosure
also gives $\pi^2<10$, proving the last inequality.
\end{proof}

\begin{lem}[Anchored loss and reciprocal minorants]\label{lem:mp-anchor}
Let $\bsp^0$ be a mixed reference spectrum with
$\int_0^\infty L_{\bsp^0}^\circ(z)\,\dif z<\infty$.
The joint loss has the affine lower bound below. Its weight $I$ is
a positive reciprocal mixture. For a uniform anchor on $n\geqslant2$
atoms, its center value and derivative are explicit, and tilting any
such mixture gives a reciprocal minorant with positive denominator.
\begin{equation}\label{eq:mp-40}
1-\gamma-\sfQ(\bsp)\geqslant 1-\gamma-\sfQ(\bsp^0) +\int_0^\infty L_{\bsp^0}^\circ(z)                 \frac{D_{\bsp}(z)-D_{\bsp^0}(z)}{z^2}\,\dif z.
\end{equation}
\end{lem}

\begin{proof}
With $D_{\bsp}$ as in Lemma~\ref{lem:mp-transforms}, convexity of the
exponential proves \eqref{eq:mp-40}.
Set
$$
 I(y)=\int_0^\infty L_{\bsp^0}^\circ(z)b(z,y)\,\dif z,\qquad f(y)=y^2I(y).
$$
The normalized kernel is the one in Lemma~\ref{lem:mp-transforms}. The reference-dependent weights then enter the three-block lower bound
\eqref{eq:mp-30} as $I$ in place of $z^2b(z,\cdot)$ after integration against $\dif z/z^2$. The constant after extracting the
contribution of $D_{\bsp^0}$ is $1-\gamma-\Q(\bsp^0)-\sum_i(p_i^0)^2 I(p_i^0)$. For a uniform
reference on $n\geqslant2$ atoms, $\Q(\bsp^0)=\log n-H_n+1$ exactly. The function $z\mapsto
L_{\bsp^0}(z)b(z,y)$ is a product of nonnegative Laplace transforms, hence convex in
$z$, and its tail integral above $T$ is bounded by $n(1+T/n)^{1-n}/[2(n-1)]$.
Consequently the weights for a uniform reference admit one-dimensional quadrature bounds
that can be evaluated once and reused over the parameter domain.

Let $I_0(y)=\int(1+vy)^{-1}\dif\nu(v)$ be a reciprocal mixture
with finite positive mass. Choose $u>0$, or choose $u=0$ if
$\int v\,\dif\nu(v)<\infty$. These hypotheses ensure that
$\int v(1+vu)^{-2}\dif\nu(v)$ is finite. Tilt $\dif\nu$ by
$(1+vu)^{-1}/I_0(u)$. Under this probability law the random
denominator $(1+vy)/(1+vu)$ has mean
$1-I'_0(u)(y-u)/I_0(u)$. Jensen for the reciprocal gives, on
$y\geqslant0$, \[
 I_0(y)\geqslant \frac{I_0(u)}{1+\theta(y-u)},\qquad
 \theta=-I_0'(u)/I_0(u).
\] The denominator is positive because it is the mean of $(1+vy)/(1+vu)$. This produces
the two reciprocal terms per anchor in the pivot check. For a
uniform anchor of rank two the derivative at zero diverges, since
$\int_0^\infty z(1+z/2)^{-2}\dif z=\infty$.
The exact center $u=1/n$ is positive for every admissible uniform
anchor. At that center, \[
 I(1/n)=1-n+n^2\left(\frac{\pi^2}{6}-\sum_{j=1}^n\frac1{j^2}\right),
 \qquad I'(1/n)=n\bigl(1-2I(1/n)\bigr).
\] For the value, substitute $v=(1+z/n)^{-1}$ and integrate
$v^n\log v/(1-v)^2$ by parts against $(v^n/(1-v))'$.
The boundary contribution is $-1$; the remaining logarithmic
integral is $\int_0^1v^n(-\log v)/(1-v)\,\dif v$.
Expanding $(1-v)^{-1}$ gives the convergent reciprocal-square tail. The identity
$2I(y)+yI'(y)
=\int L_{\bsp^0}(z)/(1+zy)\,\dif z$ gives the derivative.
\end{proof}

\begin{lem}[Hessian bounds]\label{lem:mp-hessian}
Let $\bsp$ be a generalized spectrum and let $\bsh$ be a finitely supported
feasible direction in its ordinary coordinates, with diffuse compensation $-\sum_i h_i$. On every
segment on which the derivatives are defined,
\begin{equation}\label{eq:mp-62}
\begin{aligned}
 \rD^2\mathcal J_{\bsp}[\bsh,\bsh]
 &=\int_0^\infty L_{\bsp}^\circ(z)\left[
 \left(\sum_i\frac{h_i}{1+zp_i}-\sum_i h_i\right)^2
 +\sum_i\frac{h_i^2}{(1+zp_i)^2}\right]\dif z\\
 &\geqslant\sum_i\frac{h_i^2}{1+2p_i}.
\end{aligned}
\end{equation}
For an ordinary trace-zero direction the subtraction in the square
vanishes. The independent uniform estimate is
\begin{equation}\label{eq:mp-73}
 \rD^2\mathcal J_{\bsp}[\bsh,\bsh]\geqslant\frac6{\pi^2}\norm{\bsh}_2^2.
\end{equation}
Both inequalities extend to segment endpoints by continuity in their
integrated convexity form.
\end{lem}

\begin{proof}
Differentiate $\log L_{\bsp}^\circ(z)=-z+\sum_i[zp_i-\log(1+zp_i)]$
twice. The first derivative divided by $z$ is
$\sum_i h_i-\sum_i h_i/(1+zp_i)$, giving the compensated square.
Discarding that nonnegative square leaves the diagonal terms.
The $i$th integral is the expected reciprocal of
$X_{\bsp}+p_i(E+E')$, for two additional independent unit exponentials.
Its mean is $1+2p_i$, so it proves \eqref{eq:mp-62}.

For the uniform bound put
\[
 X=1+\sum_i p_i(Z_i-1),\qquad
 Y=\sum_i h_i(Z_i-1),\qquad
 G=\sum_i h_i(\log Z_i+\gamma).
\]
Independence and the first two logarithmic moments of the
exponential law and its size bias give
$\bbE[YG]=\norm{\bsh}_2^2$ and
$\bbE[XG^2]=(\pi^2/6)\norm{\bsh}_2^2$.
Cauchy--Schwarz applied to $Y/\sqrt X$ and $\sqrt XG$, together
with $\rD^2\cJ_{\bsp}[\bsh,\bsh]=\bbE[Y^2/X]$, proves \eqref{eq:mp-73}.
Local differentiation is dominated by the positive atom factors or
the diffuse exponential, as in Lemma~\ref{lem:mp-calculus}.
Approximation and continuity give the integrated inequalities at
zero coordinates and at boundary spectra.
\end{proof}

\begin{cor}[Segment curvature]\label{cor:mp-segment}
For $\bsp_v=(1-v)\bsp^0+v\bsp^1$ and $\bsh=\bsp^1-\bsp^0$, with diffuse
compensation if necessary, the remainder after the tangent at $\bsp^0$
is at least
$$
 \cJ(\bsp^1)-\cJ(\bsp^0)-\rD\cJ_{\bsp^0}[\bsh] \geqslant\frac12\sum_i \frac{h_i^2}{1+2(2p_i^0+p_i^1)/3}.
$$
An upper cap for any mean segment entry can replace that entry in
the corresponding denominator. The uniform alternative is
$3\norm{\bsh}_2^2/\pi^2$.
\end{cor}

\begin{proof}
The measure $2(1-v)\,\dif v$ on $[0,1]$ has mean $1/3$.
Integrate \eqref{eq:mp-62} and apply Jensen to each reciprocal.
Integrating \eqref{eq:mp-73} gives the uniform alternative.
\end{proof}

\begin{lem}[Group-potential penalty]\label{lem:mp-group-penalty}
Let $f(x)\geqslant f_0(x):=Mx^2/(1+vx)$ on $x\geqslant0$,
where $M,v>0$; in particular $f(x)=x^2I(x)$ is allowed for an
anchor minorant from Lemma~\ref{lem:mp-anchor}.
On slopes $q<M/v$, the convex conjugate of $f_0$ on $x\geqslant0$ is \[
 \Psi_{M,v}(q)=
 \begin{cases}
 0, &q\leqslant0,\\
 q^2/[M(1+\sqrt{1-vq/M})^2], &0<q<M/v.
 \end{cases}
\] With the marginal
notation of \eqref{eq:mp-26}, take $U,V>0$ with $r_i\leqslant U$ for
all $i>1$ and $c_j\leqslant V$ for all $j>1$. These caps bound the
marginal tail atoms. Set $n_r=(R-a^2)/U^2$ and $n_c=(C-b^2)/V^2$. Choose $h\geqslant0$ and four
parameters $\ell_1,\ell_2,m_1,m_2\in[0,h]$, with $\ell_i+m_j-h<M/v$ for all four pairs.
Then

\begin{equation}\label{eq:mp-65}
\begin{aligned}
 \sum_{ij} f(p_{ij})\ \geqslant{}&
 a\ell_1+n_r U\ell_2+b m_1+n_c V m_2-h\\
 &-\Psi_{M,v}(\ell_1+m_1-h)
   -n_c\Psi_{M,v}(\ell_1+m_2-h)\\
 &-n_r\Psi_{M,v}(\ell_2+m_1-h)
   -n_rn_c\Psi_{M,v}(\ell_2+m_2-h) .
\end{aligned}
\end{equation}

\end{lem}

\begin{proof}
For positive $q<M/v$, the conjugate formula follows by solving $q=f_0'(x)=(M/v)[1-(1+vx)^{-2}]$ and
substituting in $qx-f_0(x)$. In particular $\Psi_{M,v}(q)/q^2$ increases for $q>0$.

Assign potential $\ell_1$ to the first row and $\ell_2 r_i/U$ to the other rows,
and assign $m_1$ to the first column and $m_2 c_j/V$ to the other columns. Subtract $h$
from the sum of the two potentials in each cell, then use $f(p_{ij})\geqslant
p_{ij}q_{ij}-\Psi_{M,v}(q_{ij})$. The sum of the linear terms is the first line of
\eqref{eq:mp-65}. For a strip cell one has, for example, \[
 (\ell_2 r_i/U+m_1-h)_+
 \leqslant (r_i/U)(\ell_2+m_1-h)_+
\] because $m_1\leqslant h$. For an inner cell, put $x=r_i/U,\ y=c_j/V$. The inequality \[
 (\ell_2 x+m_2y-h)_+
 \leqslant xy(\ell_2+m_2-h)_+
\] holds on $[0,1]^2$. The difference before positive parts, $xy(\ell_2+m_2-h)-(\ell_2
x+m_2y-h)$, is affine in each coordinate and is nonnegative at all four vertices. If
$\ell_2+m_2-h<0$, the left side with positive part is zero. Apply the monotonicity of
$\Psi_{M,v}(q)/q^2$ and sum to obtain the four penalties in \eqref{eq:mp-65}.

For a diffuse label assign potential zero. Its cell slope is
nonpositive because each opposite potential is at most $h$.
Thus including diffuse cells can only lower the full linear term;
the same finite-cell inequality proves the generalized bound.
The positive-part product estimate in this proof also applies to a
quadratic penalty $(s_+)^2/(2m)$, separately in each remainder block.
\end{proof}

\begin{lem}[Finite reciprocal mixtures]\label{lem:mp-reciprocal-mixture}
The anchor weight $I$ of Lemma~\ref{lem:mp-anchor} admits finite
positive reciprocal lower sums. Rounding their masses downward and
rates upward preserves the lower bound.
\end{lem}

\begin{proof}
For each anchor in \eqref{eq:mp-40}, use the positive reciprocal mixture
\[
 I(y)=\int_0^\infty L_{p^0}^\circ(z)
       \int_0^1\frac{v\,\dif v}{1+zvy}\,\dif z.
\]
Split the $v$ integral into four equal intervals. Their masses are $(2j-1)/32$, and the
means under density proportional to $v$ are $(3j^2-3j+1)/(6(2j-1))$, for $j=1,\ldots,4$.
Jensen on each interval and the lower Gauss rule from Lemma~\ref{lem:mp-gauss} give a finite
positive reciprocal sum below $I(y)$. The Gauss sign applies for fixed $y$, since the
integral in $z$ has a completely monotone integrand. Within any group of summands,
Jensen for $v\mapsto(1+vy)^{-1}$ permits replacing the rates by their mass-weighted
average. The reciprocal sum remains a lower bound when the masses are rounded down and
the rates rounded up.
\end{proof}

\begin{remark}[Common arithmetic protocol]\label{rem:mp-arithmetic}
The binary64 certificates use round-to-nearest IEEE 754 arithmetic
with contraction and fast-math disabled. The unrestricted certificate
instead encloses every acceptance test by outward intervals.
The quadrature grids continue by factor-two intervals to $8192$.
The concentrated-band, pivot and half-cap checks start with $[0,1/8]$.
In the unrestricted check, the outward Gauss and Lobatto grid starts
with $[0,1/4]$, and the weighted Jensen grid starts with $[0,1/8]$.
Omitted tails in lower bounds are nonnegative.
In the concentrated band of Proposition~\ref{prop:mp-concentrated},
bisection uses weights $(1,1,1/4,1/4)$ and initially weighted widths
$(63/128,31/128,1/4,1/4)$. After the first split (coordinate 1),
the selection repeats the cycle $3,4,1,2$, so the maximum depth 54
allows at most $(14,13,14,13)$ splits of the respective coordinates.
All endpoints of processed boxes have denominators dividing $2^{21}$,
and their bisection sums and the sums used for region tests are
exact in binary64. The rational cutoff $1/72$ is guarded outward by
$10^{-12}$, exceeding its rounding error. The analogous bounds for
the extended band are proved in Proposition~\ref{prop:mp-extended}.
The Gauss and Lobatto weights and
the positive quadrature nodes are guarded outward by relative offsets of $10^{-13}$. The
smallest distance of a nonendpoint node from the nearer endpoint of
its normalized interval is larger than $1/16$. The Lobatto endpoints
are the interval endpoints themselves. Evaluating the interior nodes by the displayed radical formulas, with square roots
rounded to nearest, incurs relative error at most $10^{-14}$ in the nodes. For the
cube-root map the verifier additionally checks the cube of its proposed upper node
against the corresponding radicand with an outward margin.

When the tail purity is $h/2^{14}$, the number of equal atoms is obtained by integer
division. The numerator of the radicand $((m+1)h-2^{14})/(2^{14}m)$ is exact. The
rounded large normalized atom has relative error at most $6\cdot2^{-52}$, and the small
atom $1-mq$ has absolute error at most $8\cdot2^{-52}$, including the zero endpoint. For
$0\leqslant z\leqslant8192$, powering the reciprocal linear factor by repeated squaring contributes
at most $2m$ relative rounding units. The errors from the input large atom, from
powering, and from the small linear factor total less than $5\cdot10^{-11}$ in relative
error for the transform within the normal range. Any loss through underflow is below the
absolute buffer in the transform subtraction. The verifier adds the relative transform
guard to the term subtracted in $f_{p_s}$, then subtracts a further absolute $10^{-14}$
before clipping to zero.

The positive quadrature expressions used with these transforms have at most five linear factors per
denominator. Their sums and the evaluation of $G$ incur absolute error less than
$10^{-10}$ near the comparison, using the outward weights and the explicit $10^{-12}$
guard in the evaluation of $G$. The purity target and the clipped quadratic for a
normalized loss use bounded arithmetic with denominators separated from zero since
$s,t\geqslant1/128$. Their absolute errors are also below $10^{-10}$. The $10^{-8}$ slack in
every acceptance comparison therefore certifies a strict positive normalized margin whenever
these bounds and denominator restrictions hold.

Here $s,t$ denote marginal tail masses and $G$ a tangent-gradient
comparison integral. The extended band uses the same node and transform enclosures;
its additional coefficient bounds and guards are stated with its
certificate. For the pivot tests, the supporting values and
derivatives have magnitude at most $64$, reciprocal offsets are at
least $1/128$, and rates are at most $64$. Sums contain at most $288$
positive terms. The reciprocal masses are rounded down and rates
up with guards of order $10^{-11}$. A supporting-line acceptance
requires absolute margin $10^{-7}$; a feasibility cut lowers the
intercept by $10^{-7}$ before locating its zero.
\end{remark}

\subsection{The binary case}\label{the-binary-case}

\begin{prop}[Binary tables]\label{prop:mp-binary}
Every binary table has $H\geqslant0$. With both marginals mixed,
equality holds exactly at the uniform table. The negative-correlation
parameter ranges over $[0,1]$; the positive-correlation half is analytic.
\end{prop}

\begin{proof}
Row and column swaps allow nonnegative marginal biases $\widehat{a},\widehat{b}$. Write 
$$
 \bsP(c)=\frac14\begin{pmatrix}1+\widehat{a}+\widehat{b}+c&1+\widehat{a}-\widehat{b}-c\\
                  1-\widehat{a}+\widehat{b}-c&1-\widehat{a}-\widehat{b}+c\end{pmatrix}.
$$
The product has $c=k=\widehat{a}\widehat{b}$. A pure marginal implies a product state, so assume $\widehat{a},\widehat{b}<1$.
Put $A=\widehat{a}^2$, $B=\widehat{b}^2$, $d=(1-\widehat{a})(1-\widehat{b})$, and $p=(1+\widehat{a})(1+\widehat{b})$. The lower feasible endpoint is
$\widehat{a}+\widehat{b}-1=k-d$. The marginal linear entropy product is $(1-A)(1-B)/4=dp/4$. Let
$H(c)=\Delta_{\Q}(\bsP(c))-dp/12$.

The product-point calculation in \hyperref[step:binary-product]{Step~8}
of the proof of Lemma~\ref{lem:diagonal-core} gives 
$$
 H(k)=\frac{dp}{8}T,\qquad H'(k)=\frac{k}{4}J,
 \qquad D(u)=(1-Au^2)(1-Bu^2),
$$
$$
 T=\int_0^1(1-u^2)(D(u)^{-1}-1)\,\dif u,\qquad
 J=\int_0^1\frac{(1-u^2)^2}{D(u)}\,\dif u.
$$
Here $\Delta_L(\bsP(k))=dp/4$. The formulas hold on $\widehat{a},\widehat{b}<1$, including repeated product entries by continuity.

At fixed marginals the difference of the two deficits is exact:
$F(c)=H(c)-(c^2-k^2)/12$. Thus $H''=F''+1/6$ and
$H'(k)=F'(k)+k/6$. In particular, the Hessian calculation in
\hyperref[step:binary-curvature]{Step~7} gives $H''(c)\geqslant1/5$. Integration and continuity give \[
 H(c)\geqslant \frac{dp}{8}T+\frac{kJ}{4}(c-k)+\frac{(c-k)^2}{10}.
\] For $c\geqslant k$ all terms on the right are nonnegative. They all vanish only when $\widehat{a}=\widehat{b}=c=0$.

For $c<k$, set $z=(k-c)/d\in(0,1]$. Then

\begin{equation}\label{eq:mp-1}
\frac8d H(c)\geqslant pT-2kzJ+\frac45 d\,z^2.
\end{equation}

Write $w_j=2/((2j+1)(2j+3))$ and $h_j=\sum_{i=0}^j A^iB^{j-i}$. Expanding $D^{-1}$ and integrating the nonnegative series gives \[
 T=\sum_{j\geqslant1}w_jh_j,\qquad
 J=\frac8{15}+\sum_{j\geqslant1}(w_j-w_{j+1})h_j .
\] The coefficient of each $h_j$ in $pT-2kzJ$ is positive since $p\geqslant4k$, $0\leqslant z\leqslant1$,
and $w_j>w_{j+1}>0$. Keeping terms through $j=4$ thus gives the following lower bound
for \eqref{eq:mp-1}.

\begin{equation}\label{eq:mp-2}
\mathcal P(\widehat{a},\widehat{b},z)=
 p\sum_{j=1}^4 w_jh_j
 -2\widehat{a}\widehat{b}z\left(\frac8{15}+\sum_{j=1}^4(w_j-w_{j+1})h_j\right)
 +\frac45(1-\widehat{a})(1-\widehat{b})z^2.
\end{equation}

On $0\leqslant \widehat{a},\widehat{b}\leqslant1/8$, the expression in parentheses in \eqref{eq:mp-2} is at most
$J\leqslant1$, because $D(u)\geqslant(1-u^2)^2$. Completing a square gives \[
 \mathcal P\geqslant \frac2{15}(\widehat{a}^2+\widehat{b}^2)-2\widehat{a}\widehat{b}z+\frac45 d\,z^2
 \geqslant \frac2{15}(\widehat{a}^2+\widehat{b}^2)-\frac{5\widehat{a}^2\widehat{b}^2}{4d}.
\] Here $d\geqslant49/64$ and $\widehat{a}^2\widehat{b}^2\leqslant(\widehat{a}^2+\widehat{b}^2)/128$. The lower bound is strictly positive
if $(\widehat{a},\widehat{b})\ne(0,0)$. For $\widehat{a}=\widehat{b}=0$, the polynomial is $4z^2/5$.

On the rest of the unit cube, a tensor Bernstein calculation proves strict positivity.
For a coordinate of degree $n$, a monomial of degree $q$ has normalized Bernstein
coefficient $\binom{i}{q}/\binom{n}{q}$ at index $i\geqslant q$ and zero at $i<q$. Use the
product of these coefficients in three coordinates after expanding \eqref{eq:mp-2},
whose multidegree is $(9,9,2)$.

Starting with the full cube, bisect coordinates $\widehat{a},\widehat{b},z$ cyclically at the midpoint of
the current interval. A box is terminal when all coefficients are positive or when both
bias intervals lie in $[0,1/8]$. In a coordinate of degree $n$, at each step of the
subdivision replace the current list $(q_i)$ by $(q_i+q_{i+1})$. At step $s$ take the
leftmost value at index $s$ on the left box and the rightmost value at index $n-s$ on
the right box, multiplying both values by $2^{n-s}$. The new coefficients equal the
Bernstein subdivision coefficients multiplied by $2^n$.

The exact verifier \nolinkurl{binary-certificate.py} gives 15 boxes with all
coefficients positive and one box covered by the small-bias argument. The maximum
terminal depth is 8. Each Bernstein basis is nonnegative and sums to one, so these
finitely many rational sign checks cover the whole cube. Equation \eqref{eq:mp-1} proves
strictness when $c<k$. In the other case equality with both marginals mixed requires the
uniform table. There $\Delta_{\Q}=1/12$ and the marginal linear entropy product is
$1/4$, which proves sharpness.
\end{proof}

The concentrated-band and half-cap checks also use the binary
ratio on a full marginal box. For mixed binary marginals write
$d=(1-\widehat a)(1-\widehat b)$ and $p=(1+\widehat a)(1+\widehat b)$,
so the marginal linear-entropy product is $dp/4$. With the
truncations $T_4,J_4$ of the sums just used, \eqref{eq:mp-1} gives
$$
 \frac{\Delta_{\Q}(\bsP(c))}{dp/4} \geqslant \frac13+\frac{T_4}{2}   -\frac{\widehat a\widehat b}{p}J_4 z+\frac{2d}{5p}z^2
 \quad\hbox{if }c<k,\qquad z=(k-c)/d.
$$
For $c\geqslant k$ the ratio is at least $1/3+T_4/2$. Thus
minimizing the quadratic on $0\leqslant z\leqslant1$ gives a
common lower bound. Its value at $z=0$ is available in the
minimization, which handles the positive-correlation case.
In a box, the lower bias endpoints bound $T_4$ below. The
upper bias endpoints bound $(\widehat a\widehat b/p)J_4$ above
and $2d/(5p)$ below by monotonicity of $b/(1+b)$ and
$(1-b)/(1+b)$ on $[0,1]$. Finally the binary theorem allows
clipping the resulting ratio below at $1/3$.

\subsection{Tables supported on one row and one column}\label{tables-supported-on-one-row-and-one-column}

\begin{prop}[Star support]\label{prop:mp-star}
A generalized table supported in the union of one ordinary row and
one ordinary column has $H>0$ whenever both marginals are mixed. The certificate
covers $s,t,\alpha,\beta\in[0,1]$ with $s+t\leqslant1$, where
$s=1-a,t=1-b$ and $\alpha,\beta$ are full normalized tail purities.
\begin{equation}\label{eq:mp-15}
\Delta_{\Q} > \tfrac13 \sfS_L(\bsr)\sfS_L(\bsc).
\end{equation}
\end{prop}

\begin{proof}
Choose a row and a column whose union supports the table, and let
$a,b$ be their respective masses. Then $x=a+b-1\geqslant0$ is the intersection mass. Write
$U(z)=L_{r_{\rm tail}}(z)$, $V(z)=L_{c_{\rm tail}}(z)$, using the products for the
remaining row and column atoms and no normalization. Diffuse tail mass contributes its
exponential factor. The exact kernel is
$$
 G_{\bsP}(z)=\frac1{1+z}-\frac{U(z)}{1+za}
                      -\frac{V(z)}{1+zb}
                      +\frac{U(z)V(z)}{1+zx}.
$$
It decreases in each of $U,V$ over the feasible range, because
$V(z)/(1+zx)\leqslant(1+za)^{-1}$ and $U(z)/(1+zx)\leqslant(1+zb)^{-1}$. For example the first
inequality groups the $c$-tail and $x$, whose masses sum to $a$.

Consequently there is a four-real-parameter lower bound on the deficit for the entire
family, even with diffuse mass. Let $\alpha=\|r_{\rm tail}\|_2^2/(1-a)^2$ and
$\beta=\|c_{\rm tail}\|_2^2/(1-b)^2$. Replace $U,V$ by \[
 \overline U(z)=(1+\alpha(1-a)z)^{-1/\alpha},\qquad
 \overline V(z)=(1+\beta(1-b)z)^{-1/\beta},
\] with exponential limits at zero. Lemma~\ref{lem:mp-transforms}, applied after normalizing each tail,
gives these upper bounds. Each upper envelope
also stays below the single-factor transform $(1+zs)^{-1}$, hence replacing both factors
preserves the direction of the bound on $G_{\bsP}$. The marginal linear entropies are already
exact functions of $a,b,\alpha,\beta$.

This includes generalized tables with diffuse tail mass. The proof uses a polynomial
bound on each tail determinant and a validated finite quadrature.

Set $s=1-a$, $t=1-b$, and $x=1-s-t$. Let $z>0$ be the Laplace variable. For a normalized
tail with squared sum $\alpha$, let $e_j$ denote the coefficients of its determinant,
including the exponential factor for any diffuse part. Put 
$$
 d_j(\alpha)=\frac1{j!}\prod_{\ell=1}^{j-1}(1-\ell\alpha)_+, \qquad j=2,3,4.
$$
The coefficient comparison holds for every degree. For a finite probability vector,
$j!e_j$ is the probability that $j$ independent samples have distinct labels. On that
event the conditional law of the first label weights its original probability $p_i$ by
$e_{j-1}(p_{\widehat i})$. For $j\geqslant2$ and $i\ne k$, \[
 e_{j-1}(p_{\widehat i})-e_{j-1}(p_{\widehat k})
 =(p_k-p_i)e_{j-2}(p_{\widehat i,\widehat k}).
\] Thus the weight is nonincreasing in $p_i$. For $j=1$ the weight is constant. The
covariance inequality for a variable and a decreasing function of it bounds the
conditional mean of $p_i$ by its unconditional size-biased mean $\alpha$. Given $j$
distinct labels, the next sample avoids them with conditional probability at least
$(1-j\alpha)_+$ by exchangeability. Iteration proves
$k!e_k\geqslant\prod_{j=0}^{k-1}(1-j\alpha)_+$. If a preceding distinctness event has
probability zero, the recursive lower bound is also zero. Approximate diffuse mass by
equal atoms tending to zero. Each fixed coefficient converges by Newton's identities,
since the first power sum stays equal to one and all higher power sums converge.

Write \[
 D_s(z)=1+sz+(sz)^2 C_s(z),\qquad
 C_s(z)=d_2(\alpha)+d_3(\alpha)sz+d_4(\alpha)s^2z^2.
\] Use $D_t,C_t$ with $t,\beta$ for the other tail. The tail transforms obey $U\leqslant
D_s^{-1}\leqslant (1+sz)^{-1}$ and $V\leqslant D_t^{-1}\leqslant (1+tz)^{-1}$ by the nonnegative
coefficient bounds. The two monotonicities of $G_{\bsP}$ allow replacing $U,V$ by these upper
bounds. Denote \[
 A=1+az,\ B=1+bz,\ X=1+xz,\ S=1+sz,\ T=1+tz,\qquad
 d_s=C_s/(S D_s),\quad d_t=C_t/(T D_t).
\] The resulting lower bound for $G_P(z)/(st z^2)$ is the following sum of nonnegative rational functions:

\begin{equation}\label{eq:mp-16}
f(z)=
 \frac1{STXAB}
 +\frac{abz^2}{(1+z)STAB}
 +\frac{x s z^2 d_s}{A X T}
 +\frac{x t z^2 d_t}{B X S}
 +\frac{st z^2 d_s d_t}{X}.
\end{equation}

To check the formula, first take the two tail transforms $1/S,1/T$. Clearing
denominators for this binary table gives the first two terms in \eqref{eq:mp-16}, since
the numerator before dividing by $stz^2$ is proportional to $1+z+abz^2(1+xz)$. Next the
decreases of the transforms on replacing them by $1/D_s,1/D_t$ are
$s^2z^2d_s,t^2z^2d_t$, respectively. The first decrease is multiplied by
$1/A-1/(XT)=xtz^2/(AXT)$, the second by $1/B-1/(XS)=xsz^2/(BXS)$, and their product by
$1/X$. These give the remaining three terms. The target after dividing by $st$ is

\begin{equation}\label{eq:mp-17}
W=\tfrac13[2-(1+\alpha)s][2-(1+\beta)t].
\end{equation}

The normalized comparison extends continuously to $s=0$ or $t=0$ at finite $z$. The
original inequality at a pure marginal follows separately from the vanishing gap and
product.

For validation put $q=21/20$ and $z_k=q^k$, for $-180\leqslant k\leqslant130$. For each nonzero
summand $f_i$ in \eqref{eq:mp-16} the function $\log[e^y f_i(e^y)]$ has second
derivative at least $-9$. In fact the second logarithmic derivative of a nonzero
polynomial with nonnegative coefficients is the variance of its degrees under weights
proportional to its terms. This is between zero and $d^2/4$ for degree $d$, by the
variance bound on $[0,d]$. Subtract the denominator contributions. The largest possible
sum of the squared denominator degrees divided by four is $35/4$, from the last summand
with two factors of degree four and three linear factors. Zero summands obey the
resulting integral bound trivially. Writing $h=(\log q)/2$, the second derivative bound
and the midpoint value give a lower exponential model of the form $z_k
f_i(z_k)\exp(D(y-\log z_k)-9h^2/2)$ on the interval of radius $h$. The average of
$\exp(D(y-\log z_k))$ on this interval is at least one by Jensen. The intervals for
consecutive $k$ are adjacent. Using $e^{-u}\geqslant1-u$ and $2(q-1)/(q+1)\leqslant\log q\leqslant q-1$
therefore gives the lower bound

\begin{equation}\label{eq:mp-18}
\int_0^\infty f(z)\,\dif z
 \ \geqslant\ \frac{3191}{65600}\sum_{k=-180}^{130} z_k f(z_k).
\end{equation}

Dropping the unintegrated positive tails preserves the bound.

The verifier \nolinkurl{determinant-certificate.cpp} proves the right side of
\eqref{eq:mp-18} exceeds \eqref{eq:mp-17} on $s,t,\alpha,\beta\in[0,1]$, $s+t\leqslant1$. On
each dyadic box it uses the larger endpoints for denominators and smaller endpoints for
numerators, except that a factor $u/(1+uz)$ is bounded at the smaller endpoint as a
whole. It uses $x_{\rm low}=\max(0,1-s_{\rm high}-t_{\rm high})$, $x_{\rm high}=1-s_{\rm
low}-t_{\rm low}$ on boxes meeting the triangle. The coefficients $d_j$ decrease with
the corresponding purity. Formula \eqref{eq:mp-17} increases when any of the four
parameters decreases, so its upper bound uses the lower endpoints. Subdivision bisects
the largest weighted width with weights $1,1,1/4,1/4$. Boxes outside the triangle are
skipped.

The check uses IEEE 754 binary64 round-to-nearest arithmetic without
fast-math or contraction. At depth at most 48 all dyadic endpoints, their sums for bisection and for the
triangle test, the differences formed before the lower bound, and the factors
$1-\ell\alpha$, $1-\ell\beta$ for $\ell\leqslant3$ are exactly representable. Use
$b_0=1+2^{-51}$. Every positive elementary product, quotient, or sum computed from exact
operands has relative error between $b_0^{-1}$ and $b_0$. The same holds for a positive
exact dyadic difference used above. All nonzero operands here lie in the normal range.
Indeed positive dyadic factors are at least $2^{-48}$ and $2^{-14}<z_k<2^{10}$; each
expression uses only bounded-degree products of these factors and polynomial sums with
nonnegative terms. For positive operands, relative-error exponents in base $b_0$ add
under products or quotients and take their maximum under sums, followed by an increment
of one for rounding. The recurrence for $z_k$ gives exponent at most 360. A linear
denominator factor or a scaled tail mass has exponent at most 362, a tail quadratic
$C_s$ or $C_t$ at most 750, a tail determinant at most 1480, and each $d_s,d_t$ at most
2600. The individual summands in \eqref{eq:mp-18}, including multiplication by $z_k$,
have exponent below 7000. Summing 311 terms and multiplying by the quadrature weight
keeps the exponent below 20000. Since $b_0^{20000}<1+10^{-10}$, this bounds the relative
check error. The few bounded operations in $W$ give absolute error below $10^{-13}$. The
verifier requires a margin $10^{-8}$, which is more than the possible error at
$W\leqslant4/3$. Hence each accepted box gives a rigorous strict lower bound. The program
accepts 22,286,338 boxes, excludes 39 outside the triangle, and reaches maximum accepted-leaf depth
41. This proves \eqref{eq:mp-15}.
\end{proof}

\begin{prop}[Large marginal heads]\label{prop:mp-large-heads}
If both marginals are mixed and $a+b\geqslant7/4$, then $H>0$.
At these fixed marginals a star table maximizes joint subentropy.
\end{prop}

\begin{proof}
If both marginals are mixed and their largest masses $a,b$ satisfy $a+b\geqslant7/4$, the
joint subentropy is maximized, at those marginals, by a table supported in the union of
the largest row and the largest column. In consequence the strict coefficient $1/3$
holds throughout this marginal region.

In fact put $x=a+b-1$ and consider the table $\bsP^{\star}$ with $p_{11}=x$, with column-tail
entries along the first row and row-tail entries along the first column. All other cells
vanish. At fixed marginals, adding an amount $w_{ij}$ to a cell $i,j>1$ requires moving
that amount out of $p_{i1},p_{1j}$ and adding it to $p_{11}$. The concave tangent for
joint $\Q^\circ$ at $\bsP^{\star}$ bounds its change from above by 
$$
 \sum_{i,j>1}w_{ij}\bigl(g_{\bsP^{\star}}(r_i)+g_{\bsP^{\star}}(c_j)-g_{\bsP^{\star}}(x)\bigr).
$$
For a diffuse index the corresponding $g$ value is zero. The tangent formula follows
along the feasible segment by the same derivative calculation as \eqref{eq:mp-8}. Since
$g_{\bsp}(y)\leqslant y\int L_{\bsp}^\circ$, the sum is nonpositive if

\begin{equation}\label{eq:mp-20}
\frac{g_{\bsP^{\star}}(x)}{\int_0^\infty L_{\bsP^{\star}}^\circ(z)\,\dif z}\geqslant1-x.
\end{equation}

Indeed $r_i+c_j\leqslant(1-a)+(1-b)=1-x$.

To prove \eqref{eq:mp-20}, compare $\bsP^{\star}$ with the two-atom spectrum $(x,1-x)$. The ratio
of its Laplace transform to that of the two-atom spectrum is nonincreasing in $z$. This
follows by logarithmic differentiation: refining the mass $1-x$ into tail atoms and
diffuse mass increases $\sum q/(1+zq)$, counting a diffuse mass $h$ as the additive term
$h$. The average of the decreasing function $x/(1+zx)$ under the normalized Laplace
weights of $\bsP^{\star}$ is therefore at least that under the normalized two-atom weights. For
this comparison, the covariance of two nonincreasing real functions of a common random
variable is nonnegative, by expanding the covariance using two independent copies. Set
$y=1-x$. Direct integration gives
$$
 \int_0^\infty\frac{\dif z}{(1+xz)(1+yz)}  =\frac{\log(x/y)}{x-y},
 \qquad
 \int_0^\infty\frac{x\,\dif z}{(1+xz)^2(1+yz)} =\frac{x}{x-y}-\frac{xy\log(x/y)}{(x-y)^2}.
$$
The ratio is at least $y$ exactly if $\log k\leqslant k(k-1)/(2k-1)$, where $k=x/y$. For
$x\geqslant3/4$ we have $k\geqslant3$. At $k=3$, $\log3<6/5$ by the positive series for $e^{6/5}$.
The rational function has derivative $(2k^2-2k+1)/(2k-1)^2\geqslant1/2$, whereas $1/k\leqslant1/3$.
This proves \eqref{eq:mp-20}, and \eqref{eq:mp-15} gives the stated strict bound.
\end{proof}

\begin{lem}[A second-atom star criterion]\label{lem:mp-star-criterion}
If $x_*=a+b-1>0$ and the two second marginal atoms obey the
criterion below, a star table maximizes joint subentropy. Hence
$H>0$ for mixed marginals. When $x_*\geqslant1/2$, the sufficient
condition $a_2+b_2\leqslant x_*(1-x_*)$ is available.
\end{lem}

\begin{proof}
The concentrated-marginal argument also applies when the second largest atoms are small.
Write $a,b$ for the largest marginal masses, $a_2,b_2$ for the second largest, and
$x_*=a+b-1>0$. As in \eqref{eq:mp-20}, let $P^{\star}$ have only its first row and first
column occupied, with their intersection of mass $x_*$. For $0<x<1$ define
$$
 h(x)= \begin{cases}
 \displaystyle\frac{x}{\log(x/(1-x))}-\frac{x(1-x)}{2x-1},
     &x\neq1/2,\\[4pt]
 1/4,&x=1/2.
 \end{cases}
$$
Then $\bsP^{\star}$ maximizes joint subentropy for its marginals whenever

\begin{equation}\label{eq:mp-34}
a_2+b_2\leqslant h(x_*).
\end{equation}

Indeed $g_{\bsP^{\star}}(x_*)/\int_0^\infty L_{\bsP^{\star}}\geqslant h(x_*)$ by the two-atom comparison in
\eqref{eq:mp-20}. The two-atom formula there gives $h$ after taking the quotient,
including $h(1/2)=1/4$ by direct integration. Now $g_{\bsP^{\star}}(r_i)+g_{\bsP^{\star}}(c_j)\leqslant
(a_2+b_2)\int L_{\bsP^{\star}}$ for all $i,j>1$. Thus every coefficient in the supporting tangent
at $\bsP^{\star}$ is nonpositive. Equation \eqref{eq:mp-15} proves the strict coefficient $1/3$
for mixed marginals in this region. This argument also applies to generalized tables
because the tangent and the comparison in \eqref{eq:mp-20} include diffuse mass.

For $x\geqslant1/2$ an elementary sufficient condition in \eqref{eq:mp-34} is $a_2+b_2\leqslant
x_*(1-x_*)$, with $x_*\geqslant1/2$. To see this, put $k=x/(1-x)$. The inequality $h(x)\geqslant
x(1-x)$ is equivalent for $k>1$ to $\log k\leqslant (k^2-1)/(2k)$. The right side minus the
left side has derivative $(k-1)^2/(2k^2)$ and vanishes at one. At $x=1/2$ the comparison
holds by the stated value of $h$.
\end{proof}

\subsection{A nearly pure or highly mixed marginal}\label{a-nearly-pure-or-highly-mixed-marginal}

\begin{prop}[Nearly pure marginal]\label{prop:mp-near-pure}
For mixed marginals, $\max(a,b)\geqslant71/72$ implies $H>0$.
The auxiliary gradient bound holds on the full interval
$0\leqslant a\leqslant49/64$ and all feasible tail purities.
\end{prop}

\begin{proof}
If $s=1-a$, $d=R-a^2$, and
$e=(s^2-d)/2$, the determinant of the tail is at least $1+sz+ez^2$ by its first three
coefficients. The substitution $y=(1+az)^{-1}$ in $g_r(a)$ gives

\begin{equation}\label{eq:mp-9}
1-g_r(a)\geqslant D(a,d)=\int_0^1
  \left(1-\frac{1}{1+sz+ez^2}\right)\dif y,\qquad z=\frac{1-y}{ay}.
\end{equation}

Merging the column tail and evaluating the binary subentropy gives
\begin{equation}\label{eq:mp-binary-loss}
 \Q(\bsc)\geqslant \Q(1-\epsilon,\epsilon)
 \geqslant \epsilon\left(1-\frac{\epsilon\log(1/\epsilon)}{1-2\epsilon}\right)
 \quad(0<\epsilon<1/2).
\end{equation} The second step uses $-\log(1-\epsilon)\geqslant\epsilon$ in the binary spectral formula. Also $1-C\leqslant 2\epsilon$.

For $a\leqslant49/64$, the gradient in \eqref{eq:mp-8} satisfies

\begin{equation}\label{eq:mp-38}
1-g_{\bsr}(a)>\frac23(1-R)+\frac1{16}.
\end{equation}

To verify this, use the integral lower bound $D(a,d)$ of \eqref{eq:mp-9} and set
$\alpha=d/(1-a)^2$. The integrand decreases affinely inside its denominator as $\alpha$
increases, so the integral is concave in $\alpha$. The right side of \eqref{eq:mp-38} is
affine in $\alpha$. It is enough to check $\alpha=0,1$, for which the integrands, with
$X=(1-a)(1-y)/(ay)$, are respectively \[
 \frac{2X+X^2}{2+2X+X^2},\qquad \frac{X}{1+X}.
\] Both decrease in $a$ and in $y$. On each of the 392 intervals $[i/512,(i+1)/512]$,
evaluate the integrals at the right endpoint using lower right Riemann sums on 1024
equal $y$ intervals. The integer verifier \nolinkurl{pure-boundary-certificate.py}
rounds each rational summand down to a multiple of $2^{-36}$. It compares the resulting
lower bounds to $1/16$ plus the maximum of the affine-target endpoint values
$2(1-a^2)/3$ and $4a(1-a)/3$, respectively, over that interval in $a$. Both tests are
strict on every interval.

If $b\geqslant71/72$ and $a+b\geqslant7/4$, apply \eqref{eq:mp-15} and \eqref{eq:mp-20}. Otherwise
$a<55/72<49/64$, so \eqref{eq:mp-38} applies. For $\epsilon=1-b$ with
$0<\epsilon\leqslant1/72$, monotonicity of the loss in \eqref{eq:mp-binary-loss} gives \[
 \frac{\epsilon\log(1/\epsilon)}{1-2\epsilon}
 \leqslant \frac{\log72}{70} <\frac1{16}.
\] The last comparison follows from $\log72=3\log2+2\log3<43/10$. The bounds
$\log2<7/10$ and $\log3<11/10$ follow by the positive exponential series through degree
five, whose values already exceed two and three at the respective rational arguments.
Now \eqref{eq:mp-8} and \eqref{eq:mp-binary-loss} give $\Delta_{\Q}>
(2/3)\epsilon(1-R)\geqslant(1-R)(1-C)/3$. At $b=1$ both sides vanish. Exchange rows and
columns for the symmetric region.
\end{proof}

\begin{prop}[Small marginal head]\label{prop:mp-small-head}
If $\min(a,b)\leqslant1/20$ and both marginals are mixed, then
$H>0$.
\end{prop}

\begin{proof}
If $a\leqslant1/20$, retain the largest column of mass $b$. By \eqref{eq:mp-8} and by the size-biased Jensen bound on $L_{\bsr}$, 
$$
 \Q(\bsP)-\Q(\bsr)\leqslant(1-b)g_{\bsr}(a) \leqslant(1-b)\frac{a}{1-R}\leqslant\frac{1-C}{19},
$$
using $R\leqslant a$ and $C\leqslant b$. The single-state comparison $\Q(\bsc)\geqslant(2\log2-1)(1-C)$
from Proposition~\ref{prop:single-state-sharp} then gives

\begin{equation}\label{eq:mp-48}
\Delta_{\Q}\geqslant (2\log2-1-1/19)(1-C)\geqslant \frac13(1-R)(1-C).
\end{equation}

The coefficient $2\log2-1-1/19$ is strictly greater than $1/3$, as seen by keeping the
first three positive terms of the $\log 2$ series at $1/3$. This gives strictness for
mixed marginals. Exchange rows and columns for the symmetric case.
\end{proof}

\subsection{Marginal determinant and purity estimates}\label{marginal-determinant-and-purity-estimates}

\begin{lem}[Marginal purity loss]\label{lem:mp-marginal-loss}
For every generalized spectrum of purity $u\in[0,1]$, the
following polynomial bounds $\mathcal J$ above, including the
diffuse endpoint $u=0$.
\begin{equation}\label{eq:mp-22}
\sfQ^\circ(\bsp)\geqslant 1-\gamma-u/2+u^2/12-u^4/120.
\end{equation}
\end{lem}

\begin{proof}
Lemma~\ref{lem:mp-transforms} gives
$L_{\bsp}^\circ(z)\leqslant (1+uz)^{-1/u}$. The right side is the Laplace transform of a gamma
variable $X$ with shape and rate $1/u$. If $k=1/u$, its expectation $\bbE[X\log X]$ is
$\psi(k+1)-\log k$, where $\psi$ is the logarithmic derivative of the gamma function. By
the integral formula for the digamma function, \[
 \psi(k+1)-\log k
 =\int_0^\infty e^{-kv}\left(\frac1v-\frac1{e^v-1}\right)\dif v.
\] The parenthesis is at most $1/2-v/12+v^3/720$. One elementary verification is to
multiply $\coth x\geqslant1/x+x/3-x^3/45$ by $45x\sinh x$: the coefficient of $x^{2n+1}$,
times $(2n+1)!$, in the resulting difference is $16n(n-1)(n-2)(n+2)$ for $n\geqslant0$, hence
nonnegative. Take $x=v/2$ to obtain the required bound and integrate termwise. This
proves \eqref{eq:mp-22}, with continuity at $u=0$.
\end{proof}

\begin{prop}[Low marginal caps]\label{prop:mp-low-cap}
For finite tables of arbitrary marginal ranks, $a,b\leqslant43/100$
implies
\begin{equation}\label{eq:mp-46}
H(\bsP)=\Delta_{\sfQ}(\bsP)-\frac13(1-R)(1-C)>\frac1{4000}.
\end{equation}
The same uniform positive bound extends to generalized tables,
with a non-strict limiting margin.
\end{prop}

\begin{proof}
Put $A=43/100$. Work at the fixed-marginal optimizer and set $u=(R+C)/2$, so $0\leqslant u\leqslant A$. For the two
marginal entropy losses, \eqref{eq:mp-22} gives $1-\gamma-\Q(\bsp)\leqslant U(\norm{\bsp}_2^2)$, where
$U(v)=v/2-v^2/12+v^4/120$ is concave on $[0,1]$. Thus the two losses sum to at most
$2U(u)$. Also $(1-R)(1-C)\leqslant(1-u)^2$.

For the joint spectrum, \eqref{eq:mp-32} and $a,b\leqslant A$ give 
$$
 \pi_{\bsP}\geqslant \frac{(RC-(A-R)(A-C))_+^2}{RC}     \geqslant \frac{A^2(2u-A)_+^2}{u^2},
$$
with the bound interpreted as zero if $u=0$. In the second inequality use $RC\leqslant u^2$
and the identity $RC-(A-R)(A-C)=A(2u-A)$. For finite positive marginals the first
denominator is nonzero. The diffuse endpoint follows by continuity. The largest joint
atom is at most $A^2$ by \eqref{eq:mp-26}. The first two terms of the joint entropy loss
bound \eqref{eq:mp-33} show 
$$
 1-\gamma-\Q(\bsP)\geqslant k_1 \pi_P+k_2 \pi_P^2, \qquad k_1=\frac1{2(1+2A^2/3)},\quad k_2=\frac1{4(1+3A^2)^2}.
$$
Use $1-\gamma>2113/5000$ from Lemma~\ref{lem:mp-constants}.
For $u\leqslant A/2$, the resulting lower bound is the polynomial
$2113/5000-1/3-u/3-u^2/6-u^4/60$, which decreases and exceeds $1/4000$ at $u=A/2$. For
$u\in[A/2,A]$, multiply the lower bound minus $1/4000$ by $u^4$. The rational verifier
\nolinkurl{mixing-cap-certificate.py} computes its degree-eight Bernstein coefficients
on the two halves of $[A/2,A]$, and all are positive. The check uses exact rational
harmonic sums, logarithm truncation and coefficients.
\end{proof}

\begin{prop}[Coarse low-purity region]\label{prop:mp-coarse-purity}
Every finite table with $R+C\leqslant107/200$ satisfies
\begin{equation}\label{eq:mp-59}
H(\bsP)=\Delta_{\sfQ}(\bsP)-\tfrac13(1-R)(1-C)>\frac1{5000}.
\end{equation}
The bound extends to generalized tables with a non-strict limiting
margin, which remains positive.
\end{prop}

\begin{proof}
Let $\bsp$ be a probability vector with purity $u$ and largest atom $m$. Choose index $i$
with probability $p_i$ and put $Y=p_i$, so that $\bbE Y=u$ and $V_{\bsp}:=\sum_i
p_i^3-u^2=\operatorname{Var}(Y)$. Set $h_z(y)=\int_0^z(1+vy)^{-1}\,\dif v$. The exact
remainder of the reciprocal about $y=u$ gives 
$$
\bbE [h_z(Y)]-h_z(u) = \bbE\left[(Y-u)^2 \int_0^z\frac{v^2\,\dif v}{(1+vu)^2(1+vY)}\right]
 \geqslant V_{\bsp}\int_0^z\frac{v^2\,\dif v}{(1+vu)^2(1+vm)}.
$$
Since $L_{\bsp}(z)=\exp(- \bbE [h_z(Y)])$ and $\bbE [h_z(Y)]\leqslant z$, 
$$
 (1+zu)^{-1/u}-L_{\bsp}(z)  \geqslant V_{\bsp} e^{-z}\int_0^z\frac{v^2\,\dif v}{(1+vu)^2(1+vm)} .
$$
Integrate against $\dif z/z^2$ and put $v=sz$, where $0\leqslant s\leqslant1$. With $Z$ of
density $ze^{-z}$ and independent $S$ of density $3s^2$, the resulting coefficient of
$V_{\bsp}$ is bounded below by 
$$
 \tfrac13 \bbE[(1+uSZ)^{-2}(1+mSZ)^{-1}] \geqslant k(u):=\frac1{3(1+3u/2)^2(1+3\sqrt u/2)}.
$$
The product of reciprocals is log-convex, hence convex, as a function of $SZ$. Apply
Jensen using $\bbE[SZ]=3/2$ and $m\leqslant\sqrt u$. The gamma comparison in \eqref{eq:mp-22}
now gives

\begin{equation}\label{eq:mp-60}
\cJ(\bsp)=1-\gamma-\Q(p)\leqslant U(u)-k(u)V_p,\qquad  U(u)=u/2-u^2/12+u^4/120 .
\end{equation}

For fixed finite marginals use the optimizer in \eqref{eq:mp-26}. Put $q=(R+C)/2$,
$q_0=107/400$, and $m_0=207/400$. Its largest joint atom is at most $ab\leqslant\sqrt{RC}\leqslant
q_0$. Then $\cJ(\bsP)\geqslant\lambda\pi_{\bsP}$ with $\lambda=1/[2(1+2q_0/3)]$. To see
this, use $\exp(t)\geqslant 1+t$ for the logarithmic surplus $\log L_{\bsP}(z)+z$. The normalized kernel of Lemma~\ref{lem:mp-transforms} bounds the coefficient of $\pi_{\bsP}$ below by $\tfrac12
\bbE(1+TZq_0)^{-1}$, where $T$ has density $2t$ on $[0,1]$ and $Z$ is an independent
unit exponential. Jensen gives $\lambda$.

Let $K_0=[3(1+3q_0/2)^2(1+3m_0/2)]^{-1}$ and set $k_*=\sqrt{k(R)k(C)}$. The function $k$
decreases and is log-convex, and $m_0^2\geqslant q_0$. Thus $K_0\leqslant k(q)\leqslant
k_*\leqslant1/3<\lambda$. Under the cell distribution $(I,J)\sim P$, expansion of the
following square and variance cancels their inner-product terms: 
$$
\begin{aligned}
 \lambda\pi_P+k(R)V_r+k(C)V_c
 ={}& (2k_*-k_*^2/\lambda)RC
 +\lambda\|P-(k_*/\lambda)r\otimes c\|_2^2\\
 &+\operatorname{Var}_P\!\left(\sqrt{k(R)}r_I+\sqrt{k(C)}c_J\right).
\end{aligned}
$$
The function $2k-k^2/\lambda$ increases for $0\leqslant k\leqslant\lambda$. Set
$K=2K_0-K_0^2/\lambda<1/3$. By concavity of $U$ and $RC\leqslant q^2$, 
$$
 H(\bsP)\geqslant 1-\gamma-\tfrac13-q/3+(K-1/6)q^2-q^4/60 .
$$
The right-hand side decreases on $[0,q_0]$, since its derivative is at most
$-1/3+q/3<0$. Substituting $q=q_0$ and the rational bound $1-\gamma>2113/5000$ from Lemma~\ref{lem:mp-constants} gives a rational lower bound larger than $1/5000$.
\end{proof}

\begin{prop}[Enhanced low-purity region]\label{prop:mp-enhanced-purity}
Every finite table with $R+C\leqslant29/50$ has $H>1/20000$.
The bound extends to generalized tables with a non-strict positive
limiting margin.
\end{prop}

\begin{proof}
We use the marginal variance estimate from the proof of
Proposition~\ref{prop:mp-coarse-purity}, retaining the transform of the
distribution inside its Jensen remainder.

For a marginal $p$ of purity $u$ and size-biased variance $V_p$, the reciprocal identity in \eqref{eq:mp-60} gives \[
 \log(\Gamma_u(z)/L_p(z))\geqslant V_p
 \int_0^z\frac{v^2\,\dif v}{(1+uv)^2(1+\sqrt u\,v)},
 \qquad \Gamma_u(z)=(1+uz)^{-1/u}.
\] Multiplying by $L_p(z)$ and using $e^x-1\geqslant x$ bounds $\Gamma_u-L_p$ from below. The lower transform in Lemma~\ref{lem:mp-transforms} gives $L_p(z)\geqslant e^{-(1-\sqrt
u)z}/(1+\sqrt u\,z)$. Consequently \eqref{eq:mp-60} holds with the improved coefficient

\begin{equation}\label{eq:mp-69}
k(u)=\int_0^\infty \frac{z e^{-(1-\sqrt u)z}}{1+\sqrt u\,z}
      \int_0^1\frac{s^2\,\dif s}{(1+u s z)^2(1+\sqrt u\,s z)}\,\dif z.
\end{equation}

For a purity interval with endpoints $u_0,u_1$, choose rational bounds $a\leqslant\sqrt{u_0}$
and $b\geqslant\sqrt{u_1}$. The lower transform in \eqref{eq:mp-69} increases with $\sqrt u$,
so replace $\sqrt u$ there by $a$. In the remaining two denominators use $u_1,b$. The
resulting positive integral has a rational lower sum of the same type as the preceding
calculation. It uses the upper corners of rectangles in $s,z$ for all denominators and
for $e^{-(1-a)z}$, which are decreasing factors. It integrates $s^2,z$ exactly. The
lower exponential factor is computed by powering the cubic alternating-series bound at
$(1-a)/128$. The omitted $z>20$ integral is nonnegative.

Use $\lambda=1/[2(1+2q_0/3)]$, with $q_0=29/100$ and $p_{\max}\leqslant ab\leqslant\sqrt{RC}\leqslant q_0$
at the fixed-marginal optimizer. On a purity box let $k_0$ be a lower bound on
$\sqrt{k(R)k(C)}$ from \eqref{eq:mp-69}, clipped to $1/5$, and put
$K=2k_0-k_0^2/\lambda$. The identity from the proof of Proposition~\ref{prop:mp-coarse-purity}, with each marginal
coefficient reduced to its interval lower bound if needed, yields for $q=(R+C)/2$ and
$\xi=(R-C)/2$ \[
 H\geqslant 1-\gamma-\tfrac13-q/3+(K-1/6)q^2-q^4/60
                 +(\tfrac12-K-q^2/10)\xi^2-\xi^4/60 .
\] The first polynomial decreases in $q$, since $0\leqslant K<1/3$. The two $\xi$ terms are
at least $(1/2-K-7q_0^2/60)\xi^2$, with positive coefficient. Use the box's upper
endpoint for $q$ after intersecting with $q\leqslant q_0$, and its lower endpoint for
$|\xi|$. The integer verifier \nolinkurl{enhanced-purity-sum-certificate.py} checks all
boxes of side $1/256$ intersecting the triangle $R+C\leqslant29/50$, using
$1-\gamma>2113/5000$. Every box gives a strict rational bound exceeding $1/20000$.

There are exactly $5{,}625$ ordered outer grid boxes:
$0\leqslant i\leqslant j\leqslant149$ and $i+j\leqslant148$.
The marginal coefficients use $150$ purity intervals. The inner
quadrature uses $128$ intervals in $s\in[0,1]$ and $2560$ in
$z\in[0,20]$; these are integration cells, not outer boxes.
\end{proof}

\subsection{Concentrated perturbation bounds}

\begin{lem}[Concentrated tangent remainder]\label{lem:mp-concentrated-remainder}
Suppose $a\leqslant b\leqslant71/72$ and
$3/2\leqslant a+b\leqslant7/4$. 
Set $s=1-a$, $t=1-b$, $x_*=1-s-t$, and use the table $\bsP^{\star}$
supported in the first row and column with these marginals. At the fixed-marginal
maximizer $\bsP_*$, put $d=p^*_{11}-x_*$. This is precisely the mass of $\bsP_*$ outside the
first row and column. Equation \eqref{eq:mp-26} gives $0\leqslant d\leqslant st$. Define
$\bsp^{\mathrm{ref}}=(x_*,s,t)$, the spectrum formed from $\bsP^{\star}$ by separately grouping its
two strips. For respective upper bounds $u,v$ on the largest atoms of the two marginal
tails, put
$$
 G=\big[g_{\bsp^{\mathrm{ref}}}(u)+g_{\bsp^{\mathrm{ref}}}(v)  -g_{\bsp^{\mathrm{ref}}}(x_*)\big]_+.
$$
Choose $u=\min(a,s\sqrt\alpha)$ and $v=\min(b,t\sqrt\beta)$, where $\alpha,\beta$ are
the two normalized tail purities. In this region $u,v\leqslant x_*$.
The comparison is
\begin{equation}\label{eq:mp-45}
\Delta_{\sfQ}(\bsP_*)\geqslant\Delta_{\sfQ}(\bsP^{\star})- dG+\frac{d^2}{2(1+2ab)},
\end{equation}
\end{lem}

\begin{proof}
The coefficient of a moved amount in the tangent at $\bsP^{\star}$ is
$g_{\bsP^{\star}}(r_i)+g_{\bsP^{\star}}(c_j)-g_{\bsP^{\star}}(x_*)$, as in Proposition~\ref{prop:mp-large-heads}. It is at most $G$. To
prove the last bound, write its upper bound with $u,v$ as the integral of $L_{\bsP^{\star}}(z)x_*
k(z)/(1+x_*z)$, with 
$$
 k(z)=\frac{u(1+x_*z)}{x_*(1+uz)}+
       \frac{v(1+x_*z)}{x_*(1+vz)}-1.
$$
This is increasing in $z$ since $u,v\leqslant x_*$. As in \eqref{eq:mp-20}, the transform
ratio $L_{\bsP^{\star}}/L_{\bsp^{\mathrm{ref}}}$ is decreasing, by logarithmic differentiation and
separate refinement of the two strip atoms. The covariance inequality for a decreasing
function and an increasing function shows that the expectation of $k$ under the weight
for $\bsP^{\star}$, after normalizing by $g_{\bsP^{\star}}(x_*)$, is at most its expectation under the
corresponding reference weight. Also $g_{\bsP^{\star}}(x_*)\leqslant g_{\bsp^{\mathrm{ref}}}(x_*)$ by
refinement. Taking positive parts gives the stated bound.

Both $\bsP^{\star}$ and $\bsP_*$ have largest cell at most $ab$: for $\bsP^{\star}$ the largest cell is
$x_*$, and for $\bsP_*$ use \eqref{eq:mp-26}. The bound $\rD^2\cJ_{\bsP}[\bse,\bse]\geqslant \sum_i
e_i^2/(1+2p_i)$ from \eqref{eq:mp-62} proves \eqref{eq:mp-45}, since the squared norm of the direction is at least the square of its first cell
difference $d$. On dividing by $st$, the maximum possible loss is $G y-K y^2/2$ at
$y=\min(1,G/K)$, with $K=st/(1+2ab)$. If the estimated $G$ increases or the estimated
$K$ decreases, this maximized loss increases.
\end{proof}

\begin{lem}[Four-cell and tangent improvements]\label{lem:mp-four-cell}
Suppose $a\leqslant b\leqslant71/72$ and
$71/50\leqslant a+b\leqslant3/2$. For a fixed-marginal optimizer,
put $y=(p_{11}-a-b+1)/[(1-a)(1-b)]\in[0,1]$.
The curvature remainder divided by $(1-a)(1-b)$ is bounded below
as follows, and the tangent loss has the additional displayed
improvement.
\end{lem}

\begin{proof}
Set $s=1-a$, $t=1-b$ and let $\bsP^{\star}$ be the star table.
Write $x_*=1-s-t,\ x=p_{11}$ at the optimizer, $d=x-x_*$,
and $y=d/(st)\in[0,1]$. Put $\alpha=\sum_{i>1}r_i^2/s^2,\ \beta=\sum_{j>1}c_j^2/t^2$.
Let $U,V$ be upper bounds on $r_2,c_2$.

Let $w_i=\sum_{j>1}p_{ij}$ for $i>1$, and $w'_j=\sum_{i>1}p_{ij}$ for $j>1$. By \eqref{eq:mp-26}, \[
 w_i\leqslant (1-x/a)r_i=\frac{t-d}{a}\,r_i,
 \qquad
 w'_j\leqslant\frac{s-d}{b}\,c_j .
\] Since $r_2\geqslant s\alpha$ and $c_2\geqslant t\beta$, all but the second row account for at
most $st(1-\alpha)/a$ of the mass $d$ in the inner block. All but the second column
account for at most $st(1-\beta)/b$. Set $A=(1-\alpha)/a,\ B=(1-\beta)/b$. The changes
in the top two strip cells have magnitudes at least $st(y-A)_+$ and $st(y-B)_+$, and the
$(2,2)$ entry increases by at least $st(y-A-B)_+$. The $(1,1)$ entry increases by $st
y$. Bound the mean segment entries at those four locations by $x_*+st/3,\ U,\ V,\ st/3$,
respectively. Thus, after dividing by $st$, the quadratic remainder in the joint
comparison is at least

\begin{equation}\label{eq:mp-63}
\frac{st}{2}\left[
  \frac{y^2}{1+2x_*+2st/3}
 +\frac{(y-A)_+^2}{1+2U}+\frac{(y-B)_+^2}{1+2V}
 +\frac{(y-A-B)_+^2}{1+2st/3}\right].
\end{equation}

The supporting tangent in \eqref{eq:mp-45} can also be sharpened. Write $g=g_{P^{\star}}$.
Since $g'(U)\geqslant(1+2U)^{-1}$ by the same Jensen argument as in \eqref{eq:mp-62},
concavity gives $g(U)-g(r_i)\geqslant(U-r_i)/(1+2U)$ for $i>1$. Sum after weighting with $w_i$
and use $\sum_{i>1}w_i r_i\leqslant (t-d)s^2\alpha/a$. The corresponding column comparison
uses $\sum_{j>1}w'_j c_j\leqslant (s-d)t^2\beta/b$. These improve the normalized bound on the
loss by

\begin{equation}\label{eq:mp-64}
\frac{[Uy-s\alpha(1-sy)/a]_+}{1+2U}
 +\frac{[Vy-t\beta(1-ty)/b]_+}{1+2V}.
\end{equation}

The positive parts use the separate nonnegativity of the untruncated sums.
\end{proof}

\subsection{Marginals close to binary}\label{marginals-close-to-binary}

\begin{lem}[Binary-neighborhood estimate]\label{lem:mp-binary-neighborhood}
Fix two rows and two columns carrying the largest two marginal masses on each side, and
let $a,b$ be the largest marginal masses. Write $A,B$ for the two marginal masses
outside these rows and columns, $\zeta$ for the joint mass outside the four retained
cells, and $w$ for the joint mass outside both the retained rows and the retained
columns. Then $A+B=\zeta+w$, $AB\geqslant\zeta w$, and $0\leqslant w\leqslant\zeta$. Construct a binary table
$P^{\mathrm b}$ by redistributing each outside row uniformly into the two retained rows
and each outside column uniformly into the two retained columns. Let $x$ be the largest
joint atom of $P$, set $m=\max(a+A/2,b+B/2)$, and assume $m<1$. With $K=3/[4(1-m)]$, the
deficit satisfies

\begin{equation}\label{eq:mp-41}
H(\bsP)\geqslant H(\bsP^{\mathrm b})  +\zeta\left(\frac13-x-(K+\tfrac14)\zeta\right)  +w\left(\frac13-Kw-\frac\zeta3\right).
\end{equation}
\end{lem}

\begin{proof}
For the joint-entropy term, Lemma~\ref{lem:mp-calculus} bounds
every cell gradient by the largest atom. Use the
segment from $\bsP^{\mathrm b}$ to $\bsP$, with parameter $s\in[0,1]$. Each retained cell of
$\bsP^{\mathrm b}$ receives at most $\zeta/2$ from outside the four cells, since every
outside cell is split at least two ways. The largest cell of the segment table is at
most $x+(1-s)\zeta/2$. The retained cells lose total mass $\zeta$ along the segment and
the others gain it. The nonnegative gradient representative for $-\Q$ therefore gives 
$$
 \Q(\bsP^{\mathrm b})-\Q(\bsP)\geqslant-\zeta(x+\zeta/4).
$$

The row marginal of each segment table has all its atoms at most $a+A/2\leqslant m$, so $\int
L_r\leqslant1/(1-m)$ by the size-biased Jensen comparison $L_r(z)\leqslant(1+zR)^{-1/R}$, with $R\leqslant
m$. The row derivative of $\Q$ on the segment is at least
$A/2-s(A^2/2+\sum_{i>2}r_i^2)/(1-m)$: use the sum of the cell gradients equal to one,
subtract the gradients of the outside rows, and bound each of those by $sr_i/(1-m)$.
Integration bounds the row gain below by $A/2-KA^2$. The column gain is at least
$B/2-KB^2$.

The marginal linear entropy increases from the binary table by at most $A$ for rows and
$B$ for columns. For example the row increase is $A(1-A)+A^2/2-\sum_{i>2}r_i^2$, since
the retained entries of the binary marginal are increased by $A/2$ each. Each binary
linear entropy is at most $1/2$, so the product increases by at most $(A+B)/2+AB$.
Putting the bounds together and using \[
 K(A^2+B^2)+AB/3
   = K(\zeta+w)^2+(1/3-2K)AB
   \leqslant K(\zeta^2+w^2)+\zeta w/3
\] gives \eqref{eq:mp-41}. The proof for finite tables extends to generalized tables by
the diffuse approximation preserving the four retained cells. In \eqref{eq:mp-41}, the
binary biases are exactly the differences of the two largest row masses and the two
largest column masses, since redistribution adds equal amounts to the two retained
entries of each marginal. Thus \eqref{eq:mp-1}--\eqref{eq:mp-2} can supply an explicit
lower bound for $H(P^{\mathrm b})$ when a bracket in \eqref{eq:mp-41} is negative.
\end{proof}

\begin{lem}[Small marginal tails]\label{lem:mp-small-tail}
Let $A=\sum_{i>2}r_i$, $B=\sum_{j>2}c_j$, and
$\epsilon=1/25$. If $A+B\leqslant\epsilon$ and
$s=1-a,t=1-b\geqslant\epsilon$, the optimizer is bounded by
its collapsed binary table through the following three estimates.
\end{lem}

\begin{proof}
Let $A=\sum_{i>2}r_i$ and $B=\sum_{j>2}c_j$, with marginals sorted. Assume $A+B\leqslant\epsilon=1/25$ and $s=1-a,t=1-b\geqslant\epsilon$. At fixed marginals use the optimizer of \eqref{eq:mp-26}.
Collapse all but the first row into the second row, and all but the first column into
the second column. The resulting binary table $\bsP^{\mathrm c}$ has entries $x,a-x,b-x,d$,
where $d=1-a-b+x$. In the original table, only mass outside the first two rows or
columns moves. The destination in the lower-left cell is at most $A$, and the
destination in the upper-right cell is at most $B$. Write $g=g_{P^{\mathrm c}}$ as in
Lemma~\ref{lem:mp-calculus}. The supporting tangent of $\mathcal J$ at $P^{\mathrm c}$ gives

\begin{equation}\label{eq:mp-66}
\cJ(\bsP)-\cJ(\bsP^{\mathrm c}) \geqslant -A\,g(\max(b-x,d))-B\,g(\max(a-x,d)).
\end{equation}

Indeed, each cell originally outside the retained square has zero tangent weight there.
When a cell outside both its rows and columns moves to $(2,2)$, counting its mass twice
in the displayed upper loss can only enlarge that loss.

Set $s=1-a,\ t=1-b$ and $\delta_{\bsr}=s^2-\sum_{i>1}r_i^2$, $\delta_c=t^2-\sum_{j>1}c_j^2$.
In the determinant ratio for merging the row tail, expansion of its product gives

\begin{equation}\label{eq:mp-67}
\Q(\bsr)-\Q(a,s)\geqslant \frac{\delta_{\bsr}}{2}   \int_0^\infty \frac{L_{\bsr}(z)}{1+sz}\dif z.
\end{equation}

Indeed the merge increases the Laplace transform by at least
$L_{\bsr}(z)z^2\delta_r/[2(1+sz)]$. The same estimate applies to columns. For
$\epsilon=1/25$, define
$$
 I(a,s)=\int_0^\infty \frac{e^{-\epsilon z}}{(1+az)(1+(s-\epsilon)z)(1+sz)}\,\dif z .
$$
The integral in \eqref{eq:mp-67} is at least $I(a,s)$ when $s\geqslant\epsilon$. To see
this, retain the first atom and the second atom $r_2=s-A$, and replace the remaining
mass with diffuse mass, which decreases the transform. Then reduce $r_2$ to
$s-\epsilon$, again decreasing the transform.

The change in the marginal product is $2bt\,\delta_r+2as\,\delta_c+\delta_r\delta_c$.
The coefficient $I(a,s)/2-2bt/3$ of $\delta_r$ in the deficit comparison is positive.
Jensen gives $I(a,s)\geqslant1/(1+s)$, since it is the Laplace integral for a mean-$1+s$
variable. Also $2bt\leqslant1/2$. Use $2(s-\epsilon)A\leqslant\delta_r\leqslant2sA$, and the analogous
bounds for $\delta_c$. Since $4stAB\leqslant st\,\epsilon(A+B)$,
\eqref{eq:mp-66}--\eqref{eq:mp-67} give

\begin{equation}\label{eq:mp-68}
\begin{aligned}
 H(\bsP)&\geqslant H(\bsP^{\mathrm c})+A f_r+B f_c,\\
 f_r&=(s-\epsilon)(I(a,s)-4bt/3)
           -g(\max(b-x,d))-st\epsilon/3,\\
 f_c&=(t-\epsilon)(I(b,t)-4as/3)
           -g(\max(a-x,d))-st\epsilon/3 .
\end{aligned}
\end{equation}

Minimize the affine term under $A+B\leqslant\epsilon$, $A\geqslant(1-2a)_+$, and $B\geqslant(1-2b)_+$.
These lower bounds come from $r_2\leqslant a,\ c_2\leqslant b$. The minimum is the value at the two
lower bounds plus any remaining budget times $\min(0,f_r,f_c)$.
\end{proof}

\subsection{Pivot and anchor implementation}

\begin{lem}[Convex pivot implementation]\label{lem:mp-pivot}
At fixed marginal parameters, the block bounds admit convex
lower objectives on every feasible pivot subinterval. This remains
true after replacing the kernel by an anchor weight or a positive
reciprocal minorant. Supporting lines therefore give lower bounds
on the entire pivot interval.
\end{lem}

\begin{proof}
The joint Laplace lower bound can retain the pivot without adding another coordinate to
the outer subdivision. Let $x$ range over its feasible interval from \eqref{eq:mp-26}.
At fixed marginal parameters, write $d=x+1-a-b$ and $\delta=ab-x$. The first two lower
norms in \eqref{eq:mp-27} are squares of maxima of nonnegative affine pieces. For the
inner norm use $(x+u+v-1)_+^2$. On a subinterval away from zero one can also lower bound
$xuv/(ab)-\delta/x$ by its secant line, using its concavity. Taking the squared positive
part of the larger of these bounds gives a convex lower bound for $\|W\|^2$. Denote the
three resulting squared bounds by $\ell_V(x),\ell_U(x),\ell_W(x)$.

The cap bounds for $V,U$ from \eqref{eq:mp-28} are minima of affine functions on the
pivot interval. For $W$, using the minimum of $x,a_2,b_2,d$ suffices. A sharper bound on
a subinterval $[l,h]$ with $l>0$ uses the reverse rectangle inequality. Upper bound its
two strip entries by $(x a_2+\delta)/a$ and $(x b_2+\delta)/b$, multiply, and divide by
$x$. The resulting expression is affine in $x$ plus a positive multiple of $1/x$, so its
secant over $[l,h]$ is above it. Taking the minimum with the previous inner cap gives a
concave upper cap. All three chosen caps $m_J(x)$ are now concave and nonnegative on the
pivot interval.

Let $\underline{\pi}_P(x)$ be the maximum of the Cauchy bounds of \eqref{eq:mp-29} and
$x^2+\ell_V(x)+\ell_U(x)+\ell_W(x)$. At fixed $z$, $D_P(z)/z^2$ is bounded below both by
$\underline{\pi}_P(x)b(z,x)$ and by

\begin{equation}\label{eq:mp-55}
\frac{D_P(z)}{z^2}\geqslant
x^2b(z,x)+\sum_{J\in\{V,U,W\}}\ell_J(x) b(z,m_J(x)).
\end{equation}

Each bound is convex in $x$. Indeed, the positive powers in $\underline{\pi}_P(x)$ and the $\ell_J(x)$
are of the form $g(x)^2$ for a nonnegative convex $g$. The map $(g,m)\mapsto
g^2/(1+\lambda m)$ is convex on its domain, increases in $g\geqslant0$, and decreases in $m$.
Composition with the convex $g$ and concave $m$, then integration over the positive
reciprocal weights defining $b$, gives the claim. The same proof applies to an
anchored weight $I$ of \eqref{eq:mp-40} or to any positive sum of reciprocals below it.

For pivots in $[l,h]$ let $M_J$ be a constant upper bound for the chosen cap over that
interval. The correction in \eqref{eq:mp-30} can be bounded below by
$z^2\ell_J(x)c_J(z)$, where \[
 c_J(z)=\frac{z(l-M_J)_+}
       {3(1+3zM_J/4)(1+3zh/4)}.
\] Indeed $b(z,m_J)-b(z,x)$ is nonnegative and equals $z(x-m_J)\int_0^1 v^2/[(1+zv
m_J)(1+zv x)]\,\dif v$. When $l>M_J$ use the upper endpoints in both denominator factors and
Jensen with density $3v^2$, whose mean is $3/4$. This adds a convex lower correction to
$\underline{\pi}_P(x)b(z,x)$.

For the determinant check on a marginal-parameter box, restrict to the feasible interval
of $x=p_{11}$. The additional lower endpoints $\big((1-a)\alpha+a+b-1\big)/2$ and
$\big((1-b)\beta+a+b-1\big)/2$ come from the two strips. For the first, the second
largest row mass is at least $(1-a)\alpha$, since its tail's squared sum is at most its
largest atom times its mass. The first-column entry of that row is at least
$(1-a)\alpha-d$ and at most $x$. Insert $d=1-a-b+x$. The second endpoint follows by
exchanging rows and columns.

The upper caps for the strip and inner cells, and the necessary conditions used to
narrow the pivot interval, are those in \eqref{eq:mp-28} and \eqref{eq:mp-55}. In the
implementation, the lower bound on $d$ is the positive part of $x+1-a_{\rm high}-b_{\rm
high}$. Subtracting this lower bound from an upper bound on a strip mass gives a concave
upper cap on each strip's atoms. The numerator $xuv/(ab)-\delta/x$ in \eqref{eq:mp-27}
is concave as a function of positive $x$. An affine secant on the initial pivot interval
is a lower bound on that function. These choices maintain the convexity of the
supporting-line tests.

Use these reciprocal lower bounds in the joint block comparison of \eqref{eq:mp-55}, and
use \eqref{eq:mp-51} to combine the marginal entropy comparison with the target product.
The denominator parameters in each reciprocal remain concave caps, so all support lines
give joint lower bounds across the retained interval. The simpler scalar tests use
\eqref{eq:mp-29}, \eqref{eq:mp-33}, \eqref{eq:mp-41}, \eqref{eq:mp-45}, and
\eqref{eq:mp-51}. The acceptance and feasibility guards are those of
Remark~\ref{rem:mp-arithmetic}. Bisection of the convex pivot objective chooses trial points. The subsequent
endpoint evaluation and combination of supporting lines give the certified lower bound
on the entire interval.
\end{proof}

\subsection{Concentrated marginals}\label{concentrated-marginals}

\begin{prop}[Concentrated band]\label{prop:mp-concentrated}
For mixed marginals, $a+b\geqslant3/2$ implies $H>0$.
After the large-head and nearly pure reductions, the certificate
covers $a\leqslant b\leqslant71/72$ and
$3/2\leqslant a+b\leqslant7/4$, with full normalized tail purities
$\alpha,\beta\in[0,1]$.
\begin{equation}\label{eq:mp-44}
a+b\geqslant\frac32
 \quad\Longrightarrow\quad
 \Delta_{\sfQ}>\frac13 \sfS_L(\bsr)\sfS_L(\bsc).
\end{equation}
\end{prop}

\begin{proof}

It suffices to treat $a\leqslant b$, $b\leqslant71/72$, and $3/2\leqslant a+b\leqslant7/4$, by \eqref{eq:mp-20}
and \eqref{eq:mp-38}.

Use the notation and coefficient $G$ of
Lemma~\ref{lem:mp-concentrated-remainder}. Equation~\eqref{eq:mp-16}
bounds $\Delta_{\Q}(\bsP^{\star})/(st)$, now with the exact tail
transforms as in the determinant comparison of Lemma~\ref{lem:mp-canonical-tail}. The same
expansion into five positive terms applies, with $d_s(z)=f_{p_s}(sz)$ and
$d_t(z)=f_{p_t}(tz)$, where $f$ is defined in \eqref{eq:mp-39} and $p_s,p_t$ are the two
normalized tail determinant extremizers. The grouping monotonicity preceding
\eqref{eq:mp-16} shows the direction of this replacement. In particular $d_s$ decreases
with $s$ and with $\alpha$, by \eqref{eq:mp-39} and pointwise ordering of the
determinant extremizers.

The first term is completely monotone. The other four are $z^2$
times completely monotone functions, so Lemma~\ref{lem:mp-gauss}
gives their lower quadrature after $w=z^3/3$.

Use $[0,1/8]$
and successive factor-two intervals in $z$ ending at 8192. For the upper bounds in $G$,
each of the two positive integrals has remaining tail at most $1/(3x_*st\,8192^3)$,
since $u/(1+zu)\leqslant1/z$. For all lower bounds the omitted integral tails are nonnegative.

The initial box in $(s,t,\alpha,\beta)$ is $[1/128,1/2]\times[1/128,1/4]\times[0,1]^2$.
All remaining parameters lie in it because $s\geqslant t$, $s+t\leqslant1/2$, and $t\geqslant1/72$. It
omits boxes disjoint from those conditions and from $s+t\geqslant1/4$. It tries the binary
bound \eqref{eq:mp-1}--\eqref{eq:mp-2} for the partition isolating the largest marginal
atoms first. Dividing this bound by $st$ multiplies the lower binary ratio by $4ab$.
Otherwise it compares the lower \eqref{eq:mp-16} minus the loss in \eqref{eq:mp-45} to
the normalized target \eqref{eq:mp-17}. It uses upper endpoints in the positive
denominators and lower endpoints for the positive numerators. Each $d_s,d_t$ is obtained
at the upper endpoints of its parameters, and the purity can be rounded upward to a
multiple of $2^{-14}$ by monotonicity.

The check accepts 5,281,920 boxes and omits 173,374 disjoint boxes, with maximum
accepted-leaf depth 34.

The node, transform, and arithmetic enclosures are those of
Remark~\ref{rem:mp-arithmetic}. The strict acceptance margin is $10^{-8}$.
\end{proof}

\begin{prop}[Extended concentrated band]\label{prop:mp-extended}
For mixed marginals, $a+b\geqslant71/50$ implies $H>0$.
The additional certificate covers $a\leqslant b\leqslant71/72$,
$71/50\leqslant a+b\leqslant3/2$, and full normalized tail purities
$\alpha,\beta\in[0,1]$.
\end{prop}

\begin{proof}
Use the notation of Lemma~\ref{lem:mp-four-cell}, with
$g=g_{\bsP^{\star}}$. For the tangent coefficient use an upper enclosure of $[g(U)+g(V)-g(x_*)]_+$. For each
tail, its canonical spectrum from Lemma~\ref{lem:mp-canonical-tail} gives an upper transform. Lemma~\ref{lem:mp-transforms} gives the normalized lower tail
transform $e^{-(1-\sqrt h)z}/(1+\sqrt h\,z)$ at any purity lower
bound $h$. Multiply the two tail enclosures after scaling their arguments by $s,t$, and retain the
intersection factor $(1+x_*z)^{-1}$. Use the upper rule from Lemma~\ref{lem:mp-gauss} for
$g(U),g(V)$ and the lower rule for $g(x_*)$. The lower rule is applied to the lower
enclosing transform, which is also completely monotone. The upper rule integrates the
finite canonical transform, and the three grouped reference factors in Lemma~\ref{lem:mp-concentrated-remainder}
bound its omitted tail. The exponential for the lower enclosure is bounded by an
alternating series through degree 19 at a nonnegative argument at most one, followed by
five squarings. Beyond argument 32 the code uses zero as a lower bound.

The sum of \eqref{eq:mp-63} and \eqref{eq:mp-64} is convex in $y$. On a marginal box,
use lower curvature coefficients, upper values of $A,B$, and the lower affine numerators
in \eqref{eq:mp-64}. A supporting line bounds its convex value from below on $[0,1]$.
Subtracting that line from the linear tangent upper bound gives a valid upper bound for
the loss throughout $[0,1]$. The same normalized star gap comparison
\eqref{eq:mp-16}--\eqref{eq:mp-17} as in \eqref{eq:mp-44} is then used against the
target, after subtracting this loss. If this check fails on a box at depth at least 34,
the bounds of Lemma~\ref{lem:mp-pivot} are applied at coefficient $1/3$ on that same
box.

The verifier \nolinkurl{extended-concentrated-band-certificate.cpp} covers the band with
the two tests. It uses the binary64 arithmetic specified in
Remark~\ref{rem:mp-arithmetic}, and has maximum accepted-leaf depth 48
with 53,113,510 accepted boxes and
1,653,894 boxes excluded by the analytic domain tests. The quadrature guards are those of Remark~\ref{rem:mp-arithmetic}. In \eqref{eq:mp-63}--\eqref{eq:mp-64} the divisions are by
terms bounded below by one except for $a,b$, both at least $21/50$ on the band. The
lower and upper coefficient evaluations have absolute errors smaller than $10^{-11}$;
that amount is added to the loss upper bound and every acceptance uses another $10^{-8}$
of slack. The concentrated band \eqref{eq:mp-44} and the pure-boundary
estimate \eqref{eq:mp-38} complete the case.

The enclosing root in $(s,t,\alpha,\beta)$ is
$[1/128,21/32]\times[1/128,21/64]\times[0,1]^2$.
It retains $s\geqslant t\geqslant1/72$ and
$1/2\leqslant s+t\leqslant29/50$.
The root upper endpoints exceed $29/50-1/72$ and $29/100$,
respectively, so they enclose the required band. The subdivision selects
a coordinate of maximal weighted width, with respective weights
$1,1,1/4,1/4$ and ties resolved in coordinate order. The initial weighted
widths are $(83/128,41/128,1/4,1/4)$. At each subdivision, the selected
width is halved. Consequently the sequence of selected coordinates
depends only on depth.

The first two subdivisions halve the first coordinate, and the third
halves the second. At depth three, the widths are
$(83/512,41/256,1/4,1/4)$. The next four selections, in order, are
coordinates $3,4,1,2$, after which all four widths have been halved.
This four-step cycle repeats. Thus at the maximum permitted depth
$54=3+12\cdot4+3$ the respective subdivision counts are $(15,13,13,13)$.
Every endpoint of a processed box, including excluded boxes, is
therefore a dyadic rational with denominator dividing $2^{22}$.
Its bisection sums, the sums $s_0+t_0$, $s_1+t_1$, and the differences
used to form $1-s_i$, $1-t_i$ are exact in binary64 since all such
numbers have magnitude at most two. The comparisons with $1/2$ and
with the other dyadic endpoints are exact. For the rational cutoffs
$29/50$ and $1/72$, the exclusion tests use outward offsets of
$10^{-12}$, larger than the absolute rounding errors in forming
these cutoffs (less than $10^{-15}$). Hence every omitted box is
disjoint from the required band. The source
\nolinkurl{pivot-certificate.cpp} supplies the supporting-line check
from Lemma~\ref{lem:mp-pivot}.
\end{proof}

\subsection{Small marginal-tail certificate}

\begin{prop}[Certified small marginal tails]\label{prop:mp-small-tails}
If $A=\sum_{i>2}r_i$, $B=\sum_{j>2}c_j$, and $A+B\leqslant1/25$,
then $H\geqslant0$, with equality for mixed marginals only at the
uniform binary table. Outside the concentrated region, the
certificate covers $a\leqslant b$, $a,b\geqslant12/25$,
$a+b\leqslant71/50$, and $x\in[(a+b-1)_+,ab]$.
\end{prop}

\begin{proof}
Put $s=1-a$, $t=1-b$, $\epsilon=1/25$, and use
Lemma~\ref{lem:mp-small-tail}. Proposition~\ref{prop:mp-extended}
covers $a+b\geqslant71/50$. In the remaining region, $a,b\geqslant12/25$ and
$s,t\geqslant3/50>\epsilon$. A three-dimensional covering calculation in $a,b,x$ uses
\eqref{eq:mp-68}, and checks \eqref{eq:mp-41} first near the binary uniform point.

Use the signed biases $\alpha_0=2a-1$, $\beta_0=2b-1$ of the
collapsed binary table, and put $\delta=ab-x\geqslant0$. The actual
correlation coordinate is $1-2a-2b+4x$, so its displacement from the
product correlation $\alpha_0\beta_0$ is $-4\delta$. The product-point
value and derivative in Section~\ref{the-binary-case} apply in
signed coordinates. A single swap negates both the correlation and
one bias, which accounts for the sign of the derivative. At the
product point the value is $2abst\,T$ and the derivative in
correlation is $\alpha_0\beta_0 J/4$. The curvature lower bound
$H''\geqslant1/5$ gives
$$
 H(\bsP^{\mathrm c})
 \geqslant 2abst\,T-\alpha_0\beta_0\delta J+\tfrac85\delta^2 .
$$
Write $T_4,J_4$ for the sums in Section~\ref{the-binary-case}
truncated at $j=4$, with $h_j$ evaluated on $\alpha_0^2,\beta_0^2$.
The coefficient discarded at each higher $j$ is
$$
 2abst\,w_j-\alpha_0\beta_0\delta(w_j-w_{j+1}).
$$
If the biases have opposite signs or either is zero, this is
nonnegative. If both biases are positive, then
$\alpha_0\beta_0\leqslant ab$ and $\delta\leqslant st$.
If both are negative, their product is at most $st$ and
$\delta\leqslant ab$. In both cases
$\alpha_0\beta_0\delta\leqslant abst$, giving the same sign.
Thus the following lower bound holds on both sides of either
half-mass boundary:
\begin{equation}\label{eq:mp-tail-binary-signed}
 H(\bsP^{\mathrm c})\geqslant
 2abst\,T_4-\alpha_0\beta_0\delta J_4+\tfrac85\delta^2.
\end{equation}

On a parameter box the program uses the lower constant $L$ and
the upper slope $U$ given by
\[
 L=2a_{\rm low}b_{\rm low}s_{\rm low}t_{\rm low}(T_4)_{\rm low},
 \qquad
 U=(|\alpha_0|)_{\rm high}(|\beta_0|)_{\rm high}(J_4)_{\rm high}.
\]
Both truncated sums have nonnegative coefficients in the squared
biases. For a pivot interval $[x_{\rm low},x_{\rm high}]$ the
deficit is in the interval
\[
 [\,\max(0,a_{\rm low}b_{\rm low}-x_{\rm high}),
       \max(0,a_{\rm high}b_{\rm high}-x_{\rm low})\,].
\]
The quadratic $L-U\delta+(8/5)\delta^2$ has its minimum at
$5U/16$ projected onto that interval.
The code takes the maximum of that lower bound and zero, using the
already established nonnegativity of every binary table.

The upper bounds for $g$ insert lower
endpoints for each of the four cell factors and an upper endpoint for its argument.
Positive Gauss--Lobatto quadrature from Lemma~\ref{lem:mp-gauss} bounds the integral above. Two
positive lower cell sizes $v,w$ give the tail bound $1/(2vw\,8192^2)$. The lower bound
$I$ uses Gauss--Legendre quadrature, with the lower exponential as in Proposition~\ref{prop:mp-extended}. All comparisons use binary64, disabling contraction, with the same outward
quadrature margins and an absolute acceptance margin $10^{-8}$.

The verifier \nolinkurl{small-marginal-tails-certificate.cpp} covers the remaining band
with 11,032 accepted boxes, 846 excluded boxes and maximum accepted-leaf depth 22.

The enclosing root is
$[15/32,23/32]\times[15/32,61/64]\times[0,129/250]$ in $(a,b,x)$.
Indeed feasibility and $a\leqslant b$, $a+b\leqslant71/50$,
$a,b\geqslant12/25$ imply
$a\leqslant71/100$, $b\leqslant47/50$, and
$x\leqslant ab\leqslant(71/100)^2$.
The program omits boxes disjoint from these conditions or the
pivot interval $[(a+b-1)_+,ab]$.
\end{proof}

\subsection{Largest marginal masses at most one half}\label{largest-marginal-masses-at-most-one-half}

\begin{prop}[Half-cap certificate]\label{prop:mp-half-cap}
If $a,b\leqslant1/2$, then $H\geqslant0$, with equality for
mixed marginals only at the uniform binary table. The certificate
covers the residual region after the small-head, low-purity,
small-tail, and low-cap exclusions.
\end{prop}

\begin{proof}
Suppose both largest marginal atom masses satisfy $a,b\leqslant1/2$. Then $H\geqslant0$. Use the
earlier exact exclusions for the small marginal atoms in \eqref{eq:mp-48}, the purity
sum in \eqref{eq:mp-69}, the small marginal tails of \eqref{eq:mp-68}, and
\eqref{eq:mp-46}. For the remaining tables the covering uses the two largest atoms and
the purity of each marginal.

For the row head $a$, let $\alpha=(R-a^2)/(1-a)^2$ and $t=r_2/(1-a)$. The possible interval for $t$ is \[
 [t_0(\alpha),\ \min\{\sqrt\alpha,a/(1-a)\}],
\] where $t_0(\alpha)$ is the largest atom of the canonical tail of purity $\alpha$ from Lemma~\ref{lem:mp-canonical-tail}. For the lower endpoint, use the capped purity bound \eqref{eq:mp-35}.
The two upper bounds come from the largest tail atom's squared mass and the ordering of
the full marginal. Parametrize $t$ as the linear interpolation between these endpoints,
with an independent interpolation parameter in $[0,1]$. Use the same parametrization for
columns. The system of upper and lower second-atom bounds in
\eqref{eq:mp-27}--\eqref{eq:mp-30} then applies on each box. For the normalized purity
of the part beyond the first two atoms use $(\alpha-t^2)/(1-t)^2$. This increases in
$\alpha$, and at fixed $\alpha$ its maximum on $t\geqslant t_{\rm low}$ is at
$t=\max(\alpha,t_{\rm low})$ or an endpoint. The resulting upper bound with $\alpha_{\rm
high}$ is used for the canonical entropy envelope.

A sharper marginal lower bound comes from retaining the dependence on $a$. For fixed
$a,\alpha$ and a lower bound on $t$, replace the tail by the canonical one with the
smallest feasible second atom at or above that lower bound. Equation \eqref{eq:mp-54}
minimizes $\Q$ over that interval, and the replacement leaves $R$ fixed. Then increase
$\alpha$ to its upper endpoint. While the retained atom stays fixed, the canonical
remainder at the new purity majorizes its earlier value. When the retained-atom cap
ceases to be feasible, the comparison continues with the unconstrained canonical tail.
The two choices agree at the cap boundary and the latter family is also increasing in
majorization as purity increases. Thus the replacement bounds the marginal expression
below by \eqref{eq:mp-51}. The replacement tail is normalized before scaling it by
$1-a$, and its shape can be held fixed as $a$ varies on the box.

If the opposite marginal purity is at least $1/10$,
\eqref{eq:mp-73} shows that $\Q(\bsr)-\S_L(\bsr)\S_L(\bsc)/3$ is concave along the affine variation
of $a$ with fixed normalized tail. Indeed its quadratic correction has curvature at most
$(3/5)\norm{\bse}^2<(6/\pi^2)\norm{\bse}^2$. Apply this to both marginals when their box-wide lower
purities are at least $1/10$. The lower bound on the two marginal terms together is the
minimum at the four pairs of head endpoints. The check also uses the single-head and
two-head majorization bounds in \eqref{eq:mp-51} without this condition.

For the joint comparison, the determinant and pivot checks above apply with the two
largest marginal masses retained and coefficient $1/3$. On a pivot subinterval an
additional check evaluates \eqref{eq:mp-30} directly against the same anchor: lower
bound each block norm on the interval, upper bound its atom cap, and lower bound $\underline{\pi}_P$ in \eqref{eq:mp-29}. Use the lookup bounds for $I$ from \eqref{eq:mp-40}. The lower
estimate \[
 \underline{\pi}_P I(x_{\rm high})
 +\sum_J \ell_J\bigl(I(m_{J,\rm high})-I(x_{\rm high})\bigr)_+
\] is used only with upper block caps at most $x_{\rm high}$. It follows from
\eqref{eq:mp-30} since $I$ decreases. The computation uses the positive part of a lower
enclosure of each difference.

The anchor constants use Lemma~\ref{lem:mp-constants}; the quadrature
and support-line margins are those of Remark~\ref{rem:mp-arithmetic}
and Lemma~\ref{lem:mp-pivot}. The verification uses only anchor
bounds for the joint determinant.

With binary64 arithmetic and contraction disabled, the verifier
\nolinkurl{half-cap-certificate.cpp} covers $a,b\leqslant1/2$ with 7,122,331 accepted boxes,
355,283 excluded boxes, and maximum accepted-leaf depth 38.

The enclosing root has heads in $[1/32,1/2]^2$ and four
coordinates in $[0,1]$: the two full normalized tail purities and
the two interpolation parameters for the second atoms. The
small-head reduction overlaps its lower boundary. A zero remainder
has normalized purity zero. The depth limit for this check is 45.
\end{proof}

\subsection{The six-parameter verification}\label{the-six-parameter-verification}

\begin{prop}[Unrestricted residual certificate]\label{prop:mp-residual}
After the preceding exclusions, the six-coordinate certificate
covers heads $a,b\in[1/32,511/512]$ with $a\leqslant b$ and
$u,v,\alpha,\beta\in[0,1]$, subject to the necessary order and
purity constraints below. It excludes every nonpositive global
minimum with mixed marginals outside the uniform binary case.
\end{prop}

\begin{proof}
Let $a,b$ be the first marginal masses. Set $u=r_2/(1-a)$ and $v=c_2/(1-b)$, and let
$\alpha,\beta$ be the normalized purities of the remainders after the first two masses.
A zero remainder is assigned normalized purity zero. The two total purities are \[
 R=a^2+(1-a)^2[u^2+(1-u)^2\alpha],\qquad
 C=b^2+(1-b)^2[v^2+(1-v)^2\beta].
\] For the row remainder, every atom normalized by its total mass is at most
$\min(1,\sqrt\alpha,u/(1-u))$, omitting the quotient when $u=1$
because that remainder is absent. Use the same bound on the column side. The order
constraints $(1-a)u\leqslant a$ and $(1-u)\alpha\leqslant u$ follow from the largest and second
largest atom comparisons, respectively.

\subsubsection{Marginal endpoint bounds}\label{marginal-endpoint-bounds}

Lemma~\ref{lem:mp-canonical-tail}, in particular
\eqref{eq:mp-79}, supplies branch concavity in each remainder purity.
When the normalized purity range includes
zero or many reciprocal breakpoints, one may lower only the entropy term by using the
canonical remainder at the upper endpoint throughout. Retain the actual purity in the
product of linear entropies. This lower bound is affine in $\alpha$ with the other
variables fixed.

For fixed normalized remainder shape, the canonical spectrum is affine in each of $a,u$
separately. The squared norm of its velocity in either coordinate is at most two. The
Hessian estimate \eqref{eq:mp-73}, with $6/\pi^2>3/5$, shows that the functional $\Phi(\bsr,\bsc)$ has second derivative at most $2/15$ in either coordinate separately. The bound is
still valid if a canonical tail of larger purity is used in $\Q$, as its velocity has
squared norm at least as large as the velocity in the purity correction. If the opposite
purity is at least $1/10$, the same derivative bound is zero. These conclusions also
hold for the column coordinates. They hold for a diffuse remainder by finite
approximation at fixed coordinates.

On an interval of length $d$, a function with second derivative at most $2/15$ is at
least its linear interpolant minus $d^2/60$. This follows by subtracting the quadratic
of second derivative $2/15$ and applying concavity to the difference. Therefore the
minimum of $\Phi$ over a six-parameter box is bounded below by the minimum of its vertex
values minus the sum of these errors for $a,b,u,v$. For $\alpha,\beta$, include the
reciprocal purities on branches where the entropy uses the actual purity, and use only
the standard endpoints when a fixed upper purity is used for the entropy. In the
comparisons below, additional terms are affine in each separate variable. They may be
included in the same vertex test, leaving the interpolation error as above.

\subsubsection{A joint loss bound with four reference atoms}\label{a-joint-loss-bound-with-four-reference-atoms}

Fix a reference table $\bsq$ on the same row
and column labels. Set $m=3/5$. The gradient representative for $\cJ=1-\gamma-\Q$
is $g_{\bsq}$, and \eqref{eq:mp-73} gives 
$$
 \cJ(\bsP)\geqslant\cJ(\bsq)+  \sum_{ij}\left[g_{\bsq}(q_{ij})(p_{ij}-q_{ij}) +\frac m2(p_{ij}-q_{ij})^2\right].
$$
For an ordinary reference table and arbitrary row and column potentials $\lambda,\mu$, put $v_{ij}=\lambda_i+\mu_j$ and
$$
 \Psi_m(s;x)= \begin{cases} s^2/(2m),& s\geqslant -mx,\\ -sx-mx^2/2,&s<-mx. \end{cases}
$$
This function is the supremum of $sy-my^2/2$ over $y\geqslant-x$. Apply it to
$y=p_{ij}-q_{ij}$ and $s=v_{ij}-g_q(q_{ij})$. The resulting bound is

\begin{equation}\label{eq:mp-78}
H(\bsP)\geqslant
 \Phi(\bsr,\bsc)+\lambda\cdot \bsr+\mu\cdot \bsc-\Q(\bsq) -\sum_{ij}\left[v_{ij}q_{ij} +\Psi_m(v_{ij}-g_{\bsq}(q_{ij});q_{ij})\right].
\end{equation}

The ordinary reference table and potentials may be chosen freely.
The same derivation allows a positive cell-dependent curvature,
using its corresponding conjugate in each cell.

Take a reference table $q$ with nonzero ordinary entries only in the first two rows and
columns, and put any remaining probability into a diffuse term. Use \eqref{eq:mp-62}
along the segment from this reference to the joint table. The compensated identity in Lemma~\ref{lem:mp-hessian} applies;
its nonnegative square is discarded. If
a finite entry of the table is at most $d_{ij}$, its diagonal coefficient after
integration is bounded below by

\begin{equation}\label{eq:mp-80}
m_{ij}=\frac1{1+(4/3)q_{ij}+(2/3)d_{ij}}.
\end{equation}

This is Corollary~\ref{cor:mp-segment} with $p^0=q$ and
$p^1_{ij}\leqslant d_{ij}$. The first derivative is
$\sum_{ij}g_q(q_{ij})(p_{ij}-q_{ij})$, with no ordinary gradient on the diffuse
component. For the remainder blocks use upper bounds on $d_{ij}$ that depend only on
whether each index is one, two, or in the remainder.

Choose $h\geqslant0$ and three row and three column potentials $\lambda_i,\mu_j\in[0,h]$. Use
the first two row potentials on their actual labels. For any other finite row of mass
$r_i$ use $\lambda_3 r_i/[(1-a)(1-u)U]$, where $U>0$ is a normalized remainder cap valid
on the whole box. The analogous column cap is $V>0$. Assign zero potential to diffuse
marginal labels. In each cell use the sum of these potentials minus $h$. The slope on a
diffuse joint cell is nonpositive because one index is diffuse and the opposite
potential is at most $h$. Hence the finite cell linear term alone bounds the full linear
term below. The full linear term is \[
 a\lambda_1+(1-a)u\lambda_2+(1-a)(1-u)\alpha\lambda_3/U
 +b\mu_1+(1-b)v\mu_2+(1-b)(1-v)\beta\mu_3/V-h .
\] Apply \eqref{eq:mp-78} with $m$ replaced cellwise by $m_{ij}$ from \eqref{eq:mp-80} and these cell slopes. Off
the four reference cells, $\Psi_{m_{ij}}(s;0)=(s_+)^2/(2m_{ij})$. The slope estimate from
\eqref{eq:mp-65} bounds its sum over each remainder block by the corresponding group
penalty multiplied by $\alpha/U^2$ for a row remainder, $\beta/V^2$ for a column
remainder, and their product if both indices belong to remainders. These factors are the
sums of squares of the individual potential multipliers.
For each application of the vertex test, the reference $q$, the
three row and column potential parameters, $h$, all upper atom caps,
and the resulting curvature coefficients are fixed across the box.
The resulting expression after subtracting $\Phi$ is affine in each
coordinate separately. The vertex comparison therefore bounds the
same function throughout the box.

\subsubsection{Uniform references and largest-entry bounds}\label{uniform-references-and-largest-entry-bounds}

For a uniform anchor $\bsu_n$, put $z_0=1/n$ and take the
reciprocal minorant of $I$ from Lemma~\ref{lem:mp-anchor}.
Round its numerator downward and its constant and linear
denominator coefficients upward to obtain $W(y)\leqslant I(y)$.
Equation \eqref{eq:mp-40} now gives 
$$
 H(\bsP)\geqslant \Phi(\bsr,\bsc)-\Q(\bsu_n)-I(z_0)/n  +\sum_{ij}p_{ij}^2 W(p_{ij}).
$$
The verifier uses this bound twice. First it applies Lemma~\ref{lem:mp-group-penalty} to $y^2W(y)$
with the three marginal groups above and the same vertex comparison. Its scalar
conjugate is zero for nonpositive slopes. For $W(y)=M/(d+ey)$ and $0<s<M/e$, the
conjugate is $d s^2/[M(1+\sqrt{1-es/M})^2]$. This follows by solving the stationary
equation over $y\geqslant0$.

The second test retains the largest joint entry $x$. It subdivides $[(a+b-1)_+,ab]$
using outer box bounds. On each interval use \eqref{eq:mp-27}--\eqref{eq:mp-29} and
\eqref{eq:mp-50} for lower purities and upper atom caps in the three blocks. If $\underline{\pi}_{\bsP}$ is
a joint purity lower bound, $h_x$ is the upper endpoint for $x$, and $m_J\leqslant h_x$ and
$\ell_J$ bound a block's cap and purity, the sum with $W$ is at least 
$$
 \underline{\pi}_{\bsP} W(h_x)+\sum_J \ell_J [W(m_J)-W(h_x)].
$$
This follows by monotonicity of $W$, partitioning the joint squared sum into the
blocks and the pivot. The pivot test also uses the power-series lower bound
\eqref{eq:mp-33} and checks the necessary inequalities $\pi_{\bsP}\leqslant RC$ and
$\pi_{\bsP}+\frac23[(a-R)C+(b-C)R]\leqslant x$. The latter follows from \eqref{eq:mp-74} and
\eqref{eq:mp-13}.

\subsubsection{Domain, enclosures and coverage}\label{domain-enclosures-and-coverage}

The domain tests use the established bounds \eqref{eq:mp-38}, \eqref{eq:mp-48},
\eqref{eq:mp-68}, \eqref{eq:mp-69}, the sum of largest atoms at least $71/50$, and the
half-cap comparison. For the tangent test use \eqref{eq:mp-8} with a majorant of the
opposite marginal as in \eqref{eq:mp-51}. At a stationary table, if $B_2<A_1-A_2$, then
$A_i+B_j<h$ for every $i,j\geqslant2$ since $h\geqslant A_1$. The inner cells vanish, and
\eqref{eq:mp-15} excludes the table. The verifier tests this and its transpose using
\eqref{eq:mp-13} for the lower differences and $I_r\leqslant1/(1-R)$, $I_c\leqslant1/(1-C)$ for the
upper potentials.

When $a+b>1$, one further comparison uses the table supported on the first row and
column with intersection $x_*=a+b-1$. Apply \eqref{eq:mp-16} with the canonical tail
discrepancies $d_s,d_t$ in place of the polynomial discrepancies. For terms carrying
$z^2$, the integration lower bound on each subinterval is its $z^2\,\dif z$ mass times the
integrand without $z^2$ at the corresponding mean. All such integrands are completely
monotone by \eqref{eq:mp-39}. Use the three reciprocal factors with masses $x_*,1-a,1-b$
for the upper tail of the tangent integral. The possible gain over the first-row and
first-column table is bounded by the largest value of $G y-M y^2/2$ over $0\leqslant y\leqslant1$,
where $y=(x-x_*)/[(1-a)(1-b)]$, $G$ bounds the tangent gain per unit moved mass, and
$M=(1-a)(1-b)/(1+2\max(ab,r_2,c_2))$ is a lower curvature coefficient from
\eqref{eq:mp-62}. Divide the full comparison by $(1-a)(1-b)$ as in \eqref{eq:mp-17}.

All acceptance tests in the unrestricted verifier
\nolinkurl{marginal-product-certificate.cpp} use outward intervals. A heuristic search
may propose a reference or potentials, but their values need only satisfy the sign and
normalization constraints of the bound being checked. The four-point quadrature and tail
enclosures are those above. Exponentials are bounded by the alternating series on
$[0,1]$ after division of the argument by 64, followed by six squarings. At arguments at
least 64 the upper bound $2^{-64}$ suffices. The logarithms of uniform reference sizes
use the positive series in $(n-1)/(n+1)$ with its geometric remainder bound. The $\pi$ and $1-\gamma$ enclosures are supplied by
Lemma~\ref{lem:mp-constants}.

The initial box uses $[1/32,511/512]$ in the two largest masses and $[0,1]$ in the four
other coordinates. The outer cutoffs follow from \eqref{eq:mp-38} and \eqref{eq:mp-48}.
Transposition permits $a\leqslant b$. Thus the covering can run on the 36 products of head
intervals $[i/8,(i+1)/8]$ and $[j/8,(j+1)/8]$ with $0\leqslant i\leqslant j\leqslant7$. Subdividing at a
common endpoint preserves the whole box. The omitted pivot and marginal regions obey the
inequalities stated above.

The terminal record \nolinkurl{marginal-product-terminal-counts.tsv}
specifies the root boxes and their subdivision counts. The four
subrectangles of $(i,j)=(4,5)$ cover that rectangle. For $(5,5)$,
the ten subrectangles on and above its diagonal cover the ordered half.
The 48 checks terminate with 10,781,823 accepted boxes and 299,876 boxes
omitted by the domain tests. The maximum accepted-leaf depth is 34.
The terminal record also reports the number of successful four-atom
vertex tests for each root. The supplied verification directory
identifies the complete implementation and its file digests.

All omitted boxes have a specified destination. The test
$a_{\rm low}>b_{\rm high}$ discards the unordered half by
transposition. A head upper bound at most $1/20$ invokes
Proposition~\ref{prop:mp-small-head}; a head sum lower bound at
least $71/50$ invokes Proposition~\ref{prop:mp-extended}, with
the large-head and nearly pure cases already included.
Two head upper bounds at most $1/2$ invoke
Proposition~\ref{prop:mp-half-cap}; $b_{\rm low}\geqslant71/72$
invokes Proposition~\ref{prop:mp-near-pure}. Failure of the
necessary order or purity inequalities is infeasibility.
An upper bound on $(1-a)(1-u)+(1-b)(1-v)$ at most $1/25$
invokes Proposition~\ref{prop:mp-small-tails}. Low-purity and
stationarity-star tests count as acceptances rather than omissions.

\end{proof}

\subsection{Coverage and equality}\label{sec:mp-coverage}

The preceding propositions cover the full parameter space, including
its boundary seams. The unrestricted root has strict analytic overlap
at both head cutoffs. The lower overlap is $1/32<1/20$; the upper is
$511/512>71/72$.
Its $36$ ordered head-cell pairs are tiled by the $48$ terminal
records: $(4,5)$ is replaced by its four $2\times2$ subcells and
$(5,5)$ by the ten on-and-above-diagonal subcells of a $4\times4$
grid. A diagonal subcell remains a full closed square; only
$a_{\rm low}>b_{\rm high}$ is omitted. Thus the diagonal is retained.
All tests use closed domains, including
$a=1/20$, $b=71/72$, $a+b=71/50$, $a=b=1/2$,
$A+B=1/25$, and $R+C=29/50$.
Outward guards retain additional boxes at rational cutoffs.
Each bisection preserves its closed parent box by taking the
midpoint as the upper boundary of one child and the lower boundary
of the other. Relative to any one root box, a child at depth $j$
has dyadic weight $2^{-j}$. The two children have combined weight
$2^{-j}$ when their parent is at depth $j$. Thus substitution of the
children for their parent preserves total weight. When the pending
list is empty, every leaf has either an acceptance witness or an
explicit destination in the preceding domain tests. Both the
dyadic weight check and the analytic justification of each test
are used in the coverage argument. An undecided box at the depth
limit causes failure.

\begin{proof}[Completion of Proposition~\ref{prop:mp-comparison}]
We now classify the nonpositive minima. The
small-head, nearly pure, low-purity and concentrated regions use
strict lower bounds when both marginals are mixed. For the small-tail
region the direct test in \eqref{eq:mp-41} has two positive brackets.
A positive $\zeta$ then gives $H>0$. If $\zeta=0$, the table is binary
and Section~\ref{the-binary-case} leaves only the uniform table.
The other small-tail tests give strict bounds or infeasibility.

In the half-cap and unrestricted computations, boxes omitted by
an order or purity constraint are infeasible. The other omissions
are covered by the preceding analytic or verified regions. An
accepted box uses a positive lower bound, including the strict
vertex and pivot tests, or a stationarity condition forcing support
on the first row and column. In the latter case the strict bound
\eqref{eq:mp-15} applies. The half-cap pivot checks make the same
strict comparisons on each feasible pivot subinterval. Consequently
the full covering excludes a mixed-marginal minimum with
$H\leqslant0$ outside the uniform binary case.

If the inequality failed, compactness would produce a negative
generalized minimum, contrary to this classification. The diffuse
approximation then gives nonnegativity for all generalized tables.
Any countable table with mixed marginals and $H=0$ is itself a global
minimum in that class and satisfies the same first-variation tests.
In particular it has only finitely many positive rows and columns,
as proved in Proposition~\ref{prop:mp-stationarity}.
The strictness argument leaves the uniform binary case. Together
with the pure-marginal cases, this proves the asserted classification.
\end{proof}

\begin{table}[p]
\centering
\caption{Coverage of the sharp marginal-product comparison.
The heads are sorted, and transposition permits $a\leqslant b$.}
\label{tab:mp-coverage}
\begingroup
\renewcommand{\baselinestretch}{1}\footnotesize
\setlength{\tabcolsep}{4pt}
\renewcommand{\arraystretch}{1.12}
\begin{tabular}{p{2.45in}p{2.25in}p{1.35in}}
\hline
{\raggedright Region and exact domain\par} & {\raggedright Argument or verifier\par} & {\raggedright Counts; depth\par}\\
\hline
{\raggedright Pure marginal: $a=1$ or $b=1$\par} & {\raggedright Proposition~\ref{prop:mp-endpoints}\par} & {\raggedright Analytic; equality\par}\\
{\raggedright Entirely diffuse row: $a=R=0$, or its transpose; other marginal mixed\par} & {\raggedright Proposition~\ref{prop:mp-endpoints}\par} & {\raggedright Analytic; strict\par}\\
{\raggedright Binary: biases $\widehat{a},\widehat{b}\in[0,1]$; negative-correlation
$z\in[0,1]$; positive-correlation half analytic\par} & {\raggedright \texttt{binary-certificate.py}\par} & {\raggedright $15$ positive leaves; $1$ analytic; depth $8$\par}\\
{\raggedright Star: $s,t\geqslant0$, $s+t\leqslant1$;
full-tail $\alpha,\beta\in[0,1]$\par} & {\raggedright \texttt{determinant-}\newline\texttt{certificate.cpp}\par} & {\raggedright acc. $22{,}286{,}338$; omit. $39$; depth $41$\par}\\
{\raggedright Large heads: $a+b\geqslant7/4$, both mixed\par} & {\raggedright Proposition~\ref{prop:mp-large-heads}\par} & {\raggedright Analytic reduction\par}\\
{\raggedright Nearly pure: $b\geqslant71/72$ or transpose; outside large heads,
$a<55/72<49/64$\par} & {\raggedright \texttt{pure-boundary-}\newline\texttt{certificate.py}\par} & {\raggedright $392$ intervals; $2$ tests each; $1024$ cells/test\par}\\
{\raggedright Small head: $a\leqslant1/20$ or $b\leqslant1/20$\par} & {\raggedright Proposition~\ref{prop:mp-small-head}\par} & {\raggedright Analytic; strict\par}\\
{\raggedright Low cap: $a,b\leqslant43/100$;
$u=(R+C)/2\in[0,43/100]$\par} & {\raggedright \texttt{mixing-cap-}\newline\texttt{certificate.py}\par} & {\raggedright $u\leqslant43/200$ analytic; $2$ intervals; depth $1$\par}\\
{\raggedright Coarse low purity: $R+C\leqslant107/200$\par} & {\raggedright Proposition~\ref{prop:mp-coarse-purity}\par} & {\raggedright Analytic; $H>1/5000$\par}\\
{\raggedright Enhanced low purity: $R+C\leqslant29/50$;
ordered grid $0\leqslant i\leqslant j$, $i+j\leqslant148$\par} & {\raggedright \texttt{enhanced-purity-sum-}\newline\texttt{certificate.py}\par} & {\raggedright $5625$ grid boxes; $150$ coefficient intervals\par}\\
{\raggedright Concentrated: $a\leqslant b\leqslant71/72$,
$3/2\leqslant a+b\leqslant7/4$; full-tail $\alpha,\beta\in[0,1]$\par} & {\raggedright \texttt{concentrated-band-}\newline\texttt{certificate.cpp}\par} & {\raggedright acc. $5{,}281{,}920$; omit. $173{,}374$; depth $34$\par}\\
{\raggedright Extended: $a\leqslant b\leqslant71/72$,
$71/50\leqslant a+b\leqslant3/2$; same tail coordinates\par} & {\raggedright \texttt{extended-concentrated-}\newline\texttt{band-certificate.cpp},
including \nolinkurl{pivot-certificate.cpp}\par} & {\raggedright acc. $53{,}113{,}510$; omit. $1{,}653{,}894$; depth $48$\par}\\
{\raggedright Small tails: $A+B\leqslant1/25$;
residual $a\leqslant b$, $a,b\geqslant12/25$,
$a+b\leqslant71/50$, $x\in[(a+b-1)_+,ab]$\par} & {\raggedright \texttt{small-marginal-tails-}\newline\texttt{certificate.cpp}\par} & {\raggedright acc. $11{,}032$; omit. $846$; depth $22$\par}\\
{\raggedright Half cap: heads in $[1/32,1/2]^2$;
full-tail purities and second-atom interpolation parameters in $[0,1]^4$;
preceding exclusions apply\par} & {\raggedright \texttt{half-cap-certificate.cpp}\par} & {\raggedright acc. $7{,}122{,}331$; omit. $355{,}283$; depth $38$\par}\\
{\raggedright Unrestricted: $a,b\in[1/32,511/512]$, $a\leqslant b$;
$u,v,\alpha,\beta\in[0,1]$;
$(1-a)u\leqslant a$, $(1-u)\alpha\leqslant u$ and column counterparts\par} & {\raggedright \texttt{marginal-product-}\newline\texttt{certificate.cpp},
\texttt{outward-quadrature.hpp}, terminal TSV\par} & {\raggedright $48$ runs;
acc. $10{,}781{,}823$; omit. $299{,}876$; depth $34$\par}\\
\hline
\end{tabular}
\par\smallskip
\begin{minipage}{6.10in}
Counts labelled acc. and omit. refer to accepted and excluded/omitted
outer boxes. For the C++ certificates, depth means maximum accepted-leaf depth;
the counters do not separately record depths of omitted leaves.
The binary depth is the Bernstein terminal depth; the low-cap depth
is its one subdivision. The enhanced-purity grid is nonadaptive.
Integration cells are distinct from outer covering boxes.
The pure-boundary count refers to the $392$-interval proof.
The terminal record for the unrestricted check is
\nolinkurl{marginal-product-terminal-counts.tsv}.
\end{minipage}
\endgroup
\end{table}

\end{document}